\documentclass[12pt]{article}
\usepackage{amsthm,amsmath,latexsym,amssymb,amsfonts,amscd}
\usepackage{graphics,lscape,fancyhdr,array,stmaryrd,euscript,wrapfig}
\usepackage{empheq}
\usepackage{verbatim,slashed}
\numberwithin{equation}{section}
\usepackage{hyperref,setspace}
\usepackage[numbers,sort&compress]{natbib}
\usepackage[nottoc]{tocbibind}

\usepackage{tikz-cd}
\usepackage[all]{xy}
\usepackage{tikz}
\usetikzlibrary{arrows.meta}

\usepackage{rotating}

\def\R{{\mathbb{R}}}

\numberwithin{equation}{section}
\newcommand{\pl}{\partial}
\newcommand{\fud}[2]{{}^{#1}{}_{#2}\,}
\newcommand{\fdu}[2]{{}_{#1}{}^{#2}\,}

\newcommand{\bry}{{{\bar{y}}}}
\newcommand{\besubeqs}{\begin{subequations}}
\newcommand{\esubeqs}{\end{subequations}}

\begin{document}
\pagenumbering{gobble}
\hfill
\vskip 0.01\textheight
\begin{center}
{\large\bfseries 
Strong Homotopy Algebras for Chiral Higher Spin Gravity\\ [7pt] via Stokes Theorem}

\vspace{0.4cm}

\vskip 0.03\textheight
\renewcommand{\thefootnote}{\fnsymbol{footnote}}
Alexey \textsc{Sharapov}${}^{a}$\footnote{Also a visiting professor at the Federal University of ABC, Brazil. }, 
Evgeny \textsc{Skvortsov}\footnote{Research Associate of the Fund for Scientific Research -- FNRS, Belgium.}${}^{b}$\footnote{Also on leave from Lebedev Institute of Physics.} \& Richard \textsc {Van Dongen}${}^{b}$
\renewcommand{\thefootnote}{\arabic{footnote}}
\vskip 0.03\textheight

{\em ${}^{a}$Physics Faculty, Tomsk State University, \\Lenin ave. 36, Tomsk 634050, Russia}\\
\vspace*{5pt}
{\em ${}^{b}$ Service de Physique de l'Univers, Champs et Gravitation, \\ Universit\'e de Mons, 20 place du Parc, 7000 Mons, 
Belgium}\\

\end{center}

\vskip 0.02\textheight

\begin{abstract}
Chiral higher spin gravity is defined in terms of a strong homotopy algebra of pre-Calabi--Yau type (noncommutative Poisson structure). All structure maps are given by the integrals over the configuration space of concave polygons and the first two maps are related to the (Shoikhet--Tsygan--)Kontsevich Formality. As with the known formality theorems, we prove the $A_\infty$-relations via Stokes' theorem by constructing a closed form and a configuration space whose boundary components lead to the $A_\infty$-relations. This gives a new way to formulate higher spin gravities and hints at a construct encompassing the known formality theorems.
\end{abstract}

\newpage
\tableofcontents
\newpage

\section{Introduction and summary}
\pagenumbering{arabic}
\setcounter{page}{2}
The main idea behind `higher spins' is to switch on a consistent interaction with or between the fields of spin $s>2$. For massive higher spin fields effective field theories with a single field of spin-$s$ are known to be  possible and find their applications, for instance, in the gravitational wave physics, see e.g. \cite{Cangemi:2023ysz} and references therein. Theories with massless higher-spin fields, called higher spin gravities (HiSGRA) \cite{Bekaert:2022poo}, aim at constructing viable models of quantum gravity.  Each HiSGRA is strongly constrained by a certain higher spin symmetry and incorporates the graviton as a part of the (usually infinite) symmetry multiplet. Quantum gravity models still are not easy to construct along these lines: the masslessness, which simulates some features of the UV-regime already classically and higher spins in the spectrum usually come in tension with the conventional field theory paradigm, e.g. with the requirement of locality \cite{Bekaert:2015tva, Maldacena:2015iua, Sleight:2017pcz, Ponomarev:2017qab}. 

As a result, there are very few  perturbatively local theories with massless higher-spin fields: topological $3d$ models \cite{Blencowe:1988gj, Bergshoeff:1989ns, Campoleoni:2010zq, Henneaux:2010xg, Pope:1989vj, Fradkin:1989xt, Grigoriev:2019xmp, Grigoriev:2020lzu}, conformal higher spin gravity \cite{Segal:2002gd, Tseytlin:2002gz, Bekaert:2010ky, Basile:2022nou}, Chiral Theory \cite{Metsaev:1991mt, Metsaev:1991nb, Ponomarev:2016lrm, Skvortsov:2018jea, Skvortsov:2020wtf} and its contractions \cite{Ponomarev:2017nrr,  Krasnov:2021nsq} (see also \cite{Tran:2021ukl, Tran:2022tft, Adamo:2022lah}). Very close to them is the higher spin IKKT-model, see e.g. \cite{Sperling:2017dts, Tran:2021ukl, Steinacker:2022jjv, Steinacker:2023cuf}, which is also a noncommutative field theory.\footnote{Another direction in the context of holographic theories is to derive the bulk dual by massaging the CFT partition function, see e.g. \cite{deMelloKoch:2018ivk,Aharony:2020omh}, which, however, leads to higher derivative free equations, has a slightly different spectrum and nonlocal interactions, the strong point being in that it does reproduce the CFT correlation functions by construction.} There are also incomplete (formal) theories \cite{Vasiliev:1990cm, Vasiliev:1999ba,Bekaert:2013zya,Bonezzi:2016ttk,Bekaert:2017bpy,Grigoriev:2018wrx} that can only be constructed at the formal level of $L_\infty$-algebras with the associated field equations suffering from nonlocality  \cite{Boulanger:2015ova}.\footnote{It should be noted that while every classical field theory (PDE) is defined by some $L_\infty$-algebra, e.g. \cite{Grigoriev:2019ojp}, not every $L_\infty$-algebra leads to a well-defined theory. Firstly, one problem is in that any $L_\infty$-algebra is defined up to canonical automorphisms and the latter correspond to very non-local field redefinitions, in general, which are not admissible. Secondly, it is not clear how to treat genuinely nonlocal field theories in this language, for a generic HiSGRA a way out would be to define a set of observables in a more algebraic terms at the level of the given $L_\infty$-algebra, e.g. \cite{Sharapov:2020quq}, or to resort to even more general ideas, e.g. \cite{Sezgin:2011hq, Bonezzi:2016ttk,DeFilippi:2019jqq,DeFilippi:2021xon}, that operate with a differential graded Lie algebra of which a given $L_\infty$ is the minimal model.  } The general problem of constructing formal theories, i.e., $L_\infty$-algebras from any higher spin algebra, was solved in \cite{Sharapov:2019vyd}. 

The Chiral Theory was first found in the light-cone gauge in flat space \cite{Metsaev:1991mt, Metsaev:1991nb, Ponomarev:2016lrm} and later covariantized at the level of equations of motion and extended to (anti-)de Sitter space in \cite{Skvortsov:2022syz, Sharapov:2022faa, Sharapov:2022awp, Sharapov:2022wpz, Sharapov:2022nps}. The perturbatively local equations of motion have the form of a nonlinear sigma-model
\begin{align}\label{pdefdaA}
    d\Phi&=Q(\Phi)\,, && d=dx^\mu \, \pl_\mu\,,
\end{align}
where the fields $\Phi$ are maps from the spacetime $Q$-manifold $(\Omega(\mathcal{X}),d)$ (the differential graded algebra of forms on $\mathcal{X})$ to another $Q$-manifold $(\mathcal{M},Q)$, $Q^2=0$. Perturbatively around a stationary point, $Q$ determines a flat $L_\infty$-algebra. Chiral HiSGRA is a happy occasion where the structure maps of the  $L_\infty$-algebra can be fine-tuned to maintain locality of \eqref{pdefdaA}. In this case, the formal approach yields a real field theory.

It was found that the $L_\infty$-algebra underlying Chiral HiSGRA originates from an $A_\infty$-algebra through symmetrization. The $A_\infty$-algebra is a very special one \cite{Sharapov:2022wpz, Sharapov:2022nps}: it is given by the tensor product of a pre-Calabi--Yau algebra of degree $2$ and an associative algebra.  The former can be viewed as a noncommutative counterpart of a Poisson structure. 

It was also known \cite{Sharapov:2017yde, Sharapov:2020quq} that the first two `floors' of the $A_\infty$-algebra are related to the (Shoikhet--Tsygan--)Kontsevich formality theorem \cite{Kontsevich:1997vb, Tsygan, Shoikhet:2000gw}. All vertices of the theory can be represented \cite{Sharapov:2022wpz,Sharapov:2022nps} as sums over certain graphs $\Gamma$, 
\begin{align}\label{mmm}
    m_n(f_1,\ldots, f_n)&=\sum_\Gamma w_{\Gamma}\, \mathcal{W}_\Gamma (f_1\otimes   \cdots \otimes f_n)\,,
\end{align}
where $w_\Gamma$ are certain weights and $\mathcal{W}_\Gamma $ are poly-differential operators acting on the fields $f_i$. The  maps $m_n$ define the vertices in \eqref{pdefdaA} via the symmetrization. The graphs are not much different from the (Shoikhet--)Kontsevich graphs, but the weights are given by the integrals over the configuration space $C_\Gamma$ of compact concave polygons. The graphs can be summed up into simple $\exp$-like generating functions, the Moyal--Weyl product being the trivial example.

In this paper, we prove the $A_\infty$-relations (aka Stasheff's identities) the way it is usually done for the formality theorems \cite{Kontsevich:1997vb,Shoikhet:2000gw}. This gives a further evidence that there might be a bigger formality behind Chiral Theories. The $A_\infty$-relations have the form 
\begin{align}\label{SI}
    \sum_{i,j} \pm m_i (\bullet, \ldots,  m_j(\bullet, \ldots,\bullet),  \ldots, \bullet)=0
\end{align}
and are thus given by the `products' of the vertices. To verify the $A_\infty$-relations, for each number of arguments in (\ref{SI}), we construct a configuration space $C$ and a closed form $\Omega$ on $C$ such that the boundary 
\begin{align*}
    \pl C&= \sum_{\Gamma, \Gamma'} C_{\Gamma} \times C_{\Gamma'}
\end{align*}
reproduces the various products of the configuration spaces of (\ref{mmm})  and $\Omega$ restricted to the boundary gives all summands in (\ref{SI}): 
\begin{align}
   0= \int_C d\Omega&= \int_{\pl C} \Omega  && \Longleftrightarrow && \text{$A_\infty$-relations}\,.
\end{align}

The earlier papers
\cite{Sharapov:2022faa, Sharapov:2022awp,Sharapov:2022wpz,Sharapov:2022nps} rely on homological perturbation theory based on a certain multiplicative resolution of the higher spin algebra. While giving the interaction vertices in some form for this specific case, the resolution does not have any invariant meaning in itself. It is also not clear how to construct such resolutions in general. After certain nontrivial transformations the vertices were found to have a form reminiscent of formality theorems, to which the first two structure maps are directly related. This raises several questions. (i) What are the constructs in the deformation quantization and noncommutative geometry that can explain the vertices of Chiral Theory without invoking ad hoc resolutions? (ii) Is there a more general structure (formality) of which (Shoikhet--Tsygan--)Kontsevich ones are particular cases and that gives the vertices of Chiral Theory? As a step towards the answers we show in this paper that the $A_\infty$-relations can be proved via Stokes theorem. 

The outline of the paper is as follows. Sec. \ref{sec:precalau}  contains some algebraic background on $A_\infty$-, $L_\infty$-, and pre-Calabi--Yau algebras. In Sec. \ref{sec:chiral}, we recall the basics of Chiral HiSGRA. In  Sec. \ref{sec:CD}, we reformulate the vertices of \cite{Sharapov:2022wpz,Sharapov:2022nps} in a more pre-Calabi--Yau friendly way. The proof via Stokes' theorem is confined to Sec. \ref{sec:proof}. Conclusions and discussion can be found in Sec. \ref{sec:conclusions}.

\section{Pre-Calabi--Yau algebras}
\label{sec:precalau}
One can think of a pre-Calabi--Yau (pre-CY) algebra of degree $2$ as a noncommutative analogue of a (formal) Poisson structure \cite{kontsevich2021pre}, \cite{IYUDU202163}, \cite{Kontsevuch:2006jb}.  With the hope that the algebraic structures described below may also be relevant to other HiSGRA-type models and beyond we keep our discussion as general as possible. For their string-theoretic motivations and interpretations, see \cite{Kajiura:2003ax}.

We fix $\mathbb{C}$ as the ground field so that all unadorned  Hom's and $\otimes$ would  be over $\mathbb{C}$.  
Let $V$ be a complex vector space regarded as a $\mathbb{Z}$-graded space concentrated in degree zero. Define the direct sum $W=V[-1]\oplus V^\ast$ where $V^\ast$ is the vector space dual to $V$ and $[-1]$ stands for the degree shift. In other words, $W=W_{1}\oplus W_0$ with $W_{1}=V[-1]$ and $W_0=V^\ast $. We will denote the degree of a homogeneous element $a\in W$ by $|a|$. 
The natural pairing between $V$ and $V^\ast$ gives rise to a symplectic structure of degree-one on $W$:
\begin{equation}\label{ip}
    \langle a, a'\rangle =0=\langle b,b' \rangle \,,\qquad \langle a,b\rangle= a(b)=-\langle b,a\rangle 
\end{equation}
for all $a, a'\in V^\ast$ and $b,b'\in V[-1]$.  Let $T W=\bigoplus_{n> 0} T^nW$ be the (restricted)  tensor algebra of the space $W$ with $T^nW=W^{\otimes n}$.  Besides the tensor degree, the algebra $TW$ inherits the $\mathbb{Z}$-grading from $W$, so that $TW=\bigoplus_{n\geq 0}(TW)_{n}$.  To simplify exposition, we will assume the vector space $V$ to be finite-dimensional. It should be noted, however, that all the constructions below extend to the infinite-dimensional case with appropriate modifications, see \cite{kontsevich2021pre}.

We define the space of $p$-cochains $C^p$ as the dual to the space $T^{p+1}W$, that is,
\begin{equation}
    C^p=(T^{p+1}W)^\ast=\mathrm{Hom}(T^{p+1}W, \mathbb{C})\,.
\end{equation}
The direct product $C^\bullet=\prod_{p\geq 0}C^p$ contains the  subspace of {\it cyclic cochains} $C^\bullet_{\mathrm{cyc}}$  satisfying  the cyclicity condition 
\begin{equation}\label{cyc}
    f(a_0,a_1,\ldots, a_p)=(-1)^{|a_0|(|a_1|+\cdots +|a_{p}|)}f(a_1, \ldots, a_{p}, a_0)
\end{equation}
for all $a_i \in W$ and $p\geq 0$.  
Using the symplectic structure (\ref{ip}), we can write 
\begin{equation}\label{vf}
    f (a_0,\ldots, a_p)= \langle a_0, \hat f(a_{1},\ldots, a_{p})\rangle 
\end{equation}
for some homomorphism $\hat f\in \mathrm{Hom}(T^{p}W, W)$. Since the symplectic structure is nondegenerate, the last relation determines  $\hat f$ unambiguously.

Now we can endow the  space $C^\bullet_{\mathrm{cyc}}[1]$ with the structure of a graded Lie algebra. The commutator of two homogeneous cochains $f\in C^p_{\mathrm{cyc}}[1]$ and $g\in C^{n-p}_{\mathrm{cyc}}[1]$ is given by the {\it necklace bracket}
\begin{equation}\label{nb}
    [f,g] (a_0,\ldots, a_{n})=\sum_{i=0}^n(-1)^{\kappa_i}g(\hat f(a_{i+1},\ldots),\ldots,  a_{i})
\end{equation}
$$
 =\sum_{i=0}^n(-1)^{\kappa_i}  \langle\hat f(a_{i+1},\ldots),\hat g(\ldots,  a_{i})\rangle\,.
$$
Here summation is over all cyclic permutations of $a$'s and the Koszul sign 
is determined by 
$$
\kappa_i={(|a_0|+\cdots +|a_{i}|)(|a_{i+1}|+\cdots+|a_n|)}\,.
$$
By construction, the $n$-cochain $[f,g]$ satisfies the cyclicity condition (\ref{cyc}) and graded skew-symmetry,
\begin{equation}
    [f,g]=(-1)^{\bar f\bar g}[g,f]\,.
\end{equation}
Here $\bar f=|f|-1$ is the degree of $f$ as an element of the Lie algebra $C^\bullet_{\mathrm{cyc}}[1]$.

A pre-CY structure on $W$ is given now by any Maurer--Cartan  element $S$ of the graded Lie algebra $ C^\bullet_{\mathrm{cyc}}[1]$. By definition, 
\begin{equation}\label{MC}
    [S,S]=0\,.
\end{equation}
As an element of $C^\bullet_{\mathrm{cyc}}$, $S$ has degree $-2$.  Therefore, one usually refers to the pair $(W, S)$ as a pre-CY algebra of degree $2$. It also follows from the definition that $\hat S$ defines the structure of a cyclic $A_\infty$-algebra on $W$ with the cyclic structure given by the symplectic form (\ref{ip}). 

By degree reasons and cyclicity, each pre-CY structure of degree $2$ defines and is defined by a sequence of multilinear maps
\begin{equation}\label{smn}
    S_{n,m} : V[-1]\otimes T^{n}V^\ast \otimes V[-1]\otimes T^{m}V^\ast\rightarrow \mathbb{C}\,,
\end{equation}
with $n\geq m$. Then $S=\sum_{n\geq m}S_{n,m}$. We say that the pre-CY algebra is minimal if $S_{0,0}=0$. It is clear that minimal pre-CY algebras correspond to minimal $A_\infty$-algebras, hence the name. Geometrically, one can think of the $A_\infty$-structure $\hat S$ as a homological vector field on a noncommutative manifold $\mathcal{N}$ associated with $W$, see \cite{Kontsevuch:2006jb}.

Each multilinear map $\hat S_{n,m}$ has two components $\hat S_{n,m}^{(0)}$ and $\hat S^{(1)}_{n,m}$ taking values in $W_0$ and $W_{1}$, respectively.  It follows from the MC equation (\ref{MC}) that the map $\hat S_{1,0}^{(1)}$ of the minimal pre-CY algebra gives rise to an associative product on $V$, namely, 
\begin{equation}
    b_1\cdot b_2=\hat S^{(1)}_{1,0}(b_1,b_2)\,,\qquad \forall b_1, b_2\in V[-1]\,.
\end{equation}
Moreover, the map $\hat S^{(0)}_{1,0}$ makes $V^\ast$ into a bimodule over the associative algebra $V$:
\begin{equation}
    b\cdot a =\hat S^{(0)}_{1,0}(b,a)\,,\qquad  a\cdot b =-\hat S^{(0)}_{1,0}(a,b)\,,\qquad \forall b\in V[-1], \quad\forall a\in V^\ast\,.
\end{equation}
Regarding the elements of $W$ as `coordinates' on a noncommutative manifold $\mathcal{N}$ and $C^\bullet_{\mathrm{cyc}}$ as a ring of `functions' on $\mathcal{N}$, one can define a  graded-commutative 
submanifold $\mathcal{C}\subset \mathcal{N}$ by imposing graded-commutativity conditions. Technically, this implies factorization of the tensor algebra $TW$ by the two-sided ideal $I$  generated by the commutators: 
\begin{equation}
    a\otimes b-(-1)^{|a||b|}b\otimes a\,,\qquad \forall a,b\in W\,.
\end{equation}
This results in the symmetric tensor algebra $SW=TW/I$ of the graded vector space $W$. The formula \begin{equation}\label{ff}
     f_{sym}(a_0,a_1, \ldots, a_p)=\frac{1}{p!}\sum_{\sigma\in S_{p}} (-1)^{|\sigma|} f(a_0, a_{\sigma(1)},\ldots, a_{\sigma(p)})\,,
\end{equation}
where $(-1)^{|\sigma|}$ is the Koszul sign associated with the permutations of $a$'s, defines then  a map from 
$C^p_{\mathrm{cyc}}$ to $\mathrm{Hom}(S^pW,\mathbb{C})$. Identifying $\prod_{p>0}\mathrm{Hom}(S^pW,\mathbb{C})$ with the ring of functions on $\mathcal{C}$, one can think of ${f}_{sym}$ as the restriction to $\mathcal{C}$ of the function $f$ on $\mathcal{N}$. Writing (\ref{ff}) as
\begin{equation}
       f_{sym}(a_0,a_1, \ldots, a_p)= \langle a_0, \hat f_{sym}(a_{1},\ldots, a_{p})\rangle\,,
\end{equation}
we define the map $\hat f_{sym}: S^pW\rightarrow W$ by
\begin{equation}\label{ffs}
    \hat f_{sym}(a_1,\ldots, a_p)=\frac{1}{p!}\sum_{\sigma\in S_{p}} (-1)^{|\sigma|} \hat f(a_{\sigma(1)},\ldots, a_{\sigma(p)})\,.
\end{equation}
The multilinear maps (\ref{ff}) and (\ref{ffs}), being totally symmetric, are completely determined through polarization by their restriction on the diagonal, that is, by  the nonlinear maps from $W$ to itself defined as
\begin{equation}\label{fhf}
\begin{array}{l}
    \mathbf{f}(a)=f_{sym}(a,\ldots, a) = f(a,\ldots, a)\,,\\[3mm]
    \hat{\mathbf{f}}(a)=\hat f_{sym}(a,\ldots,a)=\hat f(a,\ldots, a)\,.
    \end{array}
\end{equation}
Geometrically, we can treat $\mathbf{f}$ and $\hat{\mathbf{f}}$, respectively, as a function and a vector field on the (formal) graded-commutative manifold $\mathcal{C}$ associated with $W$. 

It is known that the symmetrization of an $A_\infty$-structure gives an $L_\infty$-algebra structure on the same vector space, see e.g. \cite{Lada_commutators}. In particular, $\hat S_{sym}$ makes $W$ into a cyclic $L_\infty$-algebra, with the cyclic structure being given by the natural pairing (\ref{ip}).  The direct sum decomposition $W=V[-1]\oplus V^\ast$ allows for more refined geometric interpretation, namely, one can think of  $\mathcal{C}$ as the total space of the shifted cotangent bundle $T^\ast[-1]V^\ast$ of the formal  manifold $\mathcal{M}$ associated with the space $V^\ast$. Then each function $\mathbf{f}$, defined by (\ref{ffs}), gives rise to a polyvector field on $\mathcal{M}$. Upon the degree shift, the space of polyvector fields is known to carry the structure of graded Lie algebra w.r.t. the Schouten--Nijenhuis bracket. Let us denote this Lie algebra by $\mathcal{P}$. The assignment $f\mapsto \mathbf{f}$ defines then a homomorphism from $C_{\mathrm{cyc}}^\bullet[1]$ to $\mathcal{P}$, i.e.,
\begin{equation}
    [\mathbf{f},\mathbf{g}]_{SN} =[f,g] (a,\ldots,a)
\end{equation}
Here, the l.h.s.  is given by the SN bracket of the poly-vector fields $\mathbf{f}$ and $\mathbf{g}$, while the r.h.s. is obtained by the restriction on $\mathcal{C}$ of the necklace bracket of their preimages. 

The function $\mathbf{S}$ corresponding to the minimal pre-CY structure $S$ has the form
\begin{equation}\label{minchoiceS}
    \mathbf{S}(a,b)=\sum_{n=1}^\infty \mathbf{S}_n (b,b, a,\ldots, a)=\sum_{n\geq m} S_{n,m}(b, \overbrace{a,\ldots,a}^{n},b,\overbrace{a,\ldots,a}^m)
\end{equation}
for all $a\in V^\ast$ and $b\in V[-1]$. Being quadratic in $b$'s, the function  $\mathbf{S}$ can be considered as a bivector field on $\mathcal{M}$. Moreover, the Maurer--Cartan equation (\ref{MC}) implies that the bivector  $\mathbf{S}$ is Poisson.

We thus see that every pre-CY structure on $W=V[-1]\otimes V^\ast$ gives rise to a Poisson bivector on the formal manifold associated with $V^\ast$. From this perspective,  pre-CY structures of degree $2$ extend the notion of a Poisson structure to noncommutative setting. 

Finally, let us mention that given any $A_\infty$-algebra $\mathbb{A}$ on a graded vector space $W$ and any associative algebra $B$, one can construct a new $A_\infty$-algebra given by the tensor product $\mathbb{A}\otimes B$. Its vector space is the tensor product $W\otimes B$ and the structure maps read
\begin{align}\label{tensorproductmaps}
m_n(a_1\otimes b_1, \ldots, a_n\otimes b_n)&= m_n(a_1, \ldots, a_n) \otimes (b_1\cdots b_n)\,,     
\end{align}
where $a_i\in W$, $b_i\in B$, and $(b_1 \cdots b_n)$ is the product of $b_i$ in $B$. This construction will be used below for Chiral Theory where $\mathbb{A}$ is a pre-CY algebra. 

\section{Chiral HiSGRA: preliminaries}
\label{sec:chiral}
The discussion above has been quite general. Starting from this section we narrow it down to a specific class of models -- Chiral HiSGRA -- where all these structures are realized. The spectrum of Chiral HiSGRA corresponds, roughly speaking, to massless particles of all spins $s=0,1,2,3,\ldots$ each appearing in one copy. There are simple extensions obtained with the help of appropriate matrix factors, e.g. with supersymmetry \cite{Metsaev:2019aig, Tsulaia:2022csz} or Yang--Mills gaugings \cite{Skvortsov:2020wtf} and the truncation to even spins only. A simple way to introduce the dynamical fields is to start with the free action \cite{Krasnov:2021nsq}
\begin{align}\label{niceaction}
    S= \int \Psi^{A(2s)}\wedge e_{AB'}\wedge e\fdu{A}{B'}\wedge \nabla w_{A(2s-2)}\,,
\end{align}
which employs the field variables originating from the twistor approach \cite{Penrose:1965am,Hughston:1979tq,Eastwood:1981jy,Woodhouse:1985id}. Here, $e^{AA'}$ is the vierbein one-form and $\nabla$ is the Lorentz-covariant derivative such that $\nabla e^{AA'}=0$; $\Psi^{A(2s)}$ is a zero-form in $(2s,0)$ of the Lorentz algebra $sl(2,\mathbb{C})$; $w^{A(2s-2)}$ is a one-form in $(2s-2,0)$ of $sl(2,\mathbb{C})$. On any self-dual background ($\nabla^2 \chi^A\equiv0, \forall\chi^A$) the action is invariant under the gauge transformations
\begin{align}\label{lin-gauge}
    \delta w^{A(2s-2)}&= \nabla \xi^{A(2s-2)} +e\fud{A}{C'} \eta^{A(2s-3),C'}\,, & \delta\Psi^{A(2s)}&=0\,,
\end{align}
where the gauge parameters $\xi$'s and $\eta$'s are zero-forms. There is also a massless scalar field $\phi(x)$, which is described in the usual way. The fields $\Psi^{A(2s)}$ and $w^{A(2s-2)}$ describe, by virtue of their equations of motion, free helicity $-s$ and $+s$ particles, $s\geq1$.  

Chiral HiSGRA is a unique completion of the free theory introduced above under the assumptions of locality and Lorentz invariance, provided at least one nontrivial cubic  higher-spin interaction vertex is turned on. The equations of motion for Chiral HiSGRA have the general form
\begin{align} \label{eom}
    d\Phi =Q(\Phi)= \sum_{n \geq 2}^{\infty} l_n(\overbrace{\Phi, \dots, \Phi}^{n}) \,,
\end{align}
where the field content $\Phi(x)=\{\omega(x),C(x)\}$ is given by infinite sets of one- and zero-forms $\omega$ and $C$. By abuse of notation $\Phi=\{\omega,C\}$ are also the coordinates on the associated graded manifold. The homological vector field $Q$ determines an $L_\infty$-algebra with structure maps $l_n$.  The set $\Phi$ contains the dynamical fields, $\Psi^{A(2s)}$, $w^{A(2s-2)}$, $\phi(x)$ as well as auxiliary fields required to write down \eqref{eom}. Both $\omega$ and $C$ can be efficiently grouped together with the help of the generating functions
\begin{align*}
    \omega \equiv \sum_{m+n = \text{even}} \omega_{A(n),A'(m)}\, y^A \dots y^A\, \bry^{A'} \dots \bry^{A'}
\end{align*}
and
\begin{align*}
    C \equiv \sum_{m+n = \text{even}} C_{A(n),A'(m)}\, y^A \dots y^A\, \bry^{A'} \dots \bry^{A'} \,.
\end{align*}
Here, $y^A$ and $\bry^{A'}$ are two-component auxiliary vectors, $A,B,\ldots=1,2$, $A',B',\ldots=1,2$, in the (anti-)fundamental representation of the Lorentz algebra $sl(2,\mathbb{C})$. The physical fields are recovered in the $y$-Taylor expansion for $w(y) = \omega(y,0)$ and $\Psi(y) = C(y,0)$, in particular, $\phi=C(0,0)$. The set of auxiliary fields is the same as in \cite{Vasiliev:1986td,Vasiliev:1988sa} since it is fixed by the free theory.\footnote{To be more precise, Chiral theory has the standard set of degrees of freedom --- massless fields of all spins, and, hence, the zero-forms have to be the same as in \cite{Vasiliev:1986td,Vasiliev:1988sa}. The one-forms are just connections of the higher-spin algebra, which, as a vector space, is the same for all values of $\lambda$ (introduced below). } However, the identification of the dynamical fields and even the free equations themselves are different from \cite{Vasiliev:1986td,Vasiliev:1988sa}, see \cite{Skvortsov:2022syz,Sharapov:2022faa}. It is also useful to omit the restriction $m+n=\text{even}$ and consider all powers in $y$ and $\bry$ for the time being.

As it was already mentioned, the $L_\infty$-algebra underlying Chiral HiSGRA originates as the symmetrization of a certain $A_\infty$-algebra. Let us denote the latter by $\hat{\mathbb{A}}$. Given the degrees of $\omega$ and $C$, for the $L_\infty$-algebra we have two groups of maps: $l_n(\omega,\omega,C,\ldots, C)$ and $l_n(\omega,C,\ldots,C)$. The $A_\infty$ structure maps will be denoted by  $m_n(\bullet, \ldots, \bullet)$. As with the $L_\infty$-algebra, it is convenient to split the $m_n$'s into two sets $\mathcal{V}(C,\ldots,C,\omega, \ldots, \omega, \ldots,C)$ and $\mathcal{U}(C,\ldots ,C,\omega,C,\ldots,C)$ according to the number of arguments $\omega$ of degree $1$ they involve. The symmetrization is automatically implemented upon plugging in the fields $\omega$ and $C$. Nevertheless, it is convenient to keep track of various $A_\infty$ structure maps by the position of the arguments, 
\besubeqs
\begin{align}
    d\omega&= \mathcal{V}(\omega,\omega)+ \mathcal{V}(\omega,\omega,C)+ \mathcal{V}(\omega,C,\omega)+ \mathcal{V}(C,\omega,\omega)+...\,,\\
    dC&= \mathcal{U}(\omega,C)+\mathcal{U}(C,\omega)+...\,.
\end{align}
\esubeqs
For example, $l_2(\omega,C)=\mathcal{U}(\omega,C)+\mathcal{U}(C,\omega)$ and $l_3(\omega,\omega,C)$ has the last three terms in the first line above. 

Let us define Weyl algebra $A_\lambda[y]$ as an associative algebra of polynomial functions $f(y)\in \mathbb{C}[y^A]$ equipped with the Moyal--Weyl star-product 
\begin{align}\label{WMP}
    (f\star g)(y)&= \exp[y^A(\pl^1_A+\pl^2_A) +\lambda \,\epsilon^{AB} \pl^1_A\pl^2_B] \, f(y_1) g(y_2) \Big|_{y_{1,2}=0}\,,
\end{align}
where $\pl_A y_B=\epsilon_{AB}$ and $\lambda$ is a parameter. Here the Levi--Civita symbol is defined by\footnote{We follow the usual Penrose--Rindler rules to raise and lower indices, i.e. $y^A \epsilon_{AB}=y_B$, $\epsilon^{AB}y_B=y^A$, etc. }
\begin{align*}
    \epsilon^{AB}=\begin{pmatrix}
        0 & 1\\
        -1 & 0
    \end{pmatrix} \,.
\end{align*}
For $\lambda=0$ we have just the commutative algebra of $f(y)$ under the point-wise product. For Chiral Theory $\lambda$ is related to the cosmological constant and $\lambda=0$ means it vanishes. We define a pairing $\langle f,g \rangle$ between Weyl algebra element $f\in A_\lambda[y]$ and its dual $g\in A^\ast_\lambda[y]$  as
\begin{align} \label{pairingBB}
    \langle f, g \rangle  = \exp[\epsilon^{AB} \pl^1_A\pl^2_B]f(y_1)g(y_2)\Big|_{y_1=y_2=0}\,.
\end{align}

The $A_\infty$-algebra $\hat{\mathbb{A}}$ is a tensor product $\hat{\mathbb{A}}= \mathbb{A} \otimes B$ of a nontrivial $A_\infty$-algebra $\mathbb{A}$ and an associative algebra $B$ that absorbs the dependence of $\bry$ and of the matrix factors (Yang--Mills gaugings and supersymmetry). More explicitly, $B=A_{1}[\bry]\otimes \mathrm{Mat}_N$. The $A_\infty$-algebra $\mathbb{A}$ is a pre-CY algebra built on $A_\lambda[y]$; in the notation of Sec. \ref{sec:precalau}, $V=\mathbb{C}[y^A]$ is the space of complex polynomials in $y^A$ and $V^\ast=\mathbb{C}[[y^A]]$ is given by formal power series. Thus, the decomposable elements of $\hat{\mathbb{A}}$ have the form 
\besubeqs
\begin{align}
    \omega(y,\bry)&= \omega'(y) \otimes \omega''(\bry) \otimes \omega'''\,, && \omega'\in A_\lambda[y]\,,\quad \omega''\in A_1[\bry]\,,\quad \omega'''\in \mathrm{Mat}_N\,,\\
    C(y,\bry)&= C'(y) \otimes C''(\bry) \otimes C'''\,, && C'\in A^\ast_\lambda[y]\,,\quad C''\in A_1[\bry]\,,\quad C'''\in \mathrm{Mat}_N\,.
\end{align}
\esubeqs
Finally, the structure maps $m_n$ of $\hat{\mathbb{A}}$ are defined by the tensor product rule \eqref{tensorproductmaps}. For example, the bilinear maps $m_2$: $\mathcal{V}(\omega,\omega)$ is just the Moyal--Weyl product (\ref{WMP}) on $A_\lambda[y]$ combined with the product on $B$; $\mathcal{U}(\omega,C)$ and $\mathcal{U}(C,\omega)$ are given by the dual bimodule structure on $A_\lambda^\ast[y]$ and by the product on $B$. The $B$-factor will be omitted below.

\section{Vertices} \label{sec:CD}
As explained in Sec. \ref{sec:precalau}, every pre-CY structure of degree $2$ is defined by a sequence (\ref{smn}) of maps\footnote{It is convenient to define a slightly over-complete set of maps that are related to each other via cyclic symmetry. For example, with the help of the cyclic symmetry we can reach $k=0$ and further impose $m\geq n$, cf. discussion around \eqref{smn}.}
\begin{equation} \label{genericS}
    S_{N}(\alpha_1,\dots,\alpha_{k},a,\alpha_{k+1},\ldots, \alpha_{k+m}, b, \alpha_{k+m+1}, \dots, \alpha_{k+m+n+1})\,,
\end{equation}
where the $\alpha$'s are elements of $V^\ast$, while the arguments $a$ and $b$ live in $V[-1]$ and $k+m+n+1=N$. We recall that the graded space of a pre-CY algebra associated with Chiral HiSGRA is given by ${W}=V[-1]\oplus V^\ast$, where $V=\mathbb{C}[y^A]$ is the space of complex polynomials in $y^A$. Correspondingly, the dual space $V^\ast=\mathbb{C}[[y^A]]$ is given by formal power series in the same formal variables. Note that in this section we ignore other factors such as the dependence on $\bry^{A'}$ and matrix factors since they enter simply via the tensor product. 

\begin{figure}[h]
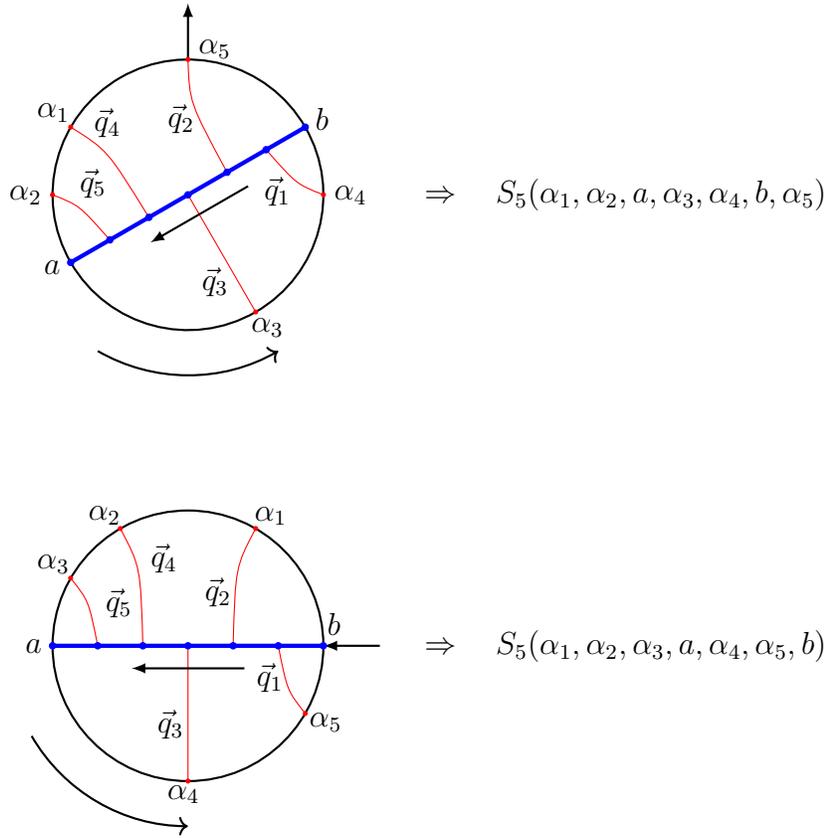

\centering

\caption{Two  disk diagrams that contribute to $S_{5}$.}\label{SD1}
\end{figure}

One can encode each map $S_N$  by disk diagrams as depicted  in Fig. \ref{SD1}. Our disk diagrams  are specific planar graphs with trivalent vertices. By definition, each disk diagram consists of a circle with a diameter, a set of (nonintersecting) lines connecting vertices on the diameter to vertices on the circle and an arrow pointing outwards from one of the $\alpha$'s in the northern semicircle, or towards the vertices $a$ or $b$, i.e. one of the points on the boundary is marked by an arrow and the direction of the arrow depends on whether the argument is in $V^\ast$ or in $V[-1]$. The direction of an arrow is to indicate the symplectic structure \eqref{ip}. No other links or vertices are allowed. As a result, each diagram is characterized by the number of vertices on the northern semicircle of the circle to the left and right of the arrow, $k$ and $n$, respectively, and the number of vertices on the southern semicircle, $m$. To avoid ambiguity we denote the set of all such  diagrams by $\mathcal{O}_{k,m,n}$. The boundary vertices are decorated either by $a$ and $b$ or by the arguments $\alpha$'s. The arguments of $S_N$ are written in the order the vertices appear on the boundary of the disk diagram in the counterclockwise direction. The starting point is taken to be to the left of the arrow. This is called the \textit{boundary ordering}. Moreover, we will always consider the semicircle that one finds when traversing from $a$ to $b$ clockwise to be the northern semicircle and its complement the southern semicircle.

To write down an analytical expression for $S_N$ we also need to decorate the red lines with $2$-vectors $\vec{q}_i = (u_i,v_i)$, one for each line. The label $i$ increases from $b$ to $a$ along the diameter. This is referred to as the \textit{bulk ordering}. The structure map $S_N$ is given then by an integral over a bounded domain $\mathbb{V}_{N-1}\subset \mathbb{R}^{2(N-1)}$ parameterized by the $u$'s and $v$'s. The definition of the integration domain involves the bulk ordering, while the definition of the integrand uses the boundary ordering. Fig. \ref{SD1} shows how labels $\alpha_i$ and vectors $\vec{q}_i$ are assigned to a disk diagram. The straight arrow illustrates the bulk ordering, while the curved arrow displays the boundary ordering.
 
\begin{figure}
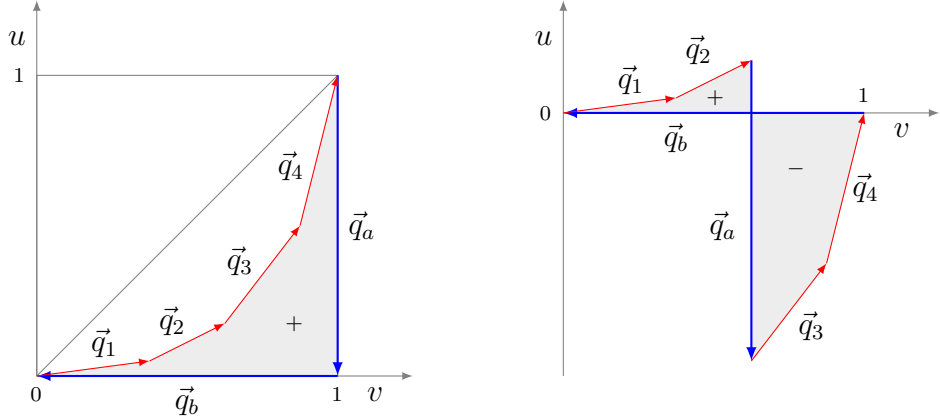

    \centering


\caption{Left panel: a swallowtail associated with the disk diagram in Fig. \ref{SD1}. The number $\frac{1}{2}|\vec{q}_a,\vec{q}_b,\vec{q}_1,\vec{q}_2,\vec{q}_3,\vec{q}_4|$ is  the area enclosed by the swallowtail. Right panel: a self-intersecting polygon  $(\vec{q}_1,\vec{q}_2,\vec{q}_a,\vec{q}_3,\vec{q}_4,\vec{q}_b)$ and its oriented area $\frac{1}{2}|\vec{q}_1,\vec{q}_2,\vec{q}_a,\vec{q}_3,\vec{q}_4,\vec{q}_b|$.}
\label{ST}
\end{figure}

\paragraph{Integration domains.} To associate an integration domain to a decorated disk diagram of $\mathcal{O}_{k,m,n}$, we treat the pairs $(u_i,v_i)$ as the coordinates of  vectors $\vec{q}_i$ on  an affine plane. We also add the pair of vectors $\vec{q}_a=(-1,0)$ and $\vec{q}_{b}=(0,-1)$ associated with the boundary vertices $a$ and $b$. The vectors are assumed to form a closed polygon chain, that is,
\begin{align}
    \vec{q}_1+\cdots +\vec{q}_N+ \vec{q}_a+\vec{q}_b=0\,,
\end{align}
or equivalently, 
\begin{align} \label{Dom1}
    u_1 + \dots + u_N = 1 =v_1 + \dots + v_N  \,.
\end{align}
We also require that 
\begin{equation}\label{Dom2}
    u_i\geq 0 \quad \mbox{and}\quad v_i\geq 0\,,\qquad i=1,\ldots,N\,.
\end{equation}
Suppose  that, moving along the diameter from $b$ to $a$ (bulk ordering) we pass through the $i$-th vertex at time $t_i=u_i/v_i$. Then the chronological ordering implies the following chain of inequalities: 
\begin{align} \label{Dom3}
   0\leq  \frac{u_1}{v_1} \leq \frac{u_2}{v_2}\leq \dots \leq \frac{u_N}{v_N} \leq \infty\,.
\end{align}
This, in turn, implies\footnote{To illustrate this, assume $u_1 > v_1$. Then, \eqref{Dom3} implies that $u_i > v_i$ for $i=1,\dots,N$, while the closure constraint \eqref{Dom1} requires $\sum_{i=1}^{N} u_i = \sum_{i=1}^{N} v_i = 1$, which leads to a contradiction. The same logic can be applied to the last time in the chain.} 
\begin{align} \label{Dom4}
    u_1 \leq v_1 \,, \quad v_N \leq u_N\,.
\end{align}
Together, Eqs. (\ref{Dom1}), (\ref{Dom2}) (\ref{Dom3}), and (\ref{Dom4}) define a bounded domain $\mathbb{V}_{N-1} \subset\mathbb{R}^{2(N-1)}$ for $N\geq 1$. In the special case that $N=0$, the domain is empty and there is no integration taking place. The integration domain admits a nice  visualization on the plane. The vectors $\vec{q}_i$ form  a maximally concave polygon inscribed into a unit square, see the left panel of Fig. \ref{ST}. The two acute angles and the right angle correspond to the fixed unit vectors $\vec{q}_a$ and $\vec{q}_{b}$; the other interior angles  of the polygon are concave.  In \cite{Sharapov:2022wpz, Sharapov:2022nps}, such polygons were called `swallowtails'. We note that $\mathbb{V}_{N-1}$ admits a $\mathbb{Z}_2$-symmetry by swapping the variables $u_i \leftrightarrow v_i$ and reversing their labels.

\paragraph{Integrands.} The integrand associated with a disk diagram of $\mathcal{O}_{k,m,n}$ is given by a family of polydifferential operators with $N+2$ arguments since ${W}=V[-1]\oplus V^\ast$, where $V=\mathbb{C}[y^A]$ and $V^\ast=\mathbb{C}[[y^A]]$. The family is parameterized by the points of the integration domain $\mathbb{V}_{N-1}$. The poly-differential operators in question are  defined through compositions of elementary endomorphisms of $TW$: 
\begin{equation}\label{pij}
\begin{array}{c}
    p_{ij}(w_1\otimes \cdots \otimes w_i\otimes \cdots\otimes  w_j\otimes \cdots \otimes w_n)\\[3mm]
  \displaystyle  = \epsilon^{AB}\, w_1\otimes \cdots \otimes \frac{\partial w_i}{\partial y^A}\otimes \cdots\otimes  \frac{\partial w_j}{\partial y^B}\otimes \cdots \otimes w_n   
    \end{array}
\end{equation}
for all $w_i\in W$ and $n>1$. 
By definition, $p_{ij}=-p_{ji}$. The operators $p_{ij}$ act by partial derivatives on the  tensor factors labelled by $i$ and $j$.  Most often we will use the individual notation for the elements of $V[-1]$ and write e.g.
\begin{equation}
    p_{a, 2}(a\otimes \alpha_1\otimes b\otimes \alpha_2)=\epsilon^{AB} \frac{\partial a}{\partial y^A}\otimes \alpha_1\otimes b\otimes \frac{\partial\alpha_2}{\partial y^B}\,,
\end{equation}
where $a,b \in V[-1]$ and $\alpha_1,\alpha_2 \in V^\ast$. Finally, we let $\mu: TW\rightarrow \mathbb{C}$ denote the multilinear map that takes each homogeneous element $w_1\otimes w_2\otimes \cdots\otimes w_n$ to the complex number $w_1(0)w_2(0)\cdots w_n(0)$, where $w(0)$ is the constant term of the power series $w(y)$. The poly-differential operators associated with disk diagrams of $\mathcal{O}_{k,m,n}$ have the following general structure:
\begin{align*}
    \begin{aligned}
        F_{uv}(&\alpha_1,\dots,\alpha_k,a,\alpha_{k+1},\ldots, \alpha_{k+m},b,\alpha_{k+m+1},\dots,\alpha_{k+m+n+1})
        =\\
        &\mu\circ I(p_{a,b}, p_{a,i}, p_{b,j})(\alpha_1 \otimes \dots \otimes \alpha_n \otimes a\otimes \alpha_{n+1}\otimes\cdots\otimes \alpha_{n+m}\otimes b \otimes \alpha_{k+m+1} \otimes \alpha_{k+m+n+1})\,.
    \end{aligned}
\end{align*}
Here $I$ is a power series in the $p$'s, $u$'s, and $v$'s. Since the operators (\ref{pij}) pairwise commute, there is no ordering ambiguity.  
Notice that the operator $I$ does not involve elementary operators $p_{ij}$ that hit the pairs of  elements $\alpha_i,\alpha_j\in V^\ast$. 
Otherwise, the operator $I$ would be ill-defined on the elements of $W$.\footnote{It may be ill-defined if $I$ depends on $p_{i,j}$ because $I$ is a power series and elements from $V^\ast$ are series as well. That $I$ does not depend on $p_{i,j}$ implies the perturbative locality of Chiral HiSGRA.} It remains to describe the construction of the operator $I$ by a given disk diagram. The construction goes as follows. 

It is convenient to organize all the vectors in a single  $2\times (N+2)$ array, $Q_D$, for a disk diagram $D\in\mathcal{O}_{k,m,n}$. The order of column vectors in the array corresponds to the boundary ordering. This reads
\begin{align} \label{QD}
    Q_D=(\vec{q}_{i_1}, \ldots, \vec{q}_{i_{N+2}}) \,,
\end{align}
with the labels on the vectors assigned according to the boundary ordering. We will often provide the matrix $Q$ that corresponds to the {\it canonical ordering} associated with $S_{N}(a,b,\alpha_1,\dots,\alpha_N)$, given by
\begin{align}\label{Q}
    Q = \begin{pmatrix}
         -1& 0& u_1 & \dots & u_N \\
         0 & -1& v_1 & \dots & v_N
    \end{pmatrix}=(\vec{q}_a, \vec{q}_b, \vec{q}_1,\ldots, \vec{q}_{N})\,.
\end{align}
From this, one can construct the matrix $Q_D$ for a particular diagram.

Together with $Q_D$, we assign the $2\times (N+2)$ array
\begin{equation}\label{QP}
    P_D=( \vec{r}_{1},\ldots , \vec{r}_{k},\vec{r}_a,\vec{r}_{k+1},\dots,\vec{r}_{k+m},\vec{r}_b,\vec{r}_{k+m+1},\dots,\vec{r}_{k+m+n+1})
\end{equation}
to the diagram $D \in \mathcal{O}_{k,m,n}$. Here the $r$-vectors are assigned according to the boundary ordering. Whenever one encounters the vertices $a$ and $b$, one fills up $P_D$ with $\vec{r}_a=(0,0)$ and $\vec{r}_b=(0,0)$, respectively. For the $\alpha$'s one inserts the vectors $\vec{r}_i=(p_{a,i},p_{b,i})$, with $i$ increasing counterclockwise. Finally, to each matrix $Q_D$ we assign a quadratic polynomial in the $u$'s and $v$'s defined by the formula
\begin{equation}
    |Q_D|=\sum_{k<l}\det (\vec{q}_{i_k}, \vec{q}_{i_l})\,.
\end{equation}
In other words, $|Q_D|$ is the sum of all $2\times 2$ minors of the matrix $Q_D$. Like the integration domain $\mathbb{V}_{N-1}$, the number $|Q_D|$ admits a simple visualisation. The column vectors of $Q_D$, ordered from left to right, define a closed polygon chain on the affine plane. In general, the polygon is self-intersecting and splits into two regions with opposite orientations, 
see Fig. \ref{ST}. Then $|Q_D|$ is nothing but two times the oriented area of the polygon.

With the data above, we define the operator $I_D$ as 
\begin{align}\label{Iddd}
    I_D=s_{D} (p_{a,b})^{N-1} \exp\Big(\text{Tr}(P_DQ_D^T) + \lambda |Q_D| p_{a,b}\Big)\,,
\end{align}
where $\lambda$ a free parameter, which is related to the cosmological constant; $s_{D}=(-1)^m$, with $m$ the number of elements $\alpha_i$ found in the southern semicircle of the corresponding disk diagram. A structure map $S(\ldots)$ of the pre-CY algebra with a given ordering of the arguments is defined by summing operators $I_D$ over all disk diagrams with this order of the arguments:
\begin{align}
    S_{N}(\alpha_1,\dots,\alpha_{k},a,\alpha_{k+1},\ldots, \alpha_{k+m}, b, \alpha_{k+m+1}, \dots, \alpha_{k+m+n+1})&= \sum_{D\in \mathcal{O}_{k,m,n}} I_D\,,
\end{align}
where the arguments $a$, $b$ and $\alpha_i$ are implicit on the right. The argument to the left of the one marked with an arrow becomes the first argument of $S$. 

\begin{figure}[h]
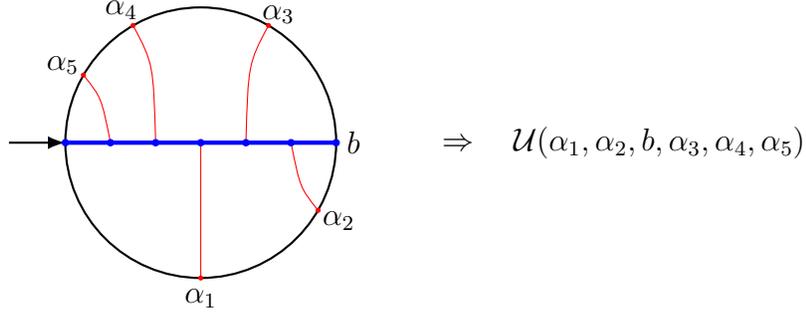

\centering
%
\caption{A disk diagram that contributes to  $\mathcal{U}$. }\label{SD2}
\end{figure}

\begin{figure}[h!]
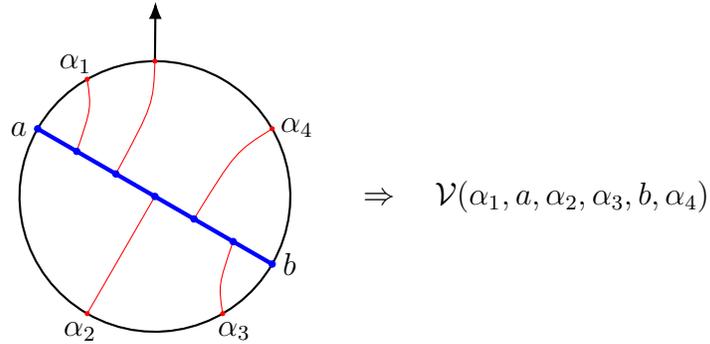

\centering
%
\caption{A disk diagram for $\mathcal{V}$. }\label{SD3}
\end{figure}

\paragraph{Vertices.}
In order to write the equations of motion  we  need the expression for the components of the structure maps $\hat{S}=(\mathcal{V}, \mathcal{U})$ defined by Eq. (\ref{vf}), using the natural pairing
\begin{align*}
    \langle\bullet,\bullet\rangle: V[-1] \otimes V^\ast \rightarrow \mathbb{C} \,,
\end{align*}
which is given by
\begin{align} \label{pairing}
    \langle f(y_1) , g(y_2) \rangle = - \langle g(y_2) , f(y_1) \rangle = \exp[p_{1,2}]f(y_1)g(y_2)|_{y_1=y_2=0}\,,
\end{align}
for $f(y_1) \in V[-1]$ and $g(y_2) \in V^\ast$. The argument on the boundary marked with an arrow is the one to be removed via the nondegenerate symplectic structure. We then find
\begin{align} \label{V}
    \begin{aligned}
        &-S_{N-1}(\alpha_1,\dots,\alpha_m,b,\alpha_{m+1},\ldots, \alpha_{m+n},a)=\\
        &=- \langle  \mathcal{U}(\alpha_{1},\ldots,\alpha_{m},b,\alpha_{m+1},\dots,\alpha_{m+n}),a\rangle = \langle a,\mathcal{U}(\alpha_{1},\ldots,\alpha_{m},b,\alpha_{m+1},\dots,\alpha_{m+n})\rangle\,,\\
        &S_{N-1}(\alpha_1,\dots,\alpha_k,a,\alpha_{k+1},\ldots, \alpha_{k+m},b)=\\
        &=\langle  \mathcal{U}(\alpha_{1},\ldots,\alpha_{k},a,\alpha_{k+1},\dots,\alpha_{k+m}),b\rangle=-\langle  b,\mathcal{U}(\alpha_{1},\ldots,\alpha_{k},a,\alpha_{k+1},\dots,\alpha_{k+m})\rangle\,,\\
        &S_{N}(\alpha_1,\dots,\alpha_k,a,\alpha_{n+1},\ldots, \alpha_{k+m},b,\alpha_{k+m+1},\dots,\alpha_{k+m+n+1})=\\
        &\langle\mathcal{V}(\alpha_1\ldots,\alpha_{k}, a,\alpha_{k+1},\ldots,\alpha_{k+m},b, \alpha_{k+m+1},\ldots,\alpha_{k+m+n}), \alpha_{k+m+n+1}\rangle\,,
    \end{aligned}
\end{align}
where we suppressed the $y$-dependence of the vertices. Notice that the minus sign was added in the first line, as to compensate for the swapping of $a$ and $b$. Using these relations, the $\mathcal{U}$-vertex and $\mathcal{V}$-vertex can be extracted by replacing
\begin{align*}
    \begin{aligned}
        p_a &\rightarrow p_0  \text{\quad or\quad}  p_b \rightarrow p_0 &\text{and} &&p_{k+m+n+1} &\rightarrow -p_0 \,, 
    \end{aligned}
\end{align*}
respectively, where $p_0^A\equiv y^A$ is the output argument. This yields
\begin{align} \label{vertices}
    \begin{aligned}
        &\mathcal{U}(\alpha_{1}\ldots,\alpha_{m}, b,\alpha_{m+1},\ldots,\alpha_{m+n}) \ni s_{D} \int_{\mathbb{V}_{m+n-1}} p_{0,b}^{m+n-1} \exp\Big(\text{Tr}(P_DQ_D^T) + \lambda |Q_D| p_{0,b}\Big) \,,\\
        &\mathcal{U}(\alpha_{1}\ldots,\alpha_{m}, a,\alpha_{m+1},\ldots,\alpha_{m+n}) \ni (-1)^{m+n}s_{D} \int_{\mathbb{V}_{m+n-1}} p_{0,a}^{m+n-1} \exp\Big(\text{Tr}(P_DQ_D^T) + \lambda |Q_D| p_{0,a}\Big) \,,\\
        &\mathcal{V}(\alpha_{1}\ldots,\alpha_{k}, a,\alpha_{k+1},\ldots,\alpha_{k+m},b, \alpha_{k+m+1},\ldots,\alpha_{k+m+n}) \ni\\
        &\qquad\qquad\qquad\qquad\qquad\qquad\qquad\ni s_{D} \int_{\mathbb{V}_{k+m+n}}p_{a,b}^{k+m+n} \exp\Big(\text{Tr}(P_DQ_D^T) + \lambda |Q_D| p_{a,b}\Big)\,,
    \end{aligned}
\end{align}
where we relabeled the $\alpha$'s in the $\mathcal{U}$-vertices for convenience. Note that we used $\ni$ instead of $=$ to indicate that what is on the r.h.s. is a specific contribution to a given vertex, while the vertex is a sum over all contributions with the same order of the arguments. The matrices $P_D$ for the $\mathcal{U}$- and $\mathcal{V}$-vertices are constructed according to the boundary ordering of the corresponding diagram from the $P$ matrices
\begin{align*}
     \begin{pmatrix}
            0 & 0 & p_{0,1} & \dots & p_{0,n+m} \\
            0 & 0 & p_{b,1} & \dots & p_{b,n+m}
        \end{pmatrix} \quad\text{and}\quad \begin{pmatrix}
            0 & 0 & p_{a,1} & \dots & p_{a,k+n+m} & p_{0,a} \\
            0 & 0 & p_{b,1} & \dots & p_{b,k+n+m} & p_{0,b}
        \end{pmatrix} \,,
\end{align*}
respectively, and $Q_D$ constructed from
\begin{align*}
    Q= (\vec{q}_a,\vec{q}_b,\vec{q}_1,\dots,\vec{q}_N)\,,
\end{align*} 
for $N=n+m$ and $N=k+n+m+1$, respectively. Notice that we provided two ways to extract a $\mathcal{U}$-vertex. Obviously, they should be identical, while at first sight they seem different. For instance, the matrix $Q_D$ is constructed differently in both cases, as the arrow is placed in a different position and therefore they have a different boundary ordering. However, if we find two diagrams that produce the same $\mathcal{U}$-vertex, while using the different methods, it is easy to see that they are just the reversed versions of each other. After renaming $b\rightarrow a$, we see that the matrix $P_D$ only differs by swapping the rows. The $\mathbb{Z}_2$-transformation on the domain allows one to exactly identify both realizations, as this is equivalent to swapping the rows in $P_D$ and reversing the order of the entries in $Q_D$. However, when $N=1$, there is no integration domain and this identification fails. Therefore, the vertices $\mathcal{U}(a,\alpha)$ and $\mathcal{U}(\alpha,a)$ are in fact different, see the examples below. This procedure of extracting vertices can also be represented diagrammatically as in Figs. \ref{SD2} and \ref{SD3}, where an element $\alpha$ or $a$, corresponding to the vertex the arrow is attached to, is removed. We refer to this arrow as the \textit{output arrow}, as it is now related to the output variables $p_0=y$. Although not necessary for most considerations, we will implicitly assume that the closure constraint is solved for the variables assigned to the output arrow. Note that while $S_N$ is a scalar, the $\mathcal{U}$- and $\mathcal{V}$-vertices are valued in $V^\ast$ and $V[-1]$, respectively.

\textit{Remark.} The vertices have an interesting property. If we symmetrize the vertices, i.e. we ignore additional tensor factors responsible for $\bry^{A'}$-dependence and for matrix extensions, and bring them to the same ordering, they satisfy 
\begin{align*}
    \begin{aligned}
        \sum_{k+m+n=N} \mathcal{V}(\alpha_{1}\ldots,\alpha_{k}, b,\alpha_{k+1},\ldots,\alpha_{k+m},a, \alpha_{k+m+1},\ldots,\alpha_{k+m+n}) &= 0 \,,\\
        \sum_{m+n=N} \mathcal{U}(\alpha_{1}\ldots,\alpha_{m}, b,\alpha_{m+1},\ldots,\alpha_{m+n}) &= 0\,,
    \end{aligned}
\end{align*}
for $\lambda = 0$. Here, the sum is over all vertices that take the same total number of arguments. In other words, if the pre-CY algebra of this section is $\mathbb{A}$ and $B$ is any associative commutative algebra then for $\lambda=0$ the $L_\infty$-algebra obtained from $\mathbb{A}\otimes B$ via the symmetrization map is trivial (all maps vanish). 

It is easy to see why this is the case for $\mathcal{V}$-vertices: after symmetrizing, the expression for zero cosmological constant changes only by a sign when flipping a red line from the northern to the southern hemisphere and vice versa. As each red line connected to an element of $V^\ast$ can be flipped, there are as many positive as negative equal contributions, which proves the above relation. Since the $\mathcal{U}$-vertices can be derived from the $\mathcal{V}$-vertices through \eqref{V} using the cyclic property of $S$, the relation is easily seen to hold for them too.

\paragraph{Examples.} Let us illustrate the general recipe above with a few examples. We start with some lowest order vertices that will be used in Sec. \ref{sec:example} to prove the $A_\infty$-relations with zero or one $\alpha$. Fig. \ref{ex1} shows the disk diagrams contributing to the expressions
\begin{align} \label{quadV}
    \begin{aligned}
    \mathcal{V}(a,b) &= \exp[p_{0,a} + p_{0,b} + \lambda p_{a,b}] \,, \\
    \mathcal{U}(b,\alpha) &= \exp[p_{0,1} + p_{b,1} + \lambda p_{0,b} ] \,, \\
    \mathcal{U}(\alpha,a) &= - \exp[p_{a,1} + p_{0,1} - \lambda p_{0,a}] \,.
    \end{aligned}
\end{align}
In order to write these vertices as in \eqref{vertices}, one associates to them the matrices
\begin{align*}
    \begin{aligned}
        Q_D &= 
 \,,
    \end{aligned}
\end{align*}
respectively. The sign in the formula for $\mathcal{U}(\alpha,a)$ consists of two factors: $(-)^{m+n}=-1$, where $m=1$, $n=0$, see the second line of \eqref{vertices}; $s_D=1$, which is the number of red lines in the southern semicircle (none). 

\begin{figure}[h]
\centering
%
\caption{Disk diagrams contributing to $\mathcal{V}(a,b)$, $\mathcal{U}(b,\alpha)$ and $\mathcal{U}(\alpha,a)$, from left to right.}\label{ex1}
\end{figure}
The disk diagram relevant for the $\mathcal{V}$-vertices at the next order are given in Fig. \ref{ex2} and correspond to the expressions
\begin{align}\label{cubicV}
    \begin{aligned}
        \mathcal{V}(a,b,\alpha) &= \int_{\mathbb{V}_{1}} p_{a,b}  \exp[u_2 p_{0,a} + v_2 p_{0,b} + u_1 p_{a,1} +v_1 p_{b,1} +\lambda A_1 p_{a,b}] \,, \\
        \mathcal{V}(a,\alpha,b) &= -  \int_{\mathbb{V}_{1}} p_{a,b} \exp[u_2 p_{0,a} + v_2 p_{0,b} + u_1 p_{a,1} +v_1 p_{b,1} +\lambda A_2 p_{a,b}] + \\ 
        &-\int_{\mathbb{V}_{1}} p_{a,b} \exp[u_1 p_{0,a} + v_1 p_{0,b} + u_2 p_{a,1} +v_2 p_{b,1} +\lambda A_3 p_{a,b}]\,,\\
        \mathcal{V}(\alpha,a,b) &=\int_{\mathbb{V}_{1}} p_{a,b} \exp[u_1 p_{0,a} + v_1 p_{0,b} + u_2 p_{a,1} +v_2 p_{b,1} +\lambda A_4 p_{a,b}] \,,
    \end{aligned}
\end{align}
with
\begin{align*}
    \begin{aligned}
        A_1 =& 1+u_1+u_2-v_1-v_2+u_1v_2-u_2v_1 \,,& A_2 =& 1-u_1+u_2-v_1-v_2+u_1v_2-u_2v_1 \,,\\
        A_3 =& 1+u_1-u_2-v_1-v_2-u_1v_2+u_2v_1 \,, & A_4 =&1+u_1-u_2-v_1+v_2-u_1v_2+u_2v_1 \,.
    \end{aligned}
\end{align*}
To these vertices one associates the matrices
\begin{align*}
    \begin{aligned}
        Q_D &= \begin{pmatrix}
            -1 & 0 & u_1 & u_2 \\
            0 & -1 & v_1 & v_2
        \end{pmatrix} \,, & P_D & = \begin{pmatrix}
            0 & 0 & p_{a,1} & p_{0,a} \\
            0 & 0 & p_{b,1} & p_{0,b}
        \end{pmatrix} \,,\\
        Q_D &= \begin{pmatrix}
            -1 & u_1 & 0 & u_2 \\
            0 & v_1 & -1 & v_2
        \end{pmatrix} \,, & P_D & = \begin{pmatrix}
            0 & p_{a,1} & 0 & p_{0,a} \\
            0 & p_{b,1} & 0 & p_{0,b}
        \end{pmatrix} \,, \\
        Q_D &= \begin{pmatrix}
            -1 & u_2 & 0 & u_1 \\
            0 & v_2 & -1 & v_1
        \end{pmatrix} \,, & P_D & = \begin{pmatrix}
            0 & p_{a,1} & 0 & p_{0,a} \\
            0 & p_{b,1} & 0 & p_{0,b}
        \end{pmatrix} \,,\\
        Q_D &= \begin{pmatrix}
            u_2 & -1 & 0 & u_1 \\
            v_2 & 0 & -1 & v_1
        \end{pmatrix} \,, & P_D & = \begin{pmatrix}
            p_{a,1} & 0 & 0 & p_{0,a} \\
            p_{b,1} & 0 & 0 & p_{0,b}
        \end{pmatrix} \,,
    \end{aligned}
\end{align*}
respectively. Here, the second and third line contribute to $\mathcal{V}(a,\alpha,b)$. The integration domain, $\mathbb{V}_{1}$, is the $2$-simplex, which is described by
\begin{align*} 
    \begin{aligned}
        0 &\leq u_1,u_2,v_1,v_2\leq 1 \,, & 0 &\leq \frac{u_1}{v_1} \leq \frac{u_2}{v_2} \leq \infty \,, & u_1+u_2=1&=v_1+v_2\,.
    \end{aligned}
\end{align*}
The `hidden constraints' are
\begin{align} \label{2-simplex2}
    0&\leq u_1\leq v_1\leq 1 \,, & 0 &\leq v_2 \leq u_2 \leq 1
\end{align}
and can be equivalently used to describe the domain $\mathbb{V}_1$.
\begin{figure}[h]
\centering

\caption{Disk diagrams contributing to $\mathcal{V}(a,b,\alpha)$, $\mathcal{V}(a,\alpha,b)$ and $\mathcal{V}(\alpha,a,b)$, from left to right.}\label{ex2}
\end{figure}

Figs. \ref{SD2} and \ref{SD3} show some more involved examples, yielding the expressions 
\begin{align*}
    \mathcal{U}(\alpha_1,\alpha_2,b,\alpha_3, \alpha_4,\alpha_5) \ni \int_{\mathbb{V}_4} p_{0,b}^4 \exp[\text{Tr}[P_DQ_D^T]+\lambda p_{0,b}|Q_D|] \,,
\end{align*}
with
\begin{align*}
    \begin{aligned}
        P_D &= %
\,.
    \end{aligned}
\end{align*}

The integration domain $\mathbb{V}_4$ is described by
\begin{align*}
    \begin{aligned}
        0&\leq u_i, v_i \leq 1 \,, & 0&\leq\frac{u_1}{v_1}\leq\frac{u_2}{v_2}\leq\frac{u_3}{v_3}\leq\frac{u_4}{v_4}\leq 1\,, & \sum_{i=1}^4 u_i &= \sum_{i=1}^4 v_i = 1\,.
    \end{aligned}
\end{align*}


\begin{figure}[h!]
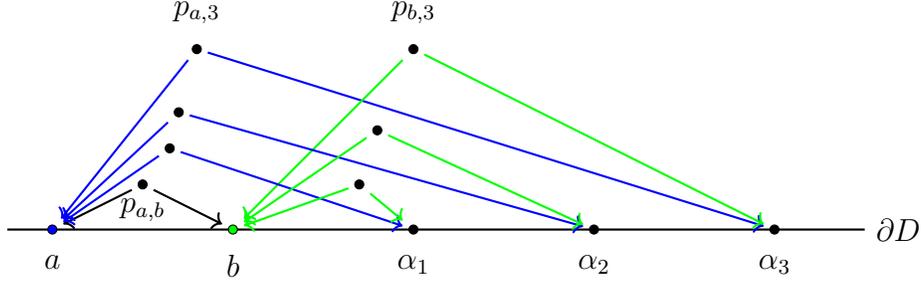

\centering

\caption{A typical Kontsevich-like graph contributing to $S(a,b,\alpha_1,\alpha_2,\alpha_3)$. $D$ is the upper half-plane and $\pl D$ is its boundary.}\label{formality}
\end{figure}

\paragraph{A more general formality?} The above construction of interaction vertices is a piece of clear evidence that (Shoikhet--Tsygan--)Kontsevich's formality can be extended further. We could only see a small piece of this hypothetical extension because our Poisson structure $\epsilon^{AB}$ is constant, nondegenerate and two-dimensional. Therefore, genuine bulk vertices of Kontsevich-like graphs are absent and all vertices have legs on the boundary. In addition, the graphs can be resumed into simple $\exp$-like generating functions $\mathcal{V}$ and $\mathcal{U}$ defined above. A generic $A_\infty$-map $m_n$, i.e. $\mathcal{V}$ or $\mathcal{U}$, can be Taylor-expanded to reveal 
\begin{align}
    m_n(f_1,\ldots,f_n)&=\sum_\Gamma w_{\Gamma}\, \mathcal{W}_\Gamma (f_1\otimes   \cdots \otimes f_n)\,, \qquad f_i\in W\,,
\end{align}
where the sum is over certain graphs $\Gamma$, $w_\Gamma$ are weights associated to $\Gamma$ and $\mathcal{W}_\Gamma$ are certain poly-differential operators (Taylor coefficients of various $I_D$, \eqref{Iddd}). Similar to the Moyal--Weyl case, the graphs $\Gamma$ are built from simple 'wedges' that represent $p_{\bullet,\bullet}$, see Fig. \ref{formality}.
What is different from the Moyal--Weyl case are the weights that are given by the integrals over the configuration space of concave polygons. The integrands are polynomials in $u_i$ and $v_i$. Also, there are no contractions between $\alpha$'s.

\section{A proof via Stokes' theorem}
\label{sec:proof}

With the tools introduced in the previous section, the master equation (\ref{MC}) can be depicted as in Fig. \ref{SD4}. Here, the blue (red) vertices represent the blue (red) lines from the previous diagrams. The summation symbol accounts for collecting contributions from cyclic permutations of both disk diagrams, where we only allow a red and blue vertex to be connected to each other by the arrow between the disks. It also sums over the number of elements of red vertices in the disk diagrams.
\begin{figure}[h]
\centering
\begin{tikzpicture}[scale=0.3]
\draw[thick](0,0) circle (4);
\draw[thick](10,0) circle (4);
\draw [thick, -Latex] (6,0) -- (4,0);
\coordinate [label=left: { $[S,S]=0\quad \Leftrightarrow\qquad\quad $}] (B) at (-6.5,0);
 \draw (-6.5,0) node[gray, scale=5]{$\Sigma$\;} ;
 \coordinate [label=below: { $\mathcal{U}_{n}$}] (B) at (0,1);
  \coordinate [label=below: { $\mathcal{V}_{m}$}] (B) at (10,1);
   \coordinate [label=below: { ${}$}] (B) at (-7.5,-2.5);
   \draw[thick, ->] (10,5) arc (90:10:5);  
   \draw[thick, ->] (0,5) arc (90:10:5);  
    \filldraw [red] (90:4)+(10,0) circle (3pt);     
  \filldraw [red] (30:4)+(10,0) circle (3pt);  
   \filldraw [blue] (60:4)+(10,0) circle (5pt);      
   \filldraw [blue] (-120:4)+(10,0) circle (5pt);     
  \filldraw [red] (0:4)+(10,0) circle (3pt);    
   \filldraw [red] (-30:4)+(10,0) circle (3pt);    
    \filldraw [red] (-75:4)+(10,0) circle (3pt);     
\filldraw [red] (-150:4)+(10,0) circle (3pt);       
    \filldraw [red] (135:4)+(10,0) circle (3pt); 
    \filldraw [red] (180:4)+(10,0) circle (3pt); 
      \filldraw [red] (90:4) circle (3pt);     
  \filldraw [red] (30:4) circle (3pt);  
   \filldraw [red] (60:4) circle (3pt);      
   \filldraw [red] (-120:4) circle (3pt);     
  \filldraw [blue] (0:4)circle (5pt);    
   \filldraw [red] (-30:4) circle (3pt);    
    \filldraw [red] (-75:4) circle (3pt);     
\filldraw [red] (-150:4) circle (3pt);       
    \filldraw [red] (135:4) circle (3pt); 
    \filldraw [blue] (180:4) circle (5pt);   
     \draw (17,0) node[black, scale=1]{$=0$} ;
    \end{tikzpicture}
    \caption{Graphical representation for the master equation. }\label{SD4}    
    \end{figure}
The master equation $[S,S]=0$ looks as Fig. \ref{SD4} and
\begin{equation} \label{master}
    [S, S](a,\ldots,b,\ldots,c, \ldots)=\sum \pm S(\ldots, \mathcal{U}, \ldots)\pm S(\ldots, \mathcal{V},\ldots)\,.
\end{equation}
Using the natural pairing \eqref{pairing}, one can extract the $A_\infty$-relations that describe Chiral HiSGRA as
\begin{equation}\label{SSJ}
    [S,S](w, \ldots)= \langle w, J(\ldots)\rangle=0\quad \Rightarrow \quad J(\ldots)=0\,.
\end{equation}
Choosing $w$ to be an element of $V$, we get an infinite set of $A_\infty$-relations of the form: 
\begin{align} \label{A}
    \begin{aligned}
         &J_{N+3}(\bullet,\ldots,\bullet, a,\bullet,\ldots,\bullet, b,\bullet,\ldots,\bullet, c,\bullet,\ldots,\bullet)=\\
        =&\sum\mathcal{V}(\bullet,\dots,\bullet,\mathcal{V}(\bullet,\dots,\bullet,a,\bullet,\dots,\bullet,b,\bullet,\dots,\bullet),\bullet,\dots,\bullet,c,\bullet,\dots,\bullet)+\\
        -&\sum\mathcal{V}(\bullet,\dots,\bullet,a,\bullet,\dots,\bullet,\mathcal{V}(\bullet,\dots,\bullet,b,\bullet,\dots,\bullet,c,\bullet,\dots,\bullet),\bullet,\dots,\bullet)+\\
        +&\sum\mathcal{V}(\bullet,\dots,\bullet,\mathcal{U}(\bullet,\dots,\bullet,a,\bullet,\dots,\bullet),\bullet,\dots,\bullet,b,\bullet,\dots,\bullet,c,\bullet,\dots,\bullet)+\\
        -&\sum\mathcal{V}(\bullet,\dots,\bullet,a,\bullet,\dots, \bullet,\mathcal{U}(\bullet,\dots,\bullet,b,\bullet,\dots,\bullet),\bullet,\dots,\bullet,c,\bullet,\dots,\bullet)+\\
        +&\sum\mathcal{V}(\bullet,\dots,\bullet,a,\bullet,\dots,\bullet,b,\bullet,\dots,\bullet,\mathcal{U}(\bullet,\dots,\bullet,c,\bullet,\dots,\bullet),\bullet,\dots,\bullet)=0\,.
    \end{aligned}
\end{align}
Here $a,b,c \in V[-1]$ and bullets stand for $N$ elements $\alpha_i \in V^\ast$. 
The summations are over all possible combinations of vertices in each term, i.e., all ordered distributions of the $\alpha$'s in the arguments and contributions from all corresponding disk diagrams. A single $A_\infty$-relation consists of all terms with the same number of elements $\alpha$ before, between and behind $a,b,c$, i.e., with the same ordering and same total number of the $\alpha$'s, $N$.  
For $w\in V[-1]$, the same Eq. (\ref{SSJ}) yields one more set of $A_\infty$-relations:
\begin{align}\label{Urelations}
    \begin{aligned}
             &J_{N+2}(\bullet,\ldots,\bullet, a,\bullet,\ldots,\bullet, b,\bullet,\ldots,\bullet)=\\       
             =&\sum \mathcal{U}(\bullet,\dots,\bullet,\mathcal{U}(\bullet,\dots,\bullet,a,\bullet,\dots,\bullet),\bullet,\dots,\bullet,b,\bullet,\dots,\bullet) +\\
        + &\sum \mathcal{U}(\bullet,\dots,\bullet,\mathcal{V}(\bullet,\dots,\bullet,a,\bullet,\dots,\bullet,b,\bullet,\dots,\bullet),\bullet,\dots,\bullet) +\\
        + &\sum \mathcal{U}(\bullet,\dots,\bullet,a,\bullet,\dots,\bullet,\mathcal{U}(\bullet,\dots,\bullet,b,\bullet,\dots,\bullet),\bullet,\dots,\bullet)=0 \,.
    \end{aligned}
\end{align}   
The $A_\infty$-relations for $\mathcal{V}$-vertices \eqref{A} will be proven via Stokes' theorem in the next section, which also implies the $A_\infty$-relations for the $\mathcal{U}$-vertices \eqref{Urelations} through the master equation \eqref{master}. Therefore, we will only focus on the former from now on. The proof will follow the scheme
\begin{align} \label{StokesToA_infty}
    \begin{aligned}
        0 = \sum \int_{\mathbb{W}_{k,l,m,n}} d\Omega_{k,l,m,n}^a(y) + d&\Omega_{k,l,m,n}^c(y) = \sum \int_{\partial\mathbb{W}_{k,l,m,n}} \Omega_{k,l,m,n}^a(y) + \Omega_{k,l,m,n}^c(y)\\
        &\Updownarrow\\
        A_{\infty}&\text{-relations}
    \end{aligned}
\end{align}
where $k+m+l+n=N$. $\Omega_{k,l,m,n}^a(y)$ and $\Omega_{k,l,m,n}^c(y)$ are closed differential forms, called \textit{potentials}, with values in multilinear maps
\begin{align*}
    \Omega_{k,l,m,n}^{a,c}(y) : T^kV^\ast \otimes V[-1] \otimes T^lV^\ast \otimes V[-1] \otimes T^mV^\ast \otimes V[-1] \otimes T^nV^\ast \rightarrow V[-1]\,.
\end{align*}
In what follows,  we will suppress the $y$-dependence of the potentials. Similar to the $A_\infty$-relation, the summation in \eqref{StokesToA_infty} is over the total number of elements of $V^\ast$ and ways of distributing them, as we will see shortly. Apart from the potentials $\Omega_{k,l,m,n}^a$ and $\Omega_{k,l,m,n}^c$, we must consider bounded domains $\mathbb{W}_{k,l,m,n}\subset \mathbb{R}^{2N+1}$ with $\text{dim}(\mathbb{W}_{k,l,m,n})=2N+1$. In Sec. $\ref{sec:recipe}$, we will explain how the potentials and domains can be constructed from disk diagrams and how to evaluate the potentials on the boundary of the domains. We will also present expressions and disk diagrams for the $A_\infty$-relations. A couple of lowest order examples of \eqref{StokesToA_infty} follow in Sec. \ref{sec:example}, while in Sec. \ref{sec:leftOrdered} we provide a proof for all orders in a particular ordering, the so-called left-ordered case. Left-ordered means that the arguments, e.g. in a disk diagram or in an $A_\infty$-relation are ordered as $a,b,c,\alpha_1,\ldots,\alpha_N$, $a,b,c\in V[-1]$, $\alpha_i\in V^\ast$.  Sec. \ref{sec:allOrderings} contains the proof for a generic ordering.

The idea of the proof follows the formality theorems: since the potentials $\Omega$ are closed forms, the l.h.s. of \eqref{StokesToA_infty} obviously vanishes. At the same time, the r.h.s. of \eqref{StokesToA_infty} is given by the sum of potentials evaluated at the boundaries of $\mathbb{W}_{k,l,m,n}$. Domain $\mathbb{W}_{k,l,m,n}$ and potentials $\Omega$ are carefully designed in such a way that this sum gives the $A_\infty$-terms plus certain other terms that vanish by themselves. The genuine $A_\infty$-terms contain vertices nested into each other. Therefore, some boundaries of $\mathbb{W}_{k,l,m,n}$ reduce to the products of two configuration spaces that define vertices and $\Omega$ evaluated on such boundaries give exactly the nested vertices.

\subsection{Recipe} \label{sec:recipe}

\begin{figure}
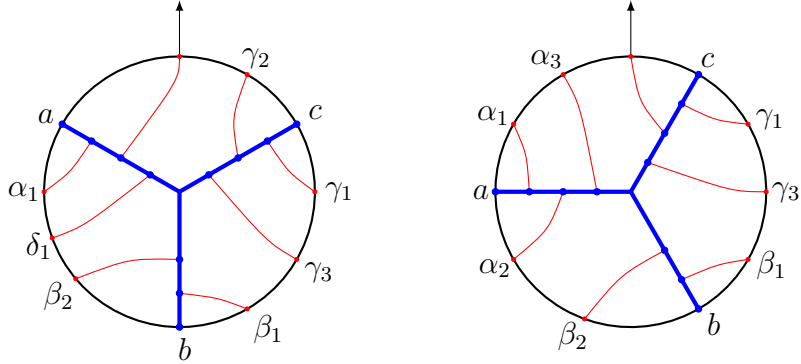

\centering

\caption{On the left a disk diagrams corresponding to $\Omega_{1,2,3,1}^a(a,\alpha_1,\delta_1,\beta_2,b,\beta_1,\gamma_3,\gamma_1,c,\gamma_2)$ and on the right a disk diagram corresponding to $\Omega_{3,2,1,1}^c(\alpha_3,\alpha_1,a,\alpha_2,\beta_2,b,\beta_1,\delta_1,\gamma_1,c)$.} \label{SD6}
\end{figure}

The data $\Omega_{k,l,m,n}^a$, $\Omega_{k,l,m,n}^c$, and $\mathbb{W}_{k,l,m,n}$, entering (\ref{StokesToA_infty}), can all be encoded by disk diagrams decorated by the variables $u_i$, $v_i$, and $w_i$ that coordinatize $\mathbb{W}_{k,l,m,n}$. To understand these disk diagrams and their properties, let us first consider disk diagrams for the scalars
\begin{align} \label{potPairing}
    \begin{aligned}
        &\langle \Omega^a_{k,l,m,n},\alpha_{k+1}\rangle & \text{and} && \langle \Omega^c_{k,l,m,n},\gamma_{m+1}\rangle \,,
    \end{aligned}
\end{align}
with $\alpha_{k+1},\gamma_{m+1} \in V^\ast$ and the natural pairing defined in \eqref{pairing}. From here the disk diagrams for the potentials can be extracted in a similar fashion as before. These disk diagrams are constructed as follows:
\begin{itemize}
    \item Consider a circle. The interior will be referred to as the \textit{bulk} and the circle as the \textit{boundary}.
    \item Choose three distinct points on the boundary and label them $a,b,c$ counterclockwise. Consider the point at the center of the bulk, now called \textit{junction}, and connect this to each of the points $a,b,c$. These points correspond to elements of $V[-1]$. The lines are called $a$\textit{-leg}, $b$\textit{-leg} and $c$\textit{-leg}, correspondingly.
    \item Draw any number of lines connecting these legs to the boundary at either side of the legs. Lines are not allowed to intersect. Their endpoints at the boundary correspond to elements of $V^\ast$.
    \item Connect an arrow to one of the vertices on the boundary of the disk between $a$ and $c$, pointing away from the disk, i.e. there has to be one marked point on the boundary. If the arrow is connected to a vertex connected to the $a$-leg by a red line, the potential that can be extracted using \eqref{potPairing} is $\Omega_{k,l,m,n}^a$, while $\Omega_{k,l,m,n}^c$ can be found when it is connected to the $c$-leg.
    \item Label the points at the boundary that are connected to red lines $\alpha_i, \beta_i, \gamma_i, \delta_i$ if the lines emanate from the $a$-leg, $b$-leg, $c$-leg or are in between the red line connected to the arrow and the junction, respectively, and $i$ increases from the boundary to the junction. This way if the arrow is attached to an argument belonging to the $a$-leg, the arguments after the arrow are labelled $\delta_i$ and those in between $a$ and including the arrow are named $\alpha_i$. The subscripts $k,l,m,n$ on the potentials count the number of points with labels $\alpha_i, \beta_i, \gamma_i, \delta_i$, respectively, disregarding the label associated with the arrow.
    \item The diagram must contain at least one element of $V^\ast$, connected to the piece of the boundary between $a$ and $c$, which can be attached to either the $a$-leg or the $c$-leg. If the diagram contains more than one element of $V^\ast$, they have to be attached to at least two different legs.
\end{itemize}
\begin{figure}[h]
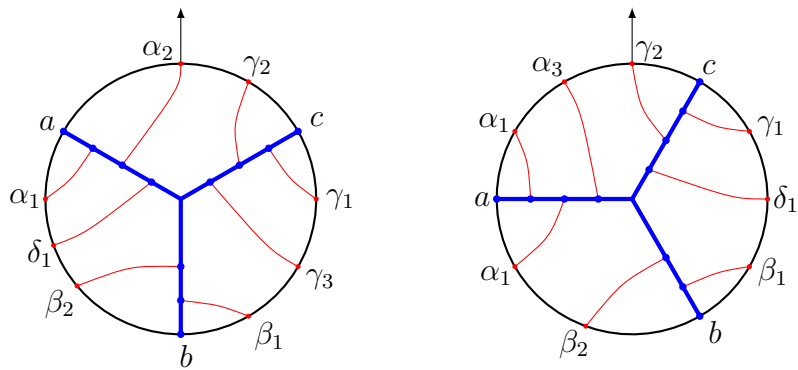

\centering
 \, 
\caption{The disk diagrams for $\langle\Omega_{1,2,3,1}^a(a,\alpha_1,\delta_1,\beta_2,b,\beta_1,\gamma_3,\gamma_1,c,\gamma_2),\alpha_2 \rangle$ and $\langle \Omega_{3,2,1,1}^c(\alpha_3,\alpha_1,a,\alpha_2,\beta_2,b,\beta_1,\delta_1,\gamma_1,c),\gamma_2\rangle$ on the left and right, respectively.}\label{SD5}
\end{figure}
Fig. \ref{SD5} shows two examples of such disk diagrams. The disk diagrams corresponding to the potentials $\Omega_{k,l,m,n}^a$ and $\Omega_{k,l,m,n}^c$ are now obtained by removing the element $\alpha_{k+1}$ or $\gamma_{m+1}$ in \eqref{potPairing}. This is visualized in the disk diagrams by removing the corresponding label, see  Fig. \ref{SD6}. 

The arrow also induces an ordering on the elements of $W$, which is counterclockwise around the circle starting from to the left of the arrow. Again, we will refer to this as the boundary ordering. The potentials $\Omega_{k,l,m,n}^{a,c}$ are poly-differential operators acting on elements of $W$. These should be read off according to the boundary ordering. Moreover, the elements of $W$ are generated by the $y_i$'s. For $a,b,c$ we write $y_a,y_b,y_c$, respectively and for elements of $V^\ast$ they are just $y_i$ with $i=1,\ldots, N$, assigned counterclockwise. Lastly, while it is cumbersome to specify the ordering of the elements of $W$ for generic potentials $\Omega_{k,l,m,n}^{a,c}$, we do specify the ordering with respect to the boundary ordering when we consider a particular potential. For general potentials we prefer to use {\it canonical ordering}, see below. The examples in Fig. \ref{SD6} correspond to poly-differential operators $\Omega_{1,2,3,1}^a(a,\alpha_1,\delta_1,\beta_2,b,\beta_1,\gamma_3,\gamma_1,c,\gamma_2)$ and $\Omega_{3,2,1,1}^c(\alpha_3,\alpha_1,a,\alpha_2,\beta_2,b,\beta_1,\delta_1,\gamma_1,c)$ acting on
\begin{align*}
    \begin{aligned}
        a(y_a) \alpha_1(y_1) \delta_1(y_2) \beta_2(y_3) b(y_b) \beta_1(y_4) \gamma_3(y_5) \gamma_1(y_6) c(y_c) \gamma_2(y_7)|_{y_{\bullet}=0} \,,\\
        \alpha_3(y_1) \alpha_1(y_2) a(y_a) \alpha_2(y_3) \beta_2(y_4) b(y_b) \beta_1(y_5) \delta_1(y_6) \gamma_1(y_7) c(y_c)|_{y_{\bullet}=0}
    \end{aligned}
\end{align*}
for the left and right disk diagrams, respectively.
The $A_\infty$-relations also consist of poly-differential operators acting on elements of $W$. The names and labels we assigned to elements of $V^\ast$ here are not the same as for the $A_\infty$-terms. This was done simply because it will be useful to keep track of what leg an element is attached to in the potentials. To relate the potentials to $A_\infty$-terms, one should rename and relabel the elements of $V^\ast$ accordingly. We now claim that an $A_\infty$-relation of a particular ordering can be rewritten as \eqref{StokesToA_infty} by considering contributions from all potentials of the same ordering, i.e. the same number of elements of $V^\ast$ before, between and after $a,b,c$. Therefore, we will no longer write the elements of $W$ as part of the expressions; they will be implied.

The procedure laid out in this section leads to two different types of disk diagrams: the arrow can be attached to the $a$-leg or $c$-leg, given rise to the disk diagrams for $\Omega_{k,l,m,n}^a$ or $\Omega_{k,l,m,n}^c$, respectively, and we refer to the diagrams as $a$-diagrams and $c$-diagrams. It is easy to see that following an $a$-diagram in the opposite boundary ordering, i.e. clockwise, and relabeling $a \leftrightarrow c$ yields a $c$-diagram. We may write
\begin{align} \label{flip}
    \Omega_{k,l,m,n}^a \leftrightarrow \Omega_{m,l,k,n}^c \,, \quad \text{with } a \leftrightarrow c 
\end{align}
and consequently, the potential $\Omega_{m,l,k,n}^c$ is accompanied by the same integration domain $\mathbb{W}_{k,l,m,n}$ as $\Omega_{k,l,m,n}^a$. A proper relabeling of the elements of $V^\ast$ according to which leg they are attached is also implied. Due to this relation between the diagrams, it will be possible to extract $c$-diagrams from $a$-diagrams and therefore we will focus mainly on the latter in the remainder of the text. We will make the above transformation more concrete, when we have all the necessary tools.

We will sometimes refer a special class of potentials as \textit{left/right-ordered}. These are the potentials with all elements of $V[-1]$ appearing before/after the elements of $V^\ast$ and they are given by $\Omega_{0,0,m,n}^a$ and $\Omega_{k,0,0,n}^c$ with the appropriate ordering, respectively. \eqref{StokesToA_infty} relates these potentials to left/right ordered $A_\infty$-terms, which are similarly defined as $A_\infty$-terms with all elements of $V[-1]$ appearing before/after the elements of $V^\ast$.

\paragraph{Domain.} When $k+l+m+n=N > 0$, the domain $\mathbb{W}_{k,l,m,n} \subset \mathbb{R}^{2N+1}$ can be read off from the disk diagrams, e.g. Fig. \ref{SD6}. For this purpose, one assigns a vector of variables $\vec{q}_{a,i}=(u^a_i,v^a_i,w^a_i)$, $\vec{q}_{b,i}=(u^b_i,v^b_i,w^b_i)$ and $\vec{q}_{c,i}=(u^c_i,v^c_i,w^c_i)$ to the red lines connected to the $a$-, $b$- and $c$-leg, respectively, and to the arrow, with $i$ increasing from boundary to junction. Additionally, we assign vectors $\vec{q}_a=(-1,0,0)$, $\vec{q}_b=(0,-1,0)$, $\vec{q}_c=(0,0,-1)$ to $a, b, c$ accordingly. Then one introduces the times $t^{uv}_i=u_i^\bullet/v_i^\bullet$, $t^{uw}_i=u_i^\bullet/w_i^\bullet$ and $t^{vw}_i=v_i^\bullet/w_i^\bullet$ and imposes chronological orderings along three different paths in the bulk to formulate the domain.
\begin{itemize}
    \item Path 1: one starts at $b$ and moves to $c$ and then from $c$ to $a$. This imposes a chronological ordering on the times $t^{uv}_i$.
    \item Path 2: one starts at $c$ and moves to $b$ and then from $b$ to $a$. This imposes a chronological ordering on the times $t^{uw}_i$.
    \item Path 3: one starts at $c$ and moves to $a$ and then from $a$ to $b$. This imposes a chronological ordering on the times $t^{vw}_i$.
\end{itemize}

\begin{figure}
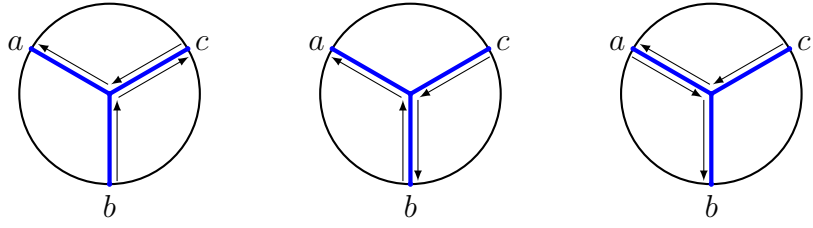

    \centering

    \caption{Paths 1 - 3 from left to right, leading to the time ordering of times $t^{uv}_i$, $t^{uw}_i$ and $t^{vw}_i$, respectively.}
    \label{paths}
\end{figure}

As an additional condition, all $u$-, $v$- and $w$-variables take values between $0$ and $1$. The domain $\mathbb{W}_{k,l,m,n}$ is then described by
\begin{align} \label{DD3}
    \begin{aligned}
        0 \leq& u_i^\bullet,v_i^\bullet,w_i^\bullet \leq 1 \,, \quad \sum (u_i^\bullet,v_i^\bullet,w_i^\bullet)=(1,1,1) \,, \quad  \frac{v_i^a}{w_i^a}\frac{w_i^b}{u_i^b}\frac{u_i^c}{v_i^c}=1\,,\\
        0 \leq& \frac{u_1^b}{v_1^b} \leq \dots \leq \frac{u_{l}^b}{v_{l}^b} \leq \frac{u_m^c}{v_m^c} = \dots = \frac{u_1^c}{v_1^c} \leq \frac{u_{k+n+1}^a}{v_{k+n+1}^a} \leq \dots \leq \frac{u_1^a}{v_1^a}\leq \infty\,,\\
       0 \leq& \frac{u_1^c}{w_1^c} \leq \dots \leq \frac{u_m^c}{w_m^c} \leq \frac{u_{l}^b}{w_{l}^b} = \dots = \frac{u_1^b}{w_1^b} \leq \frac{u_{k+n+1}^a}{w_{k+n+1}^a} \leq \dots \leq \frac{u_1^a}{w_1^a}\leq \infty\,,\\
        0 \leq& \frac{v_1^c}{w_1^c} \leq \dots \leq \frac{v_m^c}{w_m^c} \leq \frac{v_{k+n+1}^a}{w_{k+n+1}^a} = \dots = \frac{v_1^a}{w_1^a}\leq \frac{v_{l}^b}{w_{l}^b} \leq \dots \leq \frac{v_1^b}{w_1^b} \leq \infty \,.
    \end{aligned}
\end{align}
Note that the paths described above run over some legs twice. This leads to the equalities in \eqref{DD3}, since for example
\begin{align*}
    \frac{u_m^c}{v_m^c} \leq \dots \leq \frac{u_1^c}{v_1^c} \leq \frac{u_1^c}{v_1^c} \leq \dots \leq \frac{u_m^c}{v_m^c} \quad \Rightarrow \quad  \frac{u_m^c}{v_m^c} = \dots = \frac{u_1^c}{v_1^c} \,.
\end{align*}
The second equation in the first line of \eqref{DD3} is called the \textit{closure constraint} and explicitly it reads
\begin{align*}                  \sum_{i=1}^{k+n+1}\vec{q}_{a,i}+\sum_{i=1}^l\vec{q}_{b,i}+\sum_{i=1}^m\vec{q}_{c,i} +\vec{q}_a+\vec{q}_b+\vec{q}_c= (0,0,0) \,.
\end{align*}
This condition allows one to solve for one vector $\vec{q}_{\bullet,i}$ in terms of the other vectors. The third equation in the first line deserves some further explanation. To the equalities in the chains of (in)equalities one can ascribe
\begin{align*}
    \begin{aligned}
        \alpha&=\frac{u_i^c}{v_i^c} \,, & \frac{1}{\beta} &= \frac{v_i^a}{w_i^a} \,, & \gamma &= \frac{u_i^b}{w_i^b} \,.
    \end{aligned}
\end{align*}
The vectors $\vec{q}_{a,i},\vec{q}_{b,i},\vec{q}_{c,i}\in\mathbb{R}^3$ are then restricted to planes characterized by $\alpha,\frac{1}{\beta},\gamma$, as for example one may write $\vec{q}_{c,i}=(\alpha v^c_i,v^c_i,w^c_i)$ for $i=1,\dots,m$. The constraint in \eqref{DD3} tells us that the planes are related by
\begin{align} \label{abc}
    \gamma=\frac{\alpha}{\beta} \,.
\end{align}
However, this relation is only valid when $m,l,k+n \neq 0$. Otherwise, $\alpha,\beta,\gamma$ are independent. In practice, we will have
\begin{align*}
    \alpha &= \frac{u_1^c}{v_1^c} \,, & \beta &= \frac{1-\sum_{i=1}^{l} w_i^b-\sum_{i=1}^{m} w_i^c}{1-\sum_{i=1}^{l} v_i^b-\sum_{i=1}^{m} v_i^c} \,, & \gamma &= \frac{u_1^b}{w_1^b}\,.
\end{align*}
The second relation is obtained from the closure constraint for the $v$ and $w$ coordinates:
\begin{align*}
    \begin{aligned}
        \sum v_i^\bullet = 1 \quad &\Rightarrow \quad \sum_{i=1}^{k+n+1} v_i^a = 1 - \sum_{i=1}^{l} v_i^b - \sum_{i=1}^{m} v_i^c \,,\\
        \sum_{i=1}^{k+n+1} \beta v_i^a + \sum_{i=1}^{l}w_i^b + \sum_{i=1}^{m}w_i^c = 1 \quad &\Rightarrow \quad \beta = \frac{1-\sum_{i=1}^{l} w_i^b-\sum_{i=1}^{m} w_i^c}{\sum_{i=1}^{k+n+1}  v_i^a}=\frac{1-\sum_{i=1}^{l} w_i^b-\sum_{i=1}^{m} w_i^c}{1-\sum_{i=1}^{l} v_i^b-\sum_{i=1}^{m} v_i^c} \,.
    \end{aligned}
\end{align*}

The domain \eqref{DD3} contains a couple of `hidden' constraints on the variables. To illustrate this, suppose ${u_1^b}/{v_1^b} > 1$ and consider the first chain of (in)equalities, the $uv$-chain. This implies that $u_i^\bullet > v_i^\bullet$. Then, the closure constraint requires $\sum v_i^\bullet=1$ and we find $\sum u_i^\bullet > 1$, so the closure constraint cannot be satisfied for the $u$-variables. The same logic applied to the start and to the end of all three chains of (in)equalities leads to the additional constraints
\begin{align} \label{startEnd}
    \begin{aligned}
        0 \leq& u_1^b \leq v_1^b \leq 1 \,, \quad 0 \leq u_1^c \leq w_1^c \leq 1 \,, \quad 0 \leq v_1^c \leq w_1^c \leq 1 \,,\\
        0 \leq& v_1^a \leq u_1^a \leq 1 \,, \quad 0 \leq w_1^a \leq u_1^a \leq 1 \,, \quad 0 \leq w_1^b \leq v_1^b \leq 1 \,.
    \end{aligned}
\end{align}
However, one must be careful, as the domain changes significantly whenever $l=0$, in which case the first and last inequality are replaced by
\begin{align*}
    0 \leq& u_1^c \leq v_1^c \leq 1 \,, \quad 0 \leq w_1^a \leq v_1^a \leq 1\,,
\end{align*}
where in the latter we had some freedom to choose in which variables we express the inequalities. It turns out that the domain takes this form for left/right ordered $A_\infty$-relations, although not exclusively for this class. Since the start and end of the chains of (in)equalities yields constraints, the domain also takes a different form when $m=0$, in which case the second and third inequality in \eqref{startEnd} are replaced by
\begin{align} \label{startEnd2}
    0 \leq& u_l^b \leq w_l^b \leq 1 \,, \quad 0 \leq v_{k+n+1}^a \leq w_{k+n+1}^a \leq 1 \,.
\end{align}

In the special case $N=0$, the disk diagrams corresponding to $\Omega^{a,c}_{0,0,0,0}(a,b,c)$ contain no elements of $V^\ast$ and the above prescription for the domain breaks down. We manually define the domain $\mathbb{W}_{0,0,0,0} \subset \mathbb{R}$ to be described by
\begin{align*}
    0 \leq t \leq 1 \,.
\end{align*}

Furthermore, we note that the integration domain has a $\mathbb{Z}_2$-symmetry under swapping the $v$ and $w$ variables together with the labels $b$ and $c$ on all variables. Graphically, this means that swapping the $b$- and $c$-leg, together with all the lines attached to them, and renaming the elements, again yields a disk diagram to which a domain can be ascribed. This relates integration domains $\mathbb{W}_{k,l,m,n}$ and $\mathbb{W}_{k,m,l,n}$ by a coordinate transformation. 

A quick check shows that $\text{dim}(\mathbb{W}_{k,l,m,n})=2N+1$. The domain is described in \eqref{DD3} by a total of $3N+3$ coordinates. Not all coordinates are independent, as the description consists of some equalities; the chains of (in)equalities contain in total $N-2$ equalities, the closure constraint subtracts $3$ variables and, lastly, \eqref{abc} adds $1$ more equality, which justifies the dimension. In case $m=0$ or $l=0$, the chains of (in)equalities yield $N-1$ equalities, the closure constraint adds $3$ and \eqref{abc} is absent and again the right dimension is found. As the last remark, we note that the integration domain has a nice visual representation in the left-ordered case. This will be explained in Sec. \ref{sec:leftOrdered}.

\begin{figure}
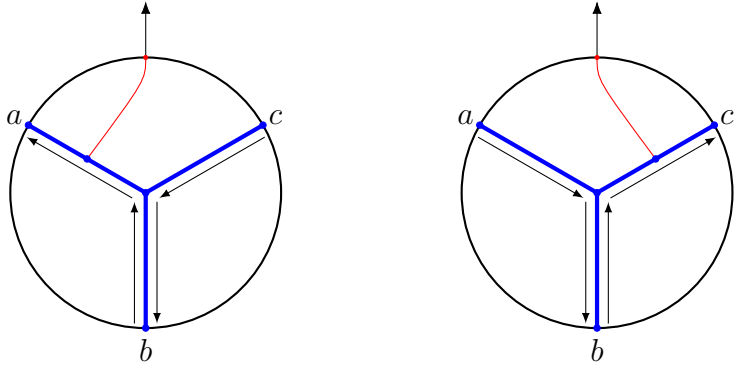

    \centering

    \caption{Canonical ordering for a potentials $\Omega_{k,l,m,n}^a$ and $\Omega_{k,l,m,n}^c$ on the left and right, respectively.}
    \label{canPot}
\end{figure}

\paragraph{Potential.} The potentials $\Omega_{k,l,m,n}^{a,c}$ are most conveniently defined through a slight detour. It is natural to define them first on a higher dimensional space and then get the actual potentials upon restricting them to $\mathbb{W}_{k,l,m,n}$. Let us first focus on the potentials $\Omega_{k,l,m,n}^a$. Consider a subspace $\mathbb{U}_{N} \subset \mathbb{R}^{3N}$ that is defined by
\begin{align} \label{masterDomain}
        0 &\leq u_i,v_i,w_i \leq 1 \,, & &\sum_{i=1}^{N+1}(u_i,v_i,w_i)=(1,1,1)\,.
\end{align}
In principle, any subspace that contains all $\mathbb{W}_{k,l,m,n}$ with dimension $2N+1$ fixed would work. We define the potential $\Omega_N^{a}$ on $\mathbb{U}_{N}$ by
\begin{align*}
    \Omega_N^{a} =\mu I_D \,,
\end{align*}
where $\mu$ is the measure and $I_D$ is the integrand. The measure reads
\begin{align} \label{measure}
    \begin{aligned}
        \mu =& \mu_1\wedge\dots\wedge\mu_N \,,\\
        \mu_{i} =&  p_{a,b} du_i \wedge dv_i + p_{a,c} du_i \wedge dw_i + p_{b,c} dv_i \wedge dw_i \,,\\
    \end{aligned}
\end{align}
and the integrand is given by
\begin{align} \label{pot}
    I_D = s_D\exp[\text{Tr}[P_D Q_D^T] + \lambda (p_{a,b}|Q_D^{12}|+p_{a,c}|Q_D^{13}|+p_{b,c}|Q_D^{23}|)] \,.
\end{align}
Here, $s_D$ is a sign that will be discussed at the end of this section. The matrix $Q_D$ is an array filled with the $q$-vectors according to the boundary ordering of the disk diagram. For example, the matrix $Q_D$ corresponding to the left disk diagram in Fig. \ref{SD6} reads
\begin{align*}
    Q_D =& (\vec{q}_a,\vec{q}_8,\vec{q}_6,\vec{q}_5,\vec{q}_b,\vec{q}_4,\vec{q}_3,\vec{q}_1,\vec{q}_c,\vec{q}_2,\vec{q}_7)\,.
\end{align*}
$|Q^{ij}_D|$ denotes the sum of all $2 \times 2$ minors composed of the $i$-th and $j$-th row of the matrix $Q_D$. A more explicit expression yields
\begin{align} \label{cosm}
    \begin{aligned}
        |Q_{D}^{12}| &= 1 + \sum_{i=1}^{N+1}(\sigma_{b,i} u_i - \sigma_{a,i} v_i + \sum_{j=1}^{N+1} \sigma_{i,j}u_i v_j) \,,\\
        |Q_{D}^{13}| &= 1 + \sum_{i=1}^{N+1}(\sigma_{c,i} u_i - \sigma_{a,i} w_i + \sum_{j=1}^{N+1}\sigma_{i,j}u_i w_j) \,,\\
        |Q_{D}^{23}| &= 1 + \sum_{i=1}^{N+1}(\sigma_{c,i} v_i - \sigma_{b,i} w_i + \sum_{j=1}^{N+1} \sigma_{i,j}v_i w_j) \,.
    \end{aligned}
\end{align}
Here, $\sigma_{I,J} = +1$ when the $J$-th element of $W$ appears before the $I$-th element in the boundary ordering and $\sigma_{I,J}=-1$ otherwise, for $I,J \in \{a,b,c,1,\dots,N+1\}$. The matrix $P_D$ is an array composed of the $r$-vectors in the following manner: going around the circle counterclockwise, starting to the left of the arrow, one fills $P_D$ with $\vec{r}_i$ for every element of $V^\ast$ or the arrow, with $i$ increasing counterclockwise, or with $\vec{r}_a=(-1,0,0)$, $\vec{r}_b=(0,-1,0)$, $\vec{r}_c=(0,0,-1)$ for the corresponding element of $V[-1]$. For example, the matrix $P_D$ corresponding to the left diagram in Fig. \ref{SD6} reads
\begin{align*}
    P_D=& (\vec{r}_a,\vec{r}_1,\vec{r}_2,\vec{r}_3,\vec{r}_b,\vec{r}_4,\vec{r}_5,\vec{r}_6,\vec{r}_c,\vec{r}_7,\vec{r}_8)\,.
\end{align*}
We will often present a matrix $Q$ when considering potentials with an unspecified ordering. We say that the entries of the matrix $Q$ are in the {\it canonical ordering}. This means that the first three entries are the vectors $\vec{q}_{a},\vec{q}_b,\vec{q}_c$, and then we insert the vectors $\vec{q}_{i}$ in the order visualized in Fig. \ref{canPot}. The matrix reads
\begin{align*}
    Q =& (\vec{q}_a,\vec{q}_b,\vec{q}_c,\vec{q}_1,\dots,\vec{q}_N) \,.
\end{align*}
This canonical ordering has the advantage to reduce to the matrix $Q_D$ for $D$ being a left ordered disk diagram with $k=l=0$. From $Q$ any matrix $Q_D$ may be constructed when the ordering is specified.

The potential $\Omega_{k,l,m,n}^{a}$ is now found by restricting $\Omega_N^{a}$ to $\mathbb{W}_{k,l,m,n} \subseteq \mathbb{U}_{N}$, i.e.,
\begin{align*}
    \Omega_{k,l,m,n}^{a} = \Omega_N^{a}\Big|_{\mathbb{W}_{k,l,m,n}} \,.
\end{align*}
Explicitly, this is achieved by renaming
\begin{align*}
    \begin{aligned}
        u_1,\dots,u_m &\rightarrow u^c_1,\dots,u_m^c \,, & u_{m+1},\dots,u_{m+l} &\rightarrow u^b_l,\dots,u^b_1 \,,\\
        u_{m+l+1},\dots,u_{k+l+m+n+1} &\rightarrow u^a_{k+n+1},\dots,u^a_{1} \,,
    \end{aligned}
\end{align*}
and similarly for the $v$- and $w$-coordinates, and setting
\begin{align} \label{R}
    \begin{aligned}
        \frac{u_1^c}{v_1^c} &= \dots = \frac{u_m^c}{v_m^c}=\alpha \,, & \frac{u_1^b}{w_1^b} &= \dots = \frac{u_l^b}{w_l^b} = \gamma\,,\\
        \frac{v_1^a}{w_1^a} &= \dots = \frac{v_{k+n+1}^a}{w_{k+n+1}^a}=\frac{1}{\beta} \,.
    \end{aligned}
\end{align}
Thus, for a potential $\Omega_{k,l,m,n}^a$ the matrix $Q$ reads
\begin{align*}
    Q =& (\vec{q}_a,\vec{q}_b,\vec{q}_c,\vec{q}_{c,1},\dots,\vec{q}_{c,m},\vec{q}_{b,l},\dots,\vec{q}_{b,1},\vec{q}_{a,k+n+1},\dots,\vec{q}_{a,1}) \,,     
\end{align*}
with the variables satisfying \eqref{R}. As an example, the left diagram in Fig. \ref{SD6} yields a potential described by the matrix
\begin{align*}
    Q_D =& (\vec{q}_a,\vec{q}_{a,1},\vec{q}_{a,3},\vec{q}_{b,2},\vec{q}_{b},\vec{q}_{b,1},\vec{q}_{c,3},\vec{q}_{c,1},\vec{q}_{c},\vec{q}_{c,2},\vec{q}_{a,2})\,,
\end{align*}
again with variables satisfying \eqref{R}. Using \eqref{cosm} and the anti-symmetric property
\begin{align*}
    \sigma_{i,j} &= - \sigma_{j,i}\,,
\end{align*}
together with the closure constraint and the Fierz identity, the potential $\Omega_N^{a}$ can easily be shown to be closed on $\mathbb{U}_{N}$, thus $\Omega_{k,l,m,n}^{a}$ is automatically closed too:
\begin{align}
    d \Omega_N^{a}&=0 && \Longrightarrow && d \Omega_{k,l,m,n}^{a}=0\,,\qquad \quad d \Omega_{k,l,m,n}^{c}=0\,.
\end{align}

As was briefly mentioned in \eqref{flip}, one can extract $\Omega_{m,l,k,n}^c$ from $\Omega_{k,l,m,n}^a$ by reversing the boundary ordering and swapping $a \leftrightarrow c$ on a disk diagram that is associated to $\Omega_{k,l,m,n}^a$. The latter implies that one has to swap $p_a \leftrightarrow p_c$ and $\vec{q}_a \leftrightarrow \vec{q}_c$ too. Reversing the boundary ordering negates the effect of swapping $\vec{q}_a$ and $\vec{q}_c$, while swapping $p_a$ and $p_c$ can be replaced by swapping the $u$ and $w$ coordinates. This tells us all we need to know to construct potentials $\Omega_{m,l,k,n}^c$ from their disk diagrams: the only difference is that the vectors 
\begin{align*}
    \begin{aligned}
        \vec{q}_{a,i}&\rightarrow\vec{q}_{c,i}^{\,\,'}=(w_i^c,v_i^c,u_i^c) \,, & \vec{q}_{b,i}&\rightarrow\vec{q}_{b,i}^{\,\,'}=(w_i^b,v_i^b,u_i^b) \,, & \vec{q}_{c,i}\rightarrow\vec{q}^{\,\,'}_{a,i}&=(w_i^a,v_i^a,u_i^a)
    \end{aligned}
\end{align*}
and that the canonical ordering for the matrix $Q$ is reversed, as can be seen in Fig. \ref{canPot}, but we keep $\vec{q}_a,\vec{q}_b,\vec{q}_c$ as its first entries. Thus, for $\Omega_{m,l,k,n}^c$ the matrix reads
\begin{align} \label{Q'}
    Q' =& (\vec{q}_a,\vec{q}_b,\vec{q}_c,\vec{q}_{c,1}^{\,\,'},\dots,\vec{q}_{c,m}^{\,\,'},\vec{q}_{b,l}^{\,\,'},\dots,\vec{q}_{b,1}^{\,\,'},\vec{q}_{a,k+n+1}^{\,\,'},\dots,\vec{q}_{a,1}^{\,\,'},)
\end{align}
and the matrix $Q'_D$ for the right disk diagram in Fig. \ref{SD6} is
\begin{align*}
    Q_D' =(\vec{q}_{c,3}^{\,\,'},\vec{q}_{c,1}^{\,\,'},\vec{q}_{a},\vec{q}_{c,2}^{\,\,'},\vec{q}_{b,2}^{\,\,'},\vec{q}_{b},\vec{q}_{b,1}^{\,\,'},\vec{q}_{a,3}^{\,\,'},\vec{q}_{a,1}^{\,\,'},\vec{q}_{c},\vec{q}_{a,2}^{\,\,'}) \,.
\end{align*}
Then, the expression for the potentials $\Omega^c_N$ becomes
\begin{align*}
    \Omega_{N}^c = \mu'I_D' \,,
\end{align*}
with 
\begin{align*}
    \begin{aligned}
        \mu' =& \mu'_1\wedge\dots\wedge\mu'_N \,,\\
        \mu'_{i} =&  p_{b,c} du_i \wedge dv_i + p_{a,c} du_i \wedge dw_i + p_{a,b} dv_i \wedge dw_i \,,\\
    \end{aligned}
\end{align*}
and 
\begin{align} \label{potc}
    I'_D = s'_D\exp[\text{Tr}[P_D Q_{D}^{\prime \, T}] + \lambda (p_{a,b}|Q_D^{\prime \, 12}|+p_{a,c}|Q_D^{\prime \, 13}|+p_{b,c}|Q_D^{\prime \, 23}|)] \,.
\end{align}
One obtains a potential $\Omega_{m,l,k,n}^c$ through
\begin{align*}
    \Omega_{m,l,k,n}^c=\Omega_{N}^c|_{\mathbb{W}_{k,l,m,n}} \,.
\end{align*}

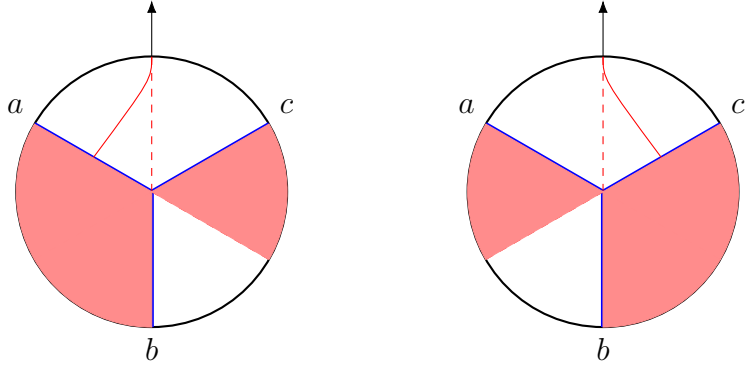
\begin{figure}
    \centering
    \begin{tikzpicture}[scale=0.3]
        \draw[thick](0,0) circle (6);
        \draw[ultra thick, blue](0,0) -- (150:6);
        \draw[ultra thick, blue](0,0) -- (30:6);
        \draw[ultra thick, blue](0,0) -- (270:6);
        \draw[red,thin] (150:3) .. controls (90:5) .. (90:6);
        \draw[dashed, red](0,0) -- (0,6);
        \draw[dashed, red](0,0) -- (-30:6);
        \draw[dashed, red](0,0) -- (210:6);
        \draw[-Latex] (0,6) -- (0,8.5);
        \filldraw [red!45!white, fill opacity=0.7] (0:0cm) -- (30:6cm) arc(30:-30:6cm) -- cycle;
        \filldraw [red!45!white, fill opacity=0.7] (0:0cm) -- (-90:6cm) arc(-90:-150:6cm) -- cycle;
        \filldraw [red!45!white, fill opacity=0.7] (0:0cm) -- (150:6cm) arc(150:210:6cm) -- cycle;
        \coordinate[label=above left : $a$] (B) at (150:6);
        \coordinate[label=above right : $c$] (B) at (30:6);
        \coordinate[label=below : $b$] (B) at (270:6);

        \begin{scope}[xshift=20cm]
        \draw[thick](0,0) circle (6);
        \draw[ultra thick, blue](0,0) -- (150:6);
        \draw[ultra thick, blue](0,0) -- (30:6);
        \draw[ultra thick, blue](0,0) -- (270:6);
        \draw[red,thin] (30:3) .. controls (90:5) .. (90:6);
        \draw[dashed, red](0,0) -- (0,6);
        \draw[dashed, red](0,0) -- (-30:6);
        \draw[dashed, red](0,0) -- (210:6);
        \draw[-Latex] (0,6) -- (0,8.5);
        \filldraw [red!45!white, fill opacity=0.7] (0:0cm) -- (-30:6cm) arc(-30:30:6cm) -- cycle;
        \filldraw [red!45!white, fill opacity=0.7] (0:0cm) -- (-30:6cm) arc(-30:-90:6cm) -- cycle;
        \filldraw [red!45!white, fill opacity=0.7] (0:0cm) -- (150:6cm) arc(150:210:6cm) -- cycle;
        \coordinate[label=above left : $a$] (B) at (150:6);
        \coordinate[label=above right : $c$] (B) at (30:6);
        \coordinate[label=below : $b$] (B) at (270:6);
        \end{scope}
    \end{tikzpicture}
    \caption{The sign $s_D$ of $\Omega_{k,l,m,n}^a$ and $\Omega_{k,l,m,n}^c$ is determined by the number of red lines in the shaded regions in the left and right diagram, respectively.}
    \label{sign}
\end{figure}

\paragraph{Signs.} The signs $s_D$ and $s'_D$ in the expression for the potentials \eqref{pot} and \eqref{potc} are determined by the orientations of the red lines, like for the vertices, where we counted the number of red lines in the southern semicircle. Here the counting rule is a bit more involved: for potentials $\Omega_{k,l,m,n}^a$, one sums the number of red lines attached to $b$- and $c$-leg in the clockwise direction and to the $a$-leg in the anticlockwise direction, see Fig. \ref{sign}. For potentials $\Omega_{k,l,m,n}^c$ this is mirrored. We call the sum $M$ and the sign is
\begin{align*}
    \begin{aligned}
        s_D &= (-1)^M  & \text{and} & & s'_D &= (-1)^{M+1}
    \end{aligned}
\end{align*}
for the potentials $\Omega_{k,l,m,n}^a$ and $\Omega_{k,l,m,n}^c$, respectively. Note that throughout the text the red lines of diagrams are not always drawn precisely in the shaded regions as in Fig. \ref{sign} for the sake of convenience, but it should be clear from the context which region they belong to. To avoid cluttering the text with minus signs, we will discuss the proof of the $A_\infty$-relations up to a sign. However, with the conventions discussed here, \eqref{StokesToA_infty} yields the correct signs.

\paragraph{Boundaries.} Stokes' theorem requires one to evaluate the potentials at the codimension-one boundary (just the boundary) of the domain $\mathbb{W}_{k,l,m,n}$. This boundary, $\partial\mathbb{W}_{k,l,m,n}$ is the union of many boundary components $P_i$, i.e., $\partial\mathbb{W}_{k,l,m,n}=\cup_i P_i$. Each $P_i$ is obtained by saturating an inequality in \eqref{DD3}-\eqref{startEnd2}. However, sometimes saturating an inequality results in a higher codimension component. By abuse of nomenclature we also refer to these as boundaries and include them into the set of $P_i$. Each boundary may have interesting characteristics, together with the potential evaluated on the boundary. In the subsequent section, we will categorize the various types of boundaries. They belong to the following classes:
\begin{itemize}
    \item At some boundaries one finds, after a change of variables, $A_\infty$-terms contributing to the proof of $\eqref{StokesToA_infty}$. The change of variables is necessary to recognize the vertices as described in Section \ref{sec:CD}.
    \item The potentials may also yield nonzero results at certain boundaries, without contributing to $A_\infty$-relations. Fortunately, this does not spoil the proof, as these terms always come in pairs and consequently cancel each other. The pairs are always formed by terms arising from different potentials and the terms are therefore called $\textit{gluing terms}$ as they `glue' together different potentials.
    \item Since potentials are differential forms, they consist of a measure and an integrand. When the measure evaluates to zero at a boundary, we refer to this as a \textit{zero measure term}.
    \item As mentioned above, we describe boundaries of $\mathbb{W}_{k,l,m,n}$ simply by saturating inequalities in \eqref{DD3}. However, this will sometimes give rise to a \textit{higher codimension boundary}, as saturating one inequality requires other inequalities to be saturated at the same time. This yields a boundary that is parameterized by less than $2N$ variables. As Stokes' theorem only requires codimension-one boundaries, higher codimension boundaries do not contribute to the proof.
\end{itemize}

So far we have established a visual representation for the potentials and their corresponding domains through disk diagrams. With some amount of hindsight, we will provide a way to visualize the evaluation of the potentials on the boundaries throughout the later sections. It is known from earlier work \cite{Sharapov:2022nps} that a parameter is associated with all line segments in the disk diagrams, except the line connected to the arrow and the segments of the legs that are directly connected to the points $a,b,c$. These parameters are coordinates of a hypercube and are related to the $u,v,w$ coordinatizing $\mathbb{W}_{k,l,m,n}$ by a smooth coordinate transformation. In fact, $\mathbb{W}_{k,l,m,n}$ is a subspace of the hypercube. We will not use the exact coordinate transformations. Still, we borrow this knowledge to realize that evaluating potentials at a boundary of $\mathbb{W}_{k,l,m,n}$ coincides with evaluating them at the upper and lower bound of these parameters. We will visualize this by drawing a green/red region on the line under evaluation for the upper/lower bound. 

\begin{figure}
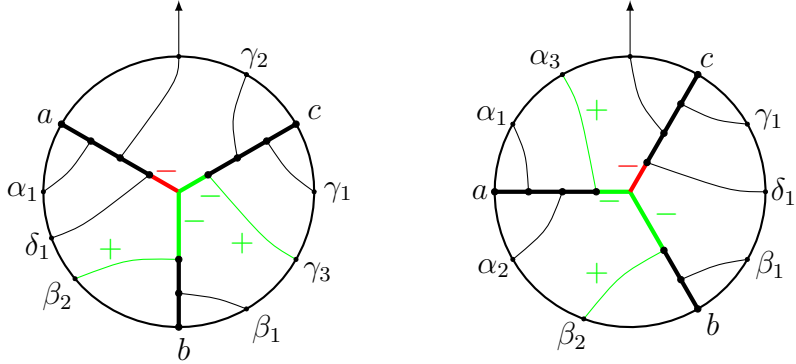

\centering

\caption{Disk diagrams that show which line segments correspond to boundaries that yield non-zero expressions for potentials of the type $\Omega_{k,l,m,n}^a$ and $\Omega_{k,l,m,n}^c$ on the left and right, respectively. The diagrams also show the signs that the boundaries are accompanied with.} \label{signs}
\end{figure}

The final result is easy to formulate. Fig. \ref{signs} shows on which lines in disk diagrams a boundary leads to a non-vanishing expression. These are the line segments connected to the junction and the bulk-to-boundary lines closest to the junction on the legs that are not connected to output arrow. The color of the line refers to type of boundary that yields this result and each line is accompanied by a sign that should be taken into account when considering the boundary. Throughout the following sections, we will show how these disk diagrams can be understood as $A_\infty$-terms or other terms.

Let us note that the pictures below are to display which part of the analytical expression for potential $\Omega$ is being affected by evaluating it at a certain boundary, i.e. the pictures are only to help visualize certain analytical manipulations. Given a disk diagram, there are, roughly speaking, two boundaries per each line. 

\paragraph{$A_\infty$-terms.} So far we have sketched the picture of how to employ Stokes' theorem to prove the $A_\infty$-relations. It only remains to provide a recipe for constructing the $A_\infty$-terms. As can be seen from \eqref{A}, the $A_\infty$-terms are nested vertices, so they can be visualized using the disk diagrams for vertices introduced in section \ref{sec:CD}. Examples of the different types of $A_\infty$-terms found in \eqref{A} are given in Fig. \ref{A_inftyDiagrams}. Here we visualize the nesting of vertices by inserting one vertex into the other through a \textit{nesting arrow}.

\begin{figure}
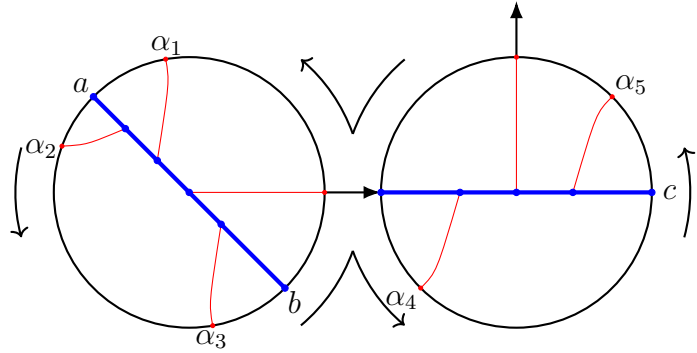

    \centering

    \caption{The nested boundary ordering for the diagram corresponding to the $A_\infty$-term $\mathcal{V}(\mathcal{V}(\alpha_1,a,\alpha_2,\alpha_3,b),\alpha_4,c,\alpha_5)$. One follows the arrow, starting from the ouput arrow. When the nesting arrow is reached, one follows the arrows around the nested vertex and when the nesting arrow is reached again, the path around the outer disk diagram is continued.}
    \label{nestedBO}
\end{figure}

To relate an expression with a nested disk diagram, one assigns vectors $\vec{q}_a=(-1,0,0),\vec{q}_b=(0,-1,0),\vec{q}_c=(0,0,-1)$ to $a,b,c\in V[-1]$, $\vec{q}_{i}^{\,\,1}=(u^1_i,v^1_i,w^1_i)$ to $\alpha_i \in V^\ast$ in the inner vertex and $\vec{q}_i^{\,\,2}=(u^2_i,v^2_i,w^2_i)$ to $\alpha_j \in V^\ast$ in the outer vertex and to the output arrow according to the bulk ordering for the two vertices separately, as was explained in section \ref{sec:CD}. For the moment, we leave the values of these vectors undefined. One also assigns vectors $\vec{r}_{a}=\vec{r}_{b}=\vec{r}_{c}=(0,0,0)$ to the elements $a,b,c$ and $\vec{r}_i=(p_{a,i},p_{b,i},p_{c,i})$ to $\alpha_i\in V^\ast$ and to the output arrow. This is done according to the \textit{nested boundary ordering}, just like the labeling of the $\alpha$'s. This ordering starts to the left of the output arrow, follows around the circle counterclockwise until it hits the nesting arrow. It then follows the nested circle counterclockwise, after which it completes the counterclockwise path around the outer circle, see Fig. \ref{nestedBO} for an example. We then construct the $3 \times (N+4)$ matrices $Q_D$ and $P_D$ by filling it up with the vectors $q$- and $r$-vectors, respectively, according to the nested boundary ordering. Like before, for the matrix $P_D$ this means that one enters the vectors $\vec{r}_i$ in increasing order, while one inserts $\vec{r}_a,\vec{r}_b,\vec{r}_c$ for the elements $a,b,c$, respectively. 

To determine the vectors $\vec{q}_i^{\,\,1}$ and $\vec{q}_i^{\,\,2}$, let us demonstrate how to insert vertices into each other at the level of expressions. Remember that $\mathcal{V}(\bullet,\dots,\bullet,a,\bullet,\dots,\bullet,b,\bullet,\dots,\bullet) \in V[-1]$ and $\mathcal{U}(\bullet,\dots,\bullet,a,\bullet,\dots,\bullet) \in V^\ast$. This means that an element of $V[-1]$ or $V^\ast$ can be replaced by a $\mathcal{V}$- or $\mathcal{U}$-vertex. For example, one can use the vertices in \eqref{quadV} and \eqref{cubicV} to compute
\begin{align*}
    \begin{aligned}
        \mathcal{V}(\mathcal{V}(a,b),c)&=\exp[p_{0,d}+p_{0,c}+\lambda p_{d,c}]\exp[y_dp_a+y_dp_b+\lambda p_{a,b}] =\\
        &= \exp[p_{0,a}+p_{0,b}+p_{0,c}+\lambda(p_{a,b}+p_{a,c}+p_{b,c})]
    \end{aligned}
\end{align*}
and (here and below, $y_d$ is an auxiliary variable to deal with the insertion of one vertex into another one, which is set to zero in the end)
\begin{align*}
    \begin{aligned}
        \mathcal{V}(a,\mathcal{V}(b,c,\alpha_1),\alpha_2) &= \int_{\mathbb{V}_1}p_{a,d}\exp[u_2^2 p_{0,a} + v_2^2 p_{0,d} + u_1^2p_{a,2} + v_1^2 p_{d,2} + \lambda p_{a,d} A_2]\times\\
        &\times\int_{\mathbb{V}_1}p_{b,c}\exp[u_2^1 y_d p_b + v_2^1 y_d p_c + u_1^1p_{b,1} + v_1^1 p_{c,1} + \lambda p_{b,c}A_1] =\\
        &=\int_{\mathbb{V}_1}\int_{\mathbb{V}_1}(u_2^1p_{a,b}+v_2^1p_{a,c})p_{b,c}\exp[u_2^2 p_{0,a} + v_2^2 u_2^1 p_{0,b} + v_2^2 v_2^1 p_{0,c} + u_1^2 p_{a,2} +\\
        &+ u_1^1 p_{b,1} + u_2^1v_1^2 p_{b,2} + v_1^1 p_{c,1} + v_2^1 v_1^2 p_{c,2} + \lambda(u_2^1 A_2 p_{a,b} + v_2^1 A_2 p_{a,c} + A_1 p_{b,c})] \,,
    \end{aligned}
\end{align*}
with $A_i = 1+u_1^i+u_2^i-v_1^i-v_2^i+u_1^iv_2^i-u_2^iv_1^i$. 
The five different types of $A_\infty$-terms are given by
\begin{align} \label{generalA_infty}
    \begin{aligned}
        &\mathcal{V}(\bullet,\dots,\bullet,\mathcal{V}(\bullet,\dots,\bullet,a,\bullet,\dots,\bullet,b,\bullet,\dots,\bullet),\bullet,\dots,\bullet,c,\bullet,\dots,\bullet) =\\
        &= s_{D_1}s_{D_2}\int_{\mathbb{V}_{s}}\int_{\mathbb{V}_{r}}p_{a,b}^{r}(u^1_t p_{a,c}+v^1_t p_{b,c})^{s}  I_1 \,,\\
        &\mathcal{V}(\bullet,\dots,\bullet,a,\bullet,\dots,\bullet,\mathcal{V}(\bullet,\dots,\bullet,b,\bullet,\dots,\bullet,c,\bullet,\dots,\bullet),\bullet,\dots,\bullet) =\\
        &=s_{D_1}s_{D_2}\int_{\mathbb{V}_{s}}\int_{\mathbb{V}_{r}}(u^1_t p_{a,b}+v^1_t p_{a,c})^{s} p_{b,c}^{r} I_2 \,,\\
        &\mathcal{V}(\bullet,\dots,\bullet,\mathcal{U}(\bullet,\dots,\bullet,a,\bullet,\dots,\bullet),\bullet,\dots,\bullet,b,\bullet,\dots,\bullet,c,\bullet,\dots,\bullet) =\\ 
        &=(-1)^{r-1}s_{D_1}s_{D_2}\int_{\mathbb{V}_{s+1}}\int_{\mathbb{V}_{r-1}}(u_{t}^2 p_{a,b}+ v_{t}^2 p_{a,c})^{r-1} p_{b,c}^{s+1} I_3 \,,\\
        &\mathcal{V}(\bullet,\dots,\bullet,a,\bullet,\dots,\bullet,\mathcal{U}(\bullet,\dots,\bullet,b,\bullet,\dots,\bullet),\bullet,\dots,\bullet,c,\bullet,\dots,\bullet) =\\ &=s_{D_1}s_{D_2}\int_{\mathbb{V}_{s+1}}\int_{\mathbb{V}_{r-1}} p_{a,c}^{s+1}(u_{t}^2 p_{a,b}-v_{t}^2 p_{b,c})^{r-1}I_4 \,,\\
        &\mathcal{V}(\bullet,\dots,\bullet,a,\bullet,\dots,\bullet,b,\bullet,\dots,\bullet,\mathcal{U}(\bullet,\dots,\bullet,c,\bullet,\dots,\bullet),\bullet,\dots,\bullet)=\\
        &=s_{D_1}s_{D_2}\int_{\mathbb{V}_{s+1}}\int_{\mathbb{V}_{r-1}} p_{a,b}^{s+1}(u_{t}^2 p_{a,c}+v_{t}^2 p_{b,c})^{r-1} I_5 \,,
    \end{aligned}
\end{align}
where $r$ and $s$ are the total number of elements of $V^\ast$ in the inner and outer vertex, respectively, and $t$ is the label of the $q$-vector corresponding to the line connected to the nesting arrow. $s_{D_1}$ and $s_{D_2}$ are the signs associated to the inner and outer vertex, respectively. The functions $I_i$ read
\begin{align*}
    I_i =& \exp[\text{Tr}[P_D(Q_{D}^i)^T]+\lambda(p_{a,b}|(Q_{D}^i)^{12}|+p_{a,c}|(Q_D^i)^{13}|+p_{b,c}|(Q_D^i)^{23}|)] \,.
\end{align*}
Like for the potentials, we provide matrices
\begin{align*}
    \begin{aligned}
        Q^i =& (
        \vec{q}_a , \vec{q}_b , \vec{q}_c , \vec{q}^{\,\,1}_1 , \dots , \vec{q}^{\,\,1}_r , \vec{q}^{\,\,2}_1 , \dots , \vec{q}^{\,\,2}_s
        )
    \end{aligned}
\end{align*}
that are said to be in the canonical ordering. For diagrams that admit a left-ordering, the matrices $Q^i$ reduce to the matrices $Q_D^i$ for left-ordered diagrams $D$. These matrices $Q$ are given by
\begin{align} \label{Qs}
    \begin{aligned}
        Q^1 =& 

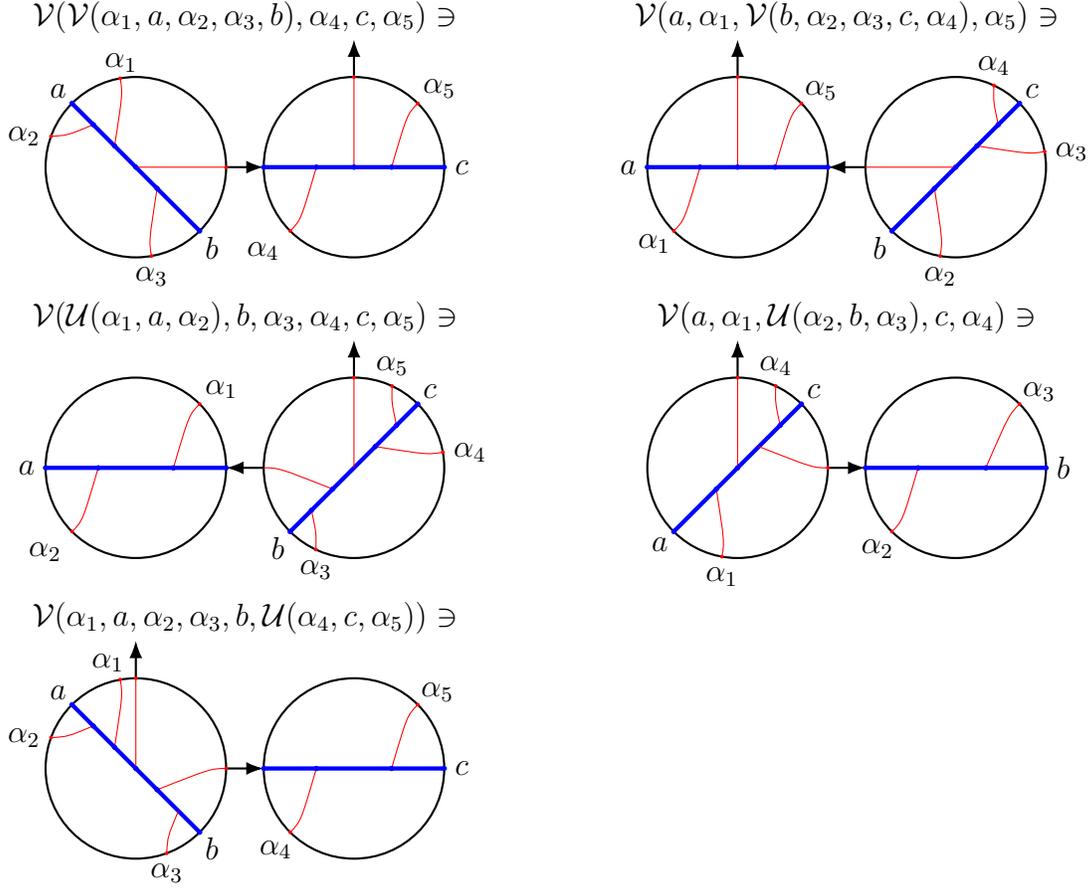
    \caption{Examples of nested disk diagrams for each type of $A_\infty$-terms. }
    \label{A_inftyDiagrams}
\end{figure}

\subsection{First examples}  \label{sec:example}
To get acquainted with the methods used in the proof that is going to follow, let us start with the lowest order examples, i.e., $N=0$ and $N=1$. 

\paragraph{$\boldsymbol{N=0}$.} This first example is perhaps a bit too simple, but it allows us to get a feel for some of the methods used for higher orders. The $A_\infty$-relation reads
\begin{align*}
    \mathcal{V}(\mathcal{V}(a,b),c) - \mathcal{V}(a,\mathcal{V}(b,c)) = 0
\end{align*}
and the only relevant vertex is the star-product $\mathcal{V}(a,b)\equiv a\star b$, given by
\begin{align*}
    \mathcal{V}(a,b) = \exp[p_{0,a}+p_{0,b}+\lambda p_{a,b}] \,.
\end{align*}

\begin{figure}[h]
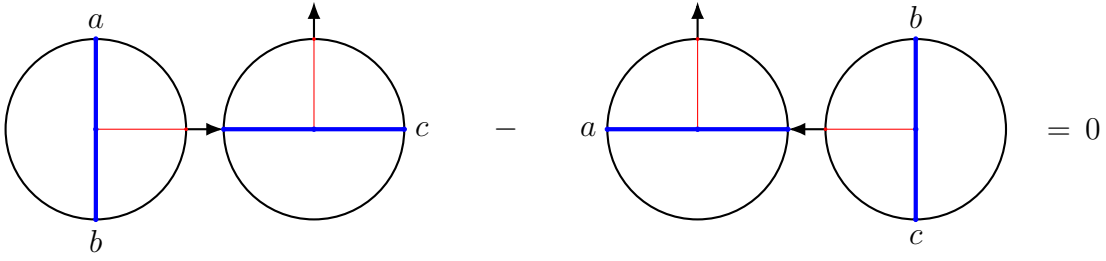

\centering


\caption{The disk diagram representation of the $A_\infty$-relation for $N=0$.}\label{nested1}
\end{figure}

\begin{figure}
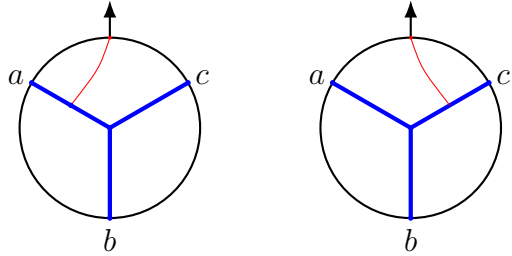

    \centering
    %
    \caption{On the left the disk diagram corresponding to the potential $\Omega_{0,0,0,0}^{a}(a,b,c)$ and on the right the one that corresponds to $\Omega_{0,0,0,0}^c(a,b,c).$}
    \label{N0}
\end{figure}

The $A_\infty$-relation is diagrammatically represented in Fig. \ref{nested1}. The relevant potentials are
\begin{align}\label{t}
    \Omega_{0,0,0,0}^a(a,b,c) = -\Omega_{0,0,0,0}^c(a,b,c) = \exp[p_{0,a} + p_{0,b} + p_{0,c} + \lambda(p_{a,b} + p_{a,c} + p_{b,c})]
\end{align}
and their disk diagrams are shown in Fig. \ref{N0}. The forms are closed, as they are constants. By definition, $\mathbb{W}_{0,0,0,0}=[0,1]$ and the boundary $\partial \mathbb{W}_{0,0,0,0}=\{0,1\}$ consists of the pair of points. This can also be written as
\begin{align*} 
    \partial\mathbb{W}_{0,0,0,0} = (\mathbb{V}_0 \times \mathbb{V}_0)\cup (\mathbb{V}_0 \times \mathbb{V}_0) \,,
\end{align*}
$\mathbb{V}_0$ being a one-point set. We stressed that the boundary components are the products of the configuration spaces of  vertices involved. The $A_\infty$-relation can now be recast in terms of Stokes' theorem \eqref{StokesToA_infty} through
\begin{align}\label{firstStokes}
    \begin{aligned}
        0 =& \int_0^1 (d\Omega_{0,0,0,0}^a(a,b,c)+d\Omega_{0,0,0,0}^c(a,b,c)) = [\Omega_{0,0,0,0}^a(a,b,c)+\Omega_{0,0,0,0}^c(a,b,c)]_0^1 \,.
    \end{aligned}
\end{align}
This 'too simple' example may look confusing since identical contributions are assigned different meaning and the integrand vanishes identically. Nevertheless, it showcases various features of the general proof. For instance, the last expression in \eqref{firstStokes} consists of four terms, while the $A_\infty$-relation contains only two terms. It turns out that whenever the proof requires contributions from multiple potentials, in this case $\Omega_{0,0,0,0}^a$ and $\Omega_{0,0,0,0}^c$, we find more terms than one would expect from the $A_\infty$-relation, but the extra terms from different potentials cancel each other. These are the gluing terms that were mentioned above.

\begin{figure}
    \centering

    \caption{$\Omega_{0,0,0,0}^a(a,b,c)$ evaluated at its boundaries. The green region highlights the boundary $t=1$, while the red region exhibits the boundary $t=0$.}
    \label{boundaries1a}
\end{figure}

\begin{figure}
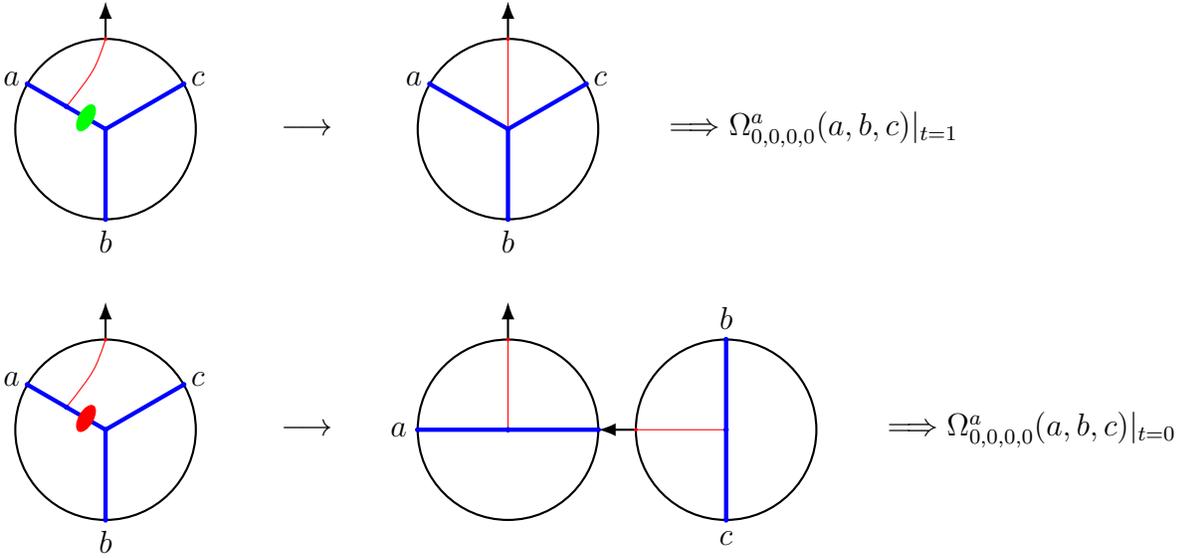

    \centering
    %
    \caption{$\Omega_{0,0,0,0}^c(a,b,c)$ evaluated at its boundaries. The green region highlights the boundary $t=1$, while the red region exhibits the boundary $t=0$.}
    \label{boundaries1c}
\end{figure}

Although the analytic expression \eqref{firstStokes} still looks manageable, the proof becomes increasingly more complicated at higher orders. It will be invaluable to have a visual representation of what happens to the disk diagram when the potentials are evaluated at a boundary. This will provide a quick way of identifying which $A_\infty$-term or gluing term is generated. Figs. \ref{boundaries1a} and \ref{boundaries1c} display how the disk diagrams can be interpreted as $A_\infty$-terms or gluing terms. The green boundary on the blue line shrinks the line to a point, giving rise to a disk diagram with a four-point vertex. This is a gluing term. The red boundary on the blue line separates the disk diagram in two disk diagrams with the blue line replaced by the nesting arrow. This is recognized as an $A_\infty$-term. Figs. \ref{boundaries1a} and \ref{boundaries1c} allow one to immediately observe that the gluing terms are identical and they will cancel each other. The remaining terms are easily read off to be the $A_\infty$-terms $\mathcal{V}(a,\mathcal{V}(b,c))$ and $\mathcal{V}(\mathcal{V}(a,b),c)$.

\paragraph{$\boldsymbol{N=1}$.} At this order there are four $A_\infty$-relations: one for each ordering of the elements of $V[-1]$ and $V^\ast$. However, some of these are related by the natural pairing and only two orderings are independent, i.e., the left-ordered case $a,b,c,\alpha$ and the almost-left-ordered case $a,b,\alpha,c$. The vertices relevant to the $A_\infty$-relations for $N=1$ are given in \eqref{quadV} and \eqref{cubicV}, while the domain is described in \eqref{2-simplex2}. The recipe of Sec. \ref{sec:recipe} tells us to use the vectors $\vec{q}_{a,1}$, $\vec{q}_{b,1}$ and/or $\vec{q}_{c,1}$ to construct expressions for the potentials. However, for notational simplicity, we will replace
\begin{align*}
    \begin{aligned}
        \vec{q}_{b,1},\vec{q}_{c,1} &\rightarrow \vec{q}_1=(u_1,v_1,w_1) \,, & \vec{q}_{a,1} &\rightarrow \vec{q}_2=(u_2,v_2,w_2)  \,.
    \end{aligned}
\end{align*}
The former makes sense, since every potential contains $\vec{q}_{b,1}$ or $\vec{q}_{c,1}$ and not both. Also note that this example will not show the full behaviour of \eqref{Qs}, as either the inner or the outer vertex has no integration domain for $N=1$, such that $u_1^1=v_1^1=1$ or $u_1^2=v_1^2=1$. As a result, the fact that some entries of the $Q$-matrices are composed of products of variables, is not visible in this example.

\paragraph{Left-ordering.} The $A_\infty$-relation for this ordering reads
\begin{align}\label{firstAinfty}
    \mathcal{V}(\mathcal{V}(a,b),c,\alpha) - \mathcal{V}(a,\mathcal{V}(b,c),\alpha) - \mathcal{V}(a,\mathcal{V}(b,c,\alpha)) + \mathcal{V}(a,b,\mathcal{U}(c,\alpha)) = 0 \,.
\end{align}

\begin{figure}
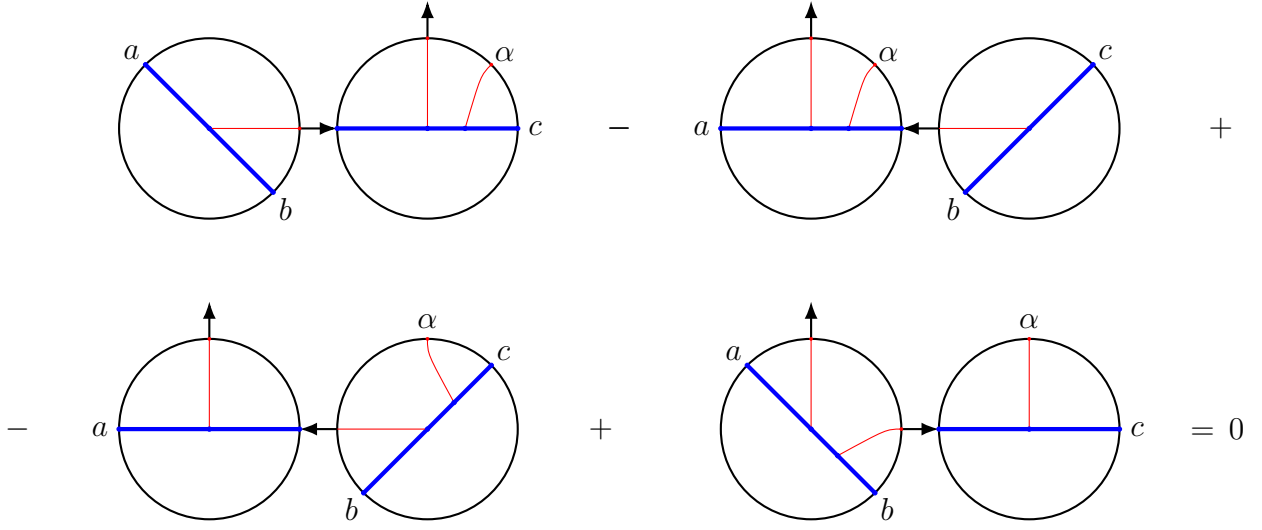

    \centering

    \caption{Graphical representation of the left-ordered $A_\infty$-relations for $N=1$.}
    \label{leftN1}
\end{figure}

This is an example of a left-ordered $A_\infty$-relation, i.e., all elements of $V[-1]$ appear before the element of $V^\ast$. A visualization in terms of disk diagrams is given in Fig. \ref{leftN1}. The order in which the elements of $W$ appear in the $A_\infty$-terms coincides with the nested boundary ordering in the disk diagram.
The left-ordered disk diagram for a potential with $N=1$ is shown in Fig. \ref{wwwCDiagram}. The potential $\Omega_{0,0,1,0}^a(a,b,c,\gamma)$ and domain $\mathbb{W}_{0,0,1,0}$ can be constructed from this diagram. We introduce the $3 \times 5$ matrices
\begin{align*}
    Q =\begin{pmatrix}
        -1 & 0 & 0 & u_1 & u_2 \\
        0 & -1 & 0 & v_1 & v_2 \\
        0 & 0 & -1 & w_1 & w_2
    \end{pmatrix} \,, \quad P &= \begin{pmatrix}
            0 & 0 & 0 & p_{a,1} & p_{0,a} \\
            0 & 0 & 0 & p_{b,1} & p_{0,b} \\
            0 & 0 & 0 & p_{c,1} & p_{0,c}
            \end{pmatrix}
\end{align*}
according to the recipe in section \ref{sec:recipe}. We also define the integrand
\begin{align*}
    I =& \exp[\text{Tr}[P Q^T]+\lambda( p_{a,b} |Q^{1,2}| +  p_{a,c} |Q^{1,3}| +  p_{b,c} |Q^{2,3}|)]
\end{align*}
and the measure
\begin{align*}
    \mu &= p_{a,b}du_1\wedge dv_1 + p_{a,c} du_1 \wedge dw_1 + p_{b,c} dv_1 \wedge dw_1\,.
\end{align*}
The potential $\Omega_{0,0,1,0}^a(a,b,c,\gamma)$ is then given by
\begin{align*}
\begin{aligned}
    \Omega_{0,0,1,0}^a(a,b,c,\gamma) =&[p_{a,b} du_1 \wedge dv_1 + p_{a,c} du_1 \wedge dw_1 + p_{b,c} dv_1 \wedge dw_1] \times \\
    \times&\exp[\text{Tr}[P Q^T]+\lambda (p_{a,b} |Q^{1,2}| + p_{a,c} |Q^{1,3}| + p_{b,c} |Q^{2,3}|)] \,.
\end{aligned}
\end{align*}
After solving the closure constraint for $\vec{q}_2$, i.e., $\vec{q}_2=1-\vec{q}_1$, we see that the potential is a closed form, since
\begin{align*}
    \begin{aligned}
    d\Omega_{0,0,1,0}^a(a,b,c,\gamma) &= (p_{a,b}p_{c,1} - p_{a,c}p_{b,1} + p_{b,c}p_{a,1} - p_{0,c}p_{a,b} + p_{0,b}p_{a,c} - p_{0,a}p_{b,c}) \times\\
    & \times du_1 \wedge dv_1 \wedge dw_1 I = 0\,,
    \end{aligned}
\end{align*}
where in the last step the Fierz identity is used twice:
\begin{align*}
    \begin{aligned}
        p_{a,b}p_{c,1} - p_{a,c}p_{b,1} + p_{b,c}p_{a,1} &= 0 \,, &
        p_{0,c}p_{a,b} - p_{0,b}p_{a,c} + p_{0,a}p_{b,c} &= 0 \,.
    \end{aligned}
\end{align*}
The terms in the $A_\infty$-relation \eqref{firstAinfty} are now conveniently written as
\begin{align} \label{A_infty terms}
    \begin{aligned}
    \mathcal{V}(\mathcal{V}(a,b),c,\alpha) &= \int_{\mathbb{V}_{1}} \, (p_{a,c} + p_{b,c}) I|_{u_\bullet=v_\bullet} \,, &
    \mathcal{V}(a,\mathcal{V}(b,c),\alpha) &= \int_{\mathbb{V}_{1}} \, (p_{a,b} + p_{a,c}) I|_{v_\bullet=w_\bullet} \,, \\
    \mathcal{V}(a,\mathcal{V}(b,c,\alpha)) &= \int_{\mathbb{V}_{1}} \, p_{b,c} I|_{u_1=0,u_2=1} \,, &
    \mathcal{V}(a,b,\mathcal{U}(c,\alpha)) &= \int_{\mathbb{V}_{1}} \, p_{a,b} I|_{w_1=1,w_2=0} \,.
    \end{aligned}
\end{align}
Here, $u_\bullet=v_\bullet$ means that the equality holds for all $u$'s and $v$'s, i.e., $u_1=v_1$ and $u_2=v_2$. 

The domain $\mathbb{W}_{0,0,1,0}$ is parameterized by the variables $u_i,v_i,w_i$, for $i=1,2$, that satisfy
\begin{align*}
    \begin{aligned}
        &0 \leq \frac{u_1}{v_1} \leq \frac{u_2}{u_2} \leq \infty \,, & 0 &\leq \frac{u_1}{w_1} \leq \frac{u_2}{w_2} \leq \infty \,, & 0 &\leq \frac{v_1}{w_1} \leq \frac{v_2}{w_2} \leq \infty \,,\\
        0 &\leq u_1,u_2,v_1,v_2 \leq 1 \,, &  u_1+u_2&=v_1+v_2=w_1+w_2=1\,.
    \end{aligned}
\end{align*}
This is a $3$-simplex and can equivalently be described by its `hidden constraints'
\begin{align} \label{simplex}
    0 \leq& u_1 \leq v_1 \leq w_1 \leq 1 \,, & 0 \leq& w_2 \leq v_2 \leq u_2 \leq 1 \,,
\end{align}
together with the closure constraint. The boundary of a $3$-simplex is composed of four $2$-simplices:
\begin{align}
    \partial \mathbb{W}_{0,0,1,0} \sim \bigcup_{i=1}^4 \mathbb{V}_1 \times \mathbb{V}_0 \,.
\end{align}%
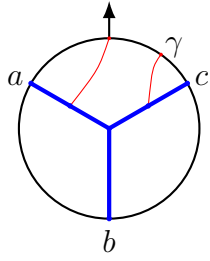
\begin{figure}
    \centering
    \begin{tikzpicture}[scale=0.2]
        \draw[thick](0,0) circle (6);
        \draw[ultra thick, blue] (0:0) -- (270:6);
        \draw[ultra thick, blue] (0:0) -- (150:6);
        \draw[ultra thick, blue] (0:0) -- (30:6);
        \draw[thick, Latex-] (0,8.5) -- (0,6);
        \draw[-][red, thin](90:6) .. controls (100: 4)..  (150:3); 
        \draw[-][red, thin](55:6) .. controls (55: 5)..  (30:3); 

        \filldraw[blue] (0,0) circle (4pt);
        \filldraw[blue] (270:6) circle (4pt);
        \filldraw[blue] (150:6) circle (4pt);
        \filldraw[blue] (30:6) circle (4pt);
        \filldraw[blue] (30:3) circle (4pt);
        \filldraw[blue] (150:3) circle (4pt);
        \filldraw[red] (90:6) circle (2.5 pt);
        \filldraw[red] (55:6) circle (2.5 pt);

        \coordinate[label=left : {$a$}] (B) at (146:6);
        \coordinate[label=below : {$b$}] (B) at (270:6);
        \coordinate[label=right : {$c$}] (B) at (34:6);
        \coordinate[label=right : {$\gamma$}] (B) at (61:6.2);
    \end{tikzpicture}
    \caption{Disk diagram corresponding to $\Omega_{0,0,1,0}^a(a,b,c,\gamma)$.}
    \label{wwwCDiagram}
\end{figure}%
The boundaries of $\mathbb{W}_{0,0,1,0}$ are reached by saturating the inequalities in \eqref{simplex}. Due to the closure conditions, saturating an inequality in the first chain of inequalities forces an inequality in the second chain to be saturated at the same time. Explicit evaluation shows
\allowdisplaybreaks{
\begin{align*}
    \begin{aligned}
        \int_{\partial\mathbb{W}_{0,0,1,0}}\Omega_{0,0,1,0}^a(a,b,c,\gamma)|_{u_1=0,u_2=1} &= \int_{\mathbb{V}_{1}}  p_{b,c} I|_{u_1=0,u_2=1} &\sim \mathcal{V}(a,\mathcal{V}(b,c,\alpha)) \,, \\[3mm]
         \int_{\partial\mathbb{W}_{0,0,1,0}}\Omega_{0,0,1,0}^a(a,b,c,\gamma)|_{u_\bullet=v_\bullet} &= \int_{\mathbb{V}_{1}}  (p_{a,c} + p_{b,c}) I|_{u_\bullet=v_\bullet} &\sim \mathcal{V}(\mathcal{V}(a,b),c,\alpha) \,, \\[3mm]
         \int_{\partial\mathbb{W}_{0,0,1,0}}\Omega_{0,0,1,0}^a(a,b,c,\gamma)|_{v_\bullet=w_\bullet} &= \int_{\mathbb{V}_{1}}  (p_{a,b} + p_{a,c}) I|_{v_\bullet=w_\bullet} &\sim \mathcal{V}(a,\mathcal{V}(b,c),\alpha) \,, \\[3mm]
        \int_{\partial\mathbb{W}_{0,0,1,0}}\Omega_{0,0,1,0}^a(a,b,c,\gamma)|_{w_1=1,w_2=0} &= \int_{\mathbb{V}_{1}}  p_{a,b} I|_{w_1=1,w_2=0} &\sim \mathcal{V}(a,b,\mathcal{U}(c,\alpha)) \,.
    \end{aligned}
\end{align*}}\noindent
For each term it is indicated which $A_\infty$-term it corresponds to, up to possible a change of integration variables and a change of the elements of $V^\ast$, as the labeling for potentials differs from the labeling for $A_\infty$-terms. The latter means in this case that $\gamma$ is replaced by $\alpha$. This proves that all $A_\infty$-terms \eqref{A_infty terms} are correctly recovered using Stokes' theorem \eqref{StokesToA_infty}. 

The above evaluation has been visualized in Fig. \ref{N0Boundaries}. Please note that the labeling of elements of $V^\ast$ is different for potentials and $A_\infty$-terms. Therefore, we change the labeling when considering a boundary, which in this case means that we replace the $\gamma$ by an $\alpha$. In the first row we again observe that a red boundary on a blue line separates the disk diagram into two disks, with the nesting arrow replacing this blue line. In the second row the green boundary shrinks the blue line to a point, like before. However, we then observe that if two legs with no red lines connected to them meet, they can be split off in a separate disk diagram. This interpretation arises from the fact that the expression for the potential produces the depicted $A_\infty$-term at this boundary. The same happens in the third row: the green boundary shrinks the blue line to a point and the $b$- and $c$-leg split off in a separate disk diagram. The red line connected to $\gamma$ is connected to the junction in the intermediate diagram and then migrates to the $a$-leg. As we will see more often, two legs with no red lines attached to them will split off as its own disk diagram. In the last row we encounter a combination we have not seen before: a green boundary on a red line. This splits off the entire leg to which this red line is attached and creates two disk diagrams, with the nesting arrow at the end of the leg in question. As turns out later, this may only happen to the last red line on a leg.

\begin{figure}[h!]
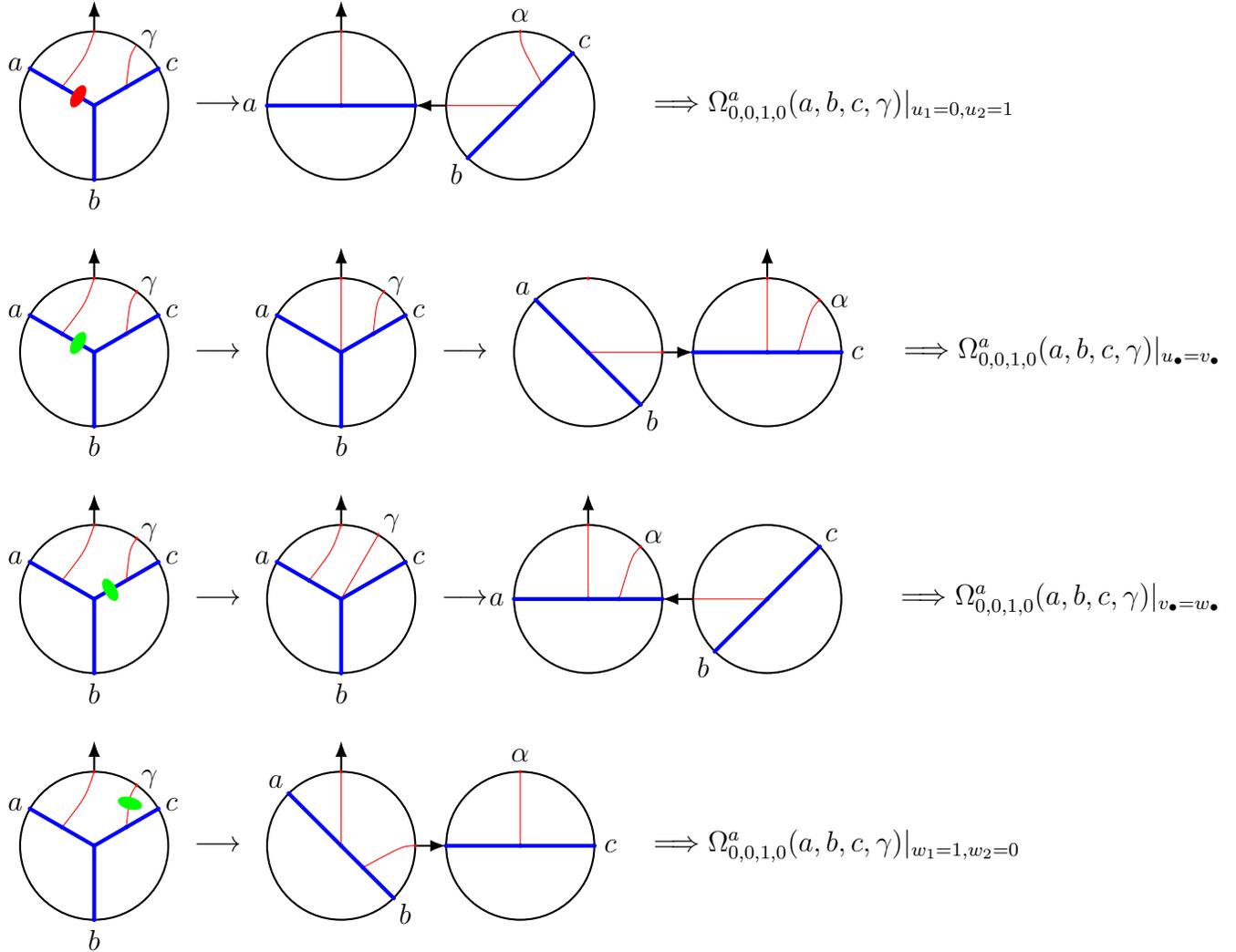

    \centering

    \caption{Visual representation of $\Omega_{0,0,1,0}^a(a,b,c,\gamma)$ evaluated at the boundaries of $\mathbb{W}_{0,0,1,0}$.} 
    \label{N0Boundaries}
\end{figure}

\paragraph{Almost-left-ordered.} 
\begin{figure}[h!]
    \centering
    %
    \caption{Graphical representation of the almost-left-ordered $A_\infty$-relations for $N=1$.}
    \label{secondOrderingN1}
\end{figure}
The $A_\infty$-relation for this ordering reads
\begin{align} \label{wwCwAinfty}
    \begin{aligned}
         \mathcal{V}(\mathcal{V}(a,b),\alpha,c)+\mathcal{V}(\mathcal{V}(a,b,\alpha),c) - \mathcal{V}(a,\mathcal{V}(b,\alpha,c)) - \mathcal{V}(a,\mathcal{U}(b,\alpha),c) + \mathcal{V}(a,b,\mathcal{U}(\alpha,c)) = 0
    \end{aligned}
\end{align}
and are visualized in Fig. \ref{secondOrderingN1}. We introduce the $3 \times 5$ matrices
\begin{align*}
    \begin{aligned}
        Q_1 =& \begin{pmatrix}
            -1 & 0 & u_1 & 0 & u_2 \\
            0 & -1 & v_1 & 0 & v_2 \\
            0 & 0 & w_1 & -1 & w_2 
        \end{pmatrix} \,, &
        Q_2 =& \begin{pmatrix}
            -1 & 0 & w_1 & 0 & w_2 \\
            0 & -1 & v_1 & 0 & v_2 \\
            0 & 0 & u_1 & -1 & u_2 
        \end{pmatrix} \,,\\
        P =& \begin{pmatrix}
            0 & 0 & p_{a,1} & 0 & p_{0,a} \\
            0 & 0 & p_{b,1} & 0 & p_{0,b} \\
            0 & 0 & p_{c,1} & 0 & p_{0,c}
        \end{pmatrix}
    \end{aligned}
\end{align*}
and define
\begin{align*}
    \begin{aligned}
        I_i =& \exp[\text{Tr}[PQ_{i}^T]+\lambda (p_{a,b} |Q_{i}^{1,2}|+ p_{a,c} |Q_i^{1,3}| + p_{b,c} |Q_{i}^{2,3}|)]\,.
    \end{aligned}
\end{align*}
The $A_\infty$-terms take the form
\allowdisplaybreaks{
\begin{align} \label{wwCwNested}
    \begin{aligned}
        \mathcal{V}(\mathcal{V}(a,b),\alpha,c) =& - (p_{a,c}+p_{b,c})(\int_0^1 dw_1\int_0^{w_1} du_1  I_1|_{u_\bullet=v_\bullet}+\int_0^1 dv_1\int_0^{v_1} du_1I_2|_{v_\bullet=w_\bullet}) \,, \\
        \mathcal{V}(\mathcal{V}(a,b,\alpha),c) =&   p_{a,b} \int_0^1 dv_1\int_0^{v_1} dw_1 I_2|_{u_1=0,u_2=1} \,,\\
        \mathcal{V}(a,\mathcal{V}(b,\alpha,c)) =& - p_{b,c} (\int_0^1 dw_1\int_0^{w_1} dv_1I_1|_{u_1=0,u_2=1}+\int_0^1 dv_1\int_0^{v_1} dw_1I_1|_{u_1=0,u_1=2}) \,,\\
        \mathcal{V}(a,\mathcal{U}(b,\alpha),c) =& -p_{a,c} (\int_0^1 dw_1\int_0^{w_1} du_1I_1|_{v_1=1,v_2=0} + \int_0^1 dw_1\int_0^{w_1} du_1I_2|_{v_1=1,v_2=0}) \,,\\
        \mathcal{V}(a,b,\mathcal{U}(\alpha,c)) =& -  p_{a,b} \int_0^1 dv_1\int_0^{v_1} du_1I_1|_{w_1=1,w_2=0} \,.    
    \end{aligned}
\end{align}}\noindent
Here, the $\mathbb{Z}_2$-symmetry of the domain $\mathbb{V}_n$ was used in several terms. The disk diagrams corresponding to potentials with the almost-left-ordering are shown in Fig. \ref{wwCwDiagram}. Please note that only one disk diagram is included with the arrow connected to the $c$-leg. Indeed, diagrams with all lines connected to a single leg are prohibited. The expressions corresponding to the disk diagrams in Fig. \ref{wwCwDiagram} are
\begin{align} \label{cubicForms}
    \begin{aligned}
    \int_{\partial\mathbb{W}_{0,0,1,0}} \Omega_{0,0,1,0}^a(a,b,\gamma,c) =& \int_{\partial\mathbb{W}_{0,0,1,0}}(p_{a,b} du_1 \wedge dv_1 + p_{a,c} du_1 \wedge dw_1 + p_{b,c} dv_1 \wedge dw_1)I_1 \,,\\
    \int_{\partial\mathbb{W}_{0,1,0,0}}\Omega_{0,1,0,0}^a(a,b,\beta,c) =& \int_{\partial\mathbb{W}_{0,1,0,0}}(p_{a,b} du_1 \wedge dv_1 + p_{a,c} du_1 \wedge dw_1 + p_{b,c} dv_1 \wedge dw_1)I_1 \,,\\
    \int_{\partial\mathbb{W}_{0,1,0,0}}\Omega_{0,1,0,0}^c(a,b,\beta,c) =& \int_{\partial\mathbb{W}_{0,1,0,0}}(-p_{b,c} du_1 \wedge dv_1 - p_{a,c} du_1 \wedge dw_1 - p_{a,b} dv_1 \wedge dw_1)I_2 \,,
    \end{aligned}
\end{align}
from left to right. The domain $\mathbb{W}_{0,0,1,0}$ is given in \eqref{simplex} and the closure constraint, while $\mathbb{W}_{0,1,0,0}$ is described by
\begin{align}\label{otherSimplex}
    \begin{aligned}
        &0 \leq \frac{u_1}{v_1} \leq \frac{u_2}{v_2} \leq \infty \,, \quad 0 \leq \frac{u_1}{w_1} \leq \frac{u_2}{w_2} \leq \infty \,, & 0 &\leq \frac{v_2}{w_2} \leq \frac{v_1}{w_1} \leq \infty \,,\\
        0 &\leq u_1,u_2,v_1,v_2 \leq 1 \,, & u_1+u_2&=v_1+v_2=w_1+w_2=1 \,,
    \end{aligned}
\end{align}
which simplifies to
\begin{align*} 
    0 \leq& u_1 \leq w_1 \leq v_1 \leq 1 \,, & 0 \leq& v_2 \leq w_2 \leq u_2 \leq 1\,.
\end{align*}
At this point, one has to be careful when describing the boundaries of $\mathbb{W}_{0,1,0,0}$. If one considers the boundary $u_1=0$, \eqref{otherSimplex} reduces to
\begin{align*}
    0 \leq& \frac{v_2}{w_2} \leq \frac{v_1}{w_1} \leq \infty \,, & u_1+u_2=v_1+v_2=w_1+w_2=1 \,,
\end{align*}
which is not the $2$-simplex in the way we usually describe it: one has to swap $v_1 \leftrightarrow v_2$ and $w_1 \leftrightarrow w_2$ to restore the correct description of the domain, which is with the labels in increasing order from left to right in the chain of inequalities and with the numerator and denominator in alphabetical order. This is related to the fact that the $q$-vectors were assigned differently for potentials than for nested vertices. Moreover, the potentials can be checked to be closed in a similar way as in the left-ordering.
\begin{figure}
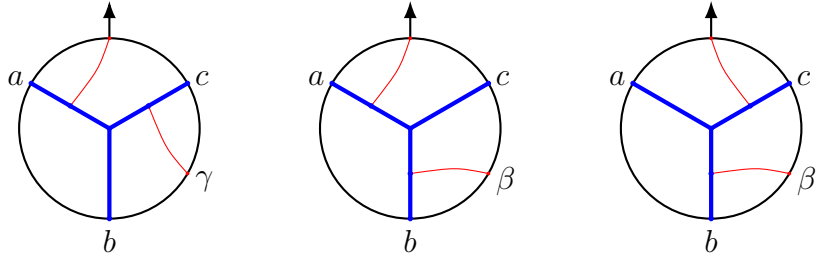

    \centering

    \caption{The disk diagrams corresponding to $\Omega_{0,0,1,0}^a(a,b,\gamma,c)$, $\Omega_{0,1,0,0}^a(a,b,\beta,c)$ and $\Omega_{0,1,0,0}^c(a,b,\beta,c)$, respectively.}
    \label{wwCwDiagram}
\end{figure}

Finally, evaluating Stokes' theorem yields the $A_\infty$-terms through
\begin{align*}
    \begin{aligned}
        \int_{\partial\mathbb{W}_{0,0,1,0}}\Omega_{0,0,1,0}^a(a,b,\gamma,c)|_{u_1=0,u_2=1} &= p_{b,c}\int_0^1dw_1\int_0^{w_1}dv_1 I_1|_{u_1=0,u_2=1} &\sim\mathcal{V}(a,\mathcal{V}(b,\alpha,c))\,, \\
        \int_{\partial\mathbb{W}_{0,0,1,0}}\Omega_{0,0,1,0}^a(a,b,\gamma,c)|_{u_\bullet=v_\bullet} &= (p_{a,c}+p_{b,c})\int_0^1dw_1\int_0^{w_1}du_1  I_1|_{u_\bullet=v_\bullet} &\sim \mathcal{V}(\mathcal{V}(a,b),\alpha,c) \,, \\ 
        \int_{\partial\mathbb{W}_{0,0,1,0}}\Omega_{0,0,1,0}^a(a,b,\gamma,c)|_{w_1=1,w_2=0} &= p_{a,b}\int_0^1dv_1\int_0^{v_1}du_1  I_1|_{w_1=1,w_2=0} &\sim \mathcal{V}(a,b,\mathcal{U}(\alpha,c)) \,, \\ 
        \int_{\partial\mathbb{W}_{0,1,0,0}}\Omega_{0,1,0,0}^a(a,b,\beta,c)|_{u_1=0,u_2=1} &= p_{b,c}\int_0^1dv_1\int_0^{v_1}dw_1 I_1|_{u_1=0,u_2=1} &\sim \mathcal{V}(a,\mathcal{V}(b,\alpha,c)) \,, \\
        \int_{\partial\mathbb{W}_{0,1,0,0}}\Omega_{0,1,0,0}^a(a,b,\beta,c)|_{v_1=1,v_2=0} &=p_{a,c}  \int_0^1dw_1\int_0^{w_1}du_1  I_1|_{v_1=1,v_2=0} &\sim \mathcal{V}(a,\mathcal{U}(b,\alpha),c) \,, \\ 
        \int_{\partial\mathbb{W}_{0,1,0,0}}\Omega_{0,1,0,0}^c(a,b,\beta,c)|_{u_1=0,u_2=1} &= - p_{a,b}\int_0^1dv_1\int_0^{v_1}dw_1  I_2|_{u_1=0,u_2=1} &\sim \mathcal{V}(\mathcal{V}(a,b,\alpha),c) \,, \\ 
        \int_{\partial\mathbb{W}_{0,1,0,0}}\Omega_{0,1,0,0}^c(a,b,\beta,c)|_{v_\bullet=w_\bullet} &= -(p_{a,c}+p_{b,c}) \int_0^1dv_1\int_0^{v_1}du_1 I_2|_{v_\bullet=w_\bullet} &\sim \mathcal{V}(\mathcal{V}(a,b),\alpha,c) \,, \\ 
        \int_{\partial\mathbb{W}_{0,1,0,0}}\Omega_{0,1,0,0}^c(a,b,\beta,c)|_{v_1=1,v_2=0} &= -p_{a,c}\int_0^1dw_1\int_0^{w_1}du_1 I_2|_{v_1=1,v_2=0} &\sim \mathcal{V}(a,\mathcal{U}(b,\alpha),c) \,.
    \end{aligned}
\end{align*}
\begin{figure}
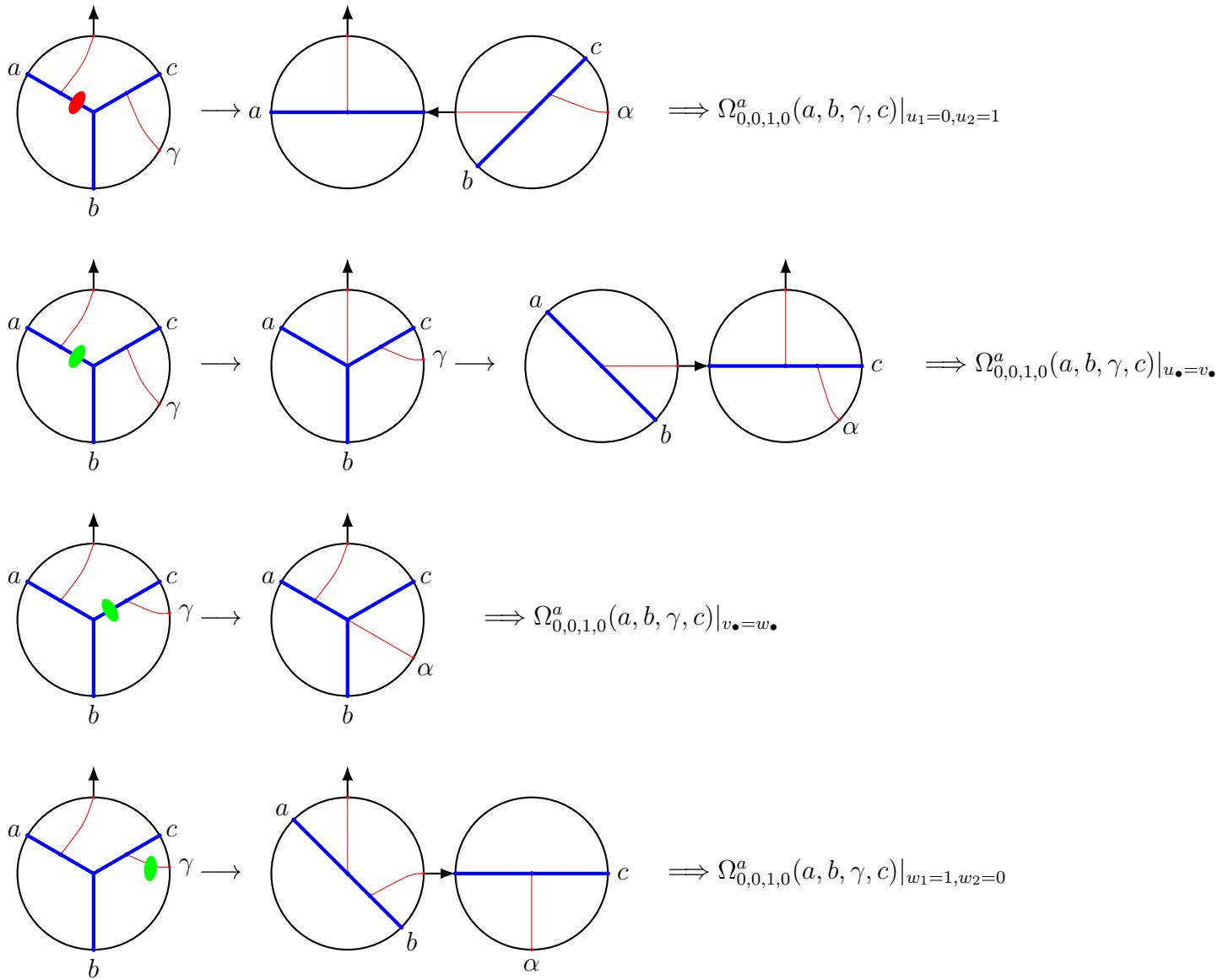

    \centering

    \caption{The disk diagrams of $\Omega_{0,0,1,0}^a(a,b,\gamma,c)$ evaluated at the boundaries of $\mathbb{W}_{0,0,1,0}$.}
    \label{1stboundaries}
\end{figure}

\begin{figure}[h!]
    \centering
    %
    \caption{The disk diagrams of $\Omega_{0,1,0,0}^a(a,b,\beta,c)$ evaluated at the boundaries of $\mathbb{W}_{0,1,0,0}$.}
    \label{2ndboundaries}
\end{figure}

\begin{figure}
    \centering
    %
    \caption{The disk diagrams of $\Omega_{0,1,0,0}^c(a,b,\beta,c)$ evaluated at the boundaries of $\mathbb{W}_{0,1,0,0}$.}
    \label{3rdboundaries}
\end{figure}

On the remaining boundaries, one finds the gluing terms
\begin{align*}
    \begin{aligned}
        \int_{\partial\mathbb{W}_{0,0,1,0}}\Omega_{0,0,1,0}^a(a,b,\gamma,c)|_{v_\bullet=w_\bullet} = (p_{a,b}+p_{a,c})\int_0^1 dw_1 \int_0^{w_1}du_1  I_1|_{v_\bullet=w_\bullet} \sim\\
        \sim\int_{\partial\mathbb{W}_{0,1,0,0}}\Omega_{0,1,0,0}^a(a,b,\beta,c)|_{v_\bullet=w_\bullet}\,,\quad\quad\hspace{9pt} \\ 
        \int_{\partial\mathbb{W}_{0,1,0,0}}\Omega_{0,1,0,0}^a(a,b,\beta,c)|_{u_\bullet=w_\bullet} = (p_{a,b}-p_{b,c})\int_0^1dv_1\int_0^{v_1}du_1 I_1|_{u_\bullet=w_\bullet} \sim\quad\quad\quad\quad\quad\hspace{2pt}\\
        \sim(p_{a,b}-p_{b,c})\int_0^1dv_1\int_0^{v_1}du_1I_2|_{u_\bullet=w_\bullet}= \int_{\partial\mathbb{W}_{0,1,0,0}}\Omega_{0,1,0,0}^c(a,b,\beta,c)|_{u_\bullet=w_\bullet} \,.\quad\quad\hspace{6pt}
    \end{aligned}
\end{align*}
All boundaries are visualized in Figs. \ref{1stboundaries} - \ref{3rdboundaries}. It is easy to see that all $A_\infty$-terms in \eqref{wwCwNested} are produced, together with gluing terms that cancel each other.

\subsection{All order generalization: left-ordered} \label{sec:leftOrdered}

It was already mentioned that the left and right-ordered cases are special: the domain is different and, in particular, for $N=1$ there are no gluing terms. Although gluing terms will appear for higher orders, there will be fewer for the left and right-ordered cases, making them easier to evaluate. In this section, we prove the $A_\infty$-relations through Stokes' theorem at all orders in the left-ordered case, from which the right-ordered case can easily be inferred.

\paragraph{$A_\infty$-terms.}
The left-ordered $A_\infty$-relations for $r+s=N \geq 1$ read
\begin{align} \label{A_infty}
    \begin{aligned}
        &\mathcal{V}(\mathcal{V}(a,b),c,\alpha_1,\dots,\alpha_{N})-\sum_{r+s=N}\mathcal{V}(a,\mathcal{V}(b,c,\alpha_1,\dots,\alpha_r),\alpha_{r+1},\dots,\alpha_{r+s})+\\
        &+\sum_{r+s=N}\mathcal{V}(a,b,\mathcal{U}(c,\alpha_1,\dots,\alpha_r),\alpha_{r+1},\dots,\alpha_{r+s})=0 \,.
    \end{aligned}
\end{align}
The individual $A_\infty$-terms then read
\begin{align} \label{nestedVertices}
    \begin{aligned}
    \mathcal{V}(\mathcal{V}(a,b),c,\alpha_1,\dots,\alpha_{N}) =& \int_{\mathbb{V}_{N}} (p_{a,c} + p_{b,c})^{N} I_1 \,,\\
    \mathcal{V}(a,\mathcal{V}(b,c,\alpha_1,\dots,\alpha_r),\alpha_{r+1},\dots,\alpha_{r+s}) =& \int_{\mathbb{V}_{s}}\int_{\mathbb{V}_{r}} (u^1_t p_{a,b}+v^1_t p_{a,c})^s p_{b,c}^r I_2 \,,\\
    \mathcal{V}(a,b,\mathcal{U}(c,\alpha_1,\dots,\alpha_r),\alpha_{r+1},\dots,\alpha_{r+s}) =& \int_{\mathbb{V}_{s+1}}\int_{\mathbb{V}_{r-1}} p_{a,b}^{s+1}(u^2_t p_{a,c} + v^2_t p_{b,c})^{r-1} I_3 \,,
    \end{aligned}
\end{align}
where
\begin{align*}
    I_i =& \exp[\text{Tr}[PQ_i^T]+\lambda (|Q_i^{12}|p_{a,b} + |Q_i^{13}| p_{a,c} + |Q_i^{23}| p_{b,c})]
\end{align*}
and
\begin{align*}
    \begin{aligned}
            Q_1 =& 

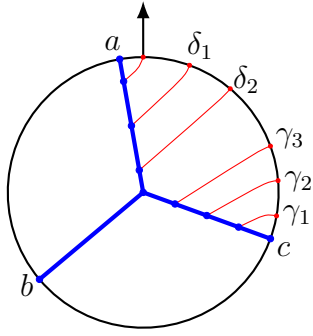
    \caption{Disk diagram corresponding to $\Omega_{0,0,3,2}^a(a,b,c,\gamma_1,\gamma_2,\gamma_3,\delta_2,\delta_1)$.}
    \label{leftOrdered}
\end{figure}

\paragraph{Domain.}
An example of a disk diagram for a potential with a left-ordering is shown in Fig. \ref{leftOrdered}. The relevant potentials and domain for $m+n=N$ are $\Omega_{0,0,m,n}^a(a,b,c,\gamma_1,\dots,\gamma_m,\delta_n,\dots,\delta_1)$ and $\mathbb{W}_{0,0,m,n}$, respectively. The domain $\mathbb{W}_{0,0,m,n}$ is described by
\begin{align} \label{leftOrderedDomain}
    \begin{aligned}
        0 &\leq u_i^\bullet,v_i^\bullet,w_i^\bullet \leq 1 \,, \quad  \sum_i(u_i^\bullet,v_i^\bullet,w_i^\bullet)=(1,1,1)\,,\\
        0 &\leq u_1^c \leq v_1^c \leq w_1^c \leq 1 \,, \quad 0 \leq w_1^a \leq v_1^a \leq u_1^a \leq 1\,,\\
        0 &\leq \frac{u_m^c}{v_m^c} = \dots = \frac{u_1^c}{v_1^c} \leq \frac{u_{n+1}^a}{v_{n+1}^a} \leq \dots \leq \frac{u_1^a}{v_1^a} \leq \infty\,,\\
        0 &\leq \frac{u_1^c}{w_1^c} \leq \dots \leq \frac{u_m^c}{w_m^c} \leq \frac{u_{n+1}^a}{w_{n+1}^a} \leq \dots \leq \frac{u_1^a}{w_1^a} \leq \infty\,,\\
        0 &\leq \frac{v_1^c}{w_1^c} \leq \dots \leq \frac{v_m^c}{w_m^c} \leq \frac{v_{n+1}^a}{w_{n+1}^a} = \dots = \frac{v_1^a}{w_1^a} \leq \infty \,.
    \end{aligned}
\end{align}
and has a special visualization in $\mathbb{R}^3$ in terms of the vectors $\vec{q}_a, \vec{q}_b, \vec{q}_c$ and $\vec{q}_{\bullet,i}$. In Fig. \ref{R3} we see that they form a closed polygon in $\mathbb{R}^3$. The domain is described by three chains of (in)equalities that obey the same chronological ordering, as the $uv$-chain starts with equalities. Therefore, the projection of the closed polygon on the $uv$-, $uw$- and $vw$-plane are swallowtails, each described by one of these chains, as shown in Fig. \ref{R3}. We refer to these polygons $(\vec{q}_{a},\vec{q}_b,\vec{q}_c,\vec{q}_{c,1},\dots,\vec{q}_{c,m},\vec{q}_{a,n+1},\dots,\vec{q}_{a,1})$ in $\mathbb{R}^3$ as maximally concave polygons. The equalities in the $uv$-plane ensure that the vectors $\vec{q}_{c,i}$ and $\vec{q}_{c,j}$ are coplanar, whereas the equalities in the $vw$-chain imply that vectors $\vec{q}_{a,i}$ and $\vec{q}_{a,j}$ are coplanar. This is depicted in Fig. \ref{coplan} for the domain $\mathbb{W}_{0,0,2,2}$. The blue arrows $\vec{q}_{c,1}$ and $\vec{q}_{c,2}$ lie in the same plane, highlighted by the blue shaded region, while the vectors $\vec{q}_{a,1}$, $\vec{q}_{a,2}$ and $\vec{q}_{a,3}$ are coplanar in the red shaded plane.

\begin{figure}
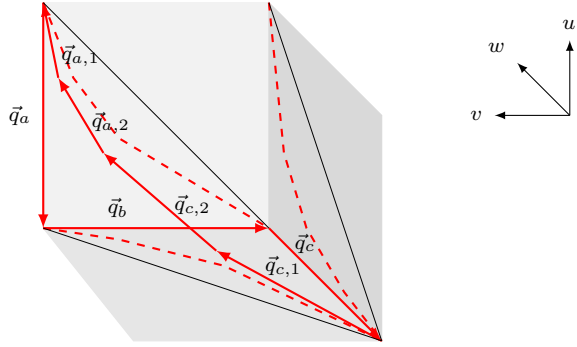

    \centering

    \caption{A visual representation of the left-ordered integration domain $\mathbb{W}_{0,0,2,1}$. The vectors $(\vec{q}_a,\vec{q}_b,\vec{q}_c,\vec{q}_{c,1},\vec{q}_{c,2},\vec{q}_{a,2},\vec{q}_{a,1})$ form a maximally concave polygon.}
    \label{R3}
\end{figure}

\begin{figure}[h!]
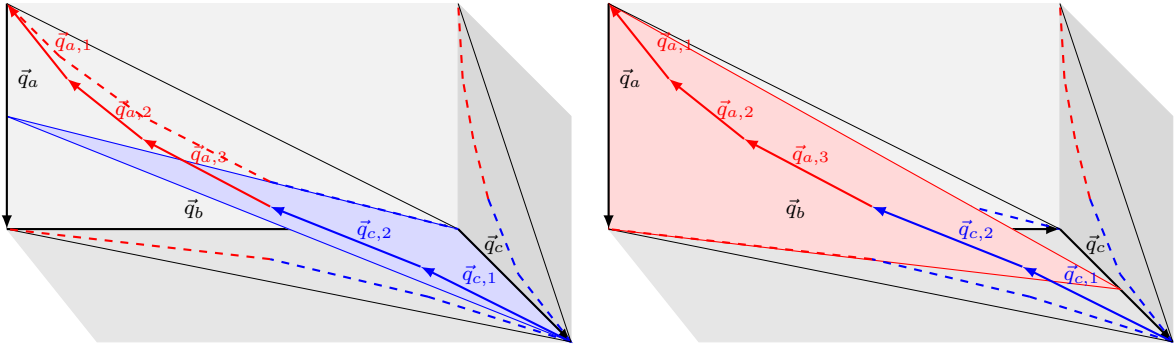


\begin{center}


%

\end{center}

\caption{On the left (right) the blue (red) shaded region depicts the plane in which the vectors $\vec{q}_{c,i}$ ($\vec{q}_{a,i}$) lie in $\mathbb{W}_{0,0,2,2}$. The scale in the $v$-direction was doubled to accentuate the details in the pictures.}\label{coplan}
\end{figure}

\paragraph{Potential.}
Following the recipe in section \ref{sec:recipe}, we construct a $2N$-form 
\begin{align*}
\Omega_{m+n}^a(a,b,c,\gamma_1,\dots,\gamma_m,\delta_n,\dots,\delta_1)
\end{align*}
on the space $\mathbb{U}_{m+n} \supseteq \mathbb{W}_{0,0,m,n}$. 
We construct a potential that reads
\begin{align*}                          \Omega_{m+n}^a(a,b,c,\gamma_1,\dots,\gamma_m,\delta_n,\dots,\delta_1) =& \mu I_{m+n} \,,
\end{align*}
with
\begin{align}
    I_{m+n} = & \exp[\text{Tr}[PQ^T] + \lambda(|Q^{12}|p_{a,b} + |Q^{13}|p_{a,c} +|Q^{23}|p_{b,c})]
\end{align}
and
\begin{align} \label{leftQ}
    Q =& (\vec{q}_{a},\vec{q}_{b},\vec{q}_{c},\vec{q}_{c,1},\dots,\vec{q}_{c,m},\vec{q}_{a,n},\dots,\vec{q}_{a,1})\,.
\end{align}
The measure $\mu$ is given in \eqref{measure}. The restriction of the potential to $\mathbb{W}_{0,0,m,n}$ is effectuated by requiring
\begin{align*}
    \begin{aligned}
        u_1,\dots,u_m &\rightarrow u_1^c,\dots,u_m^c \,, & u_{m+1},\dots,u_{m+n+1} &\rightarrow u_{n+1}^a,\dots,u_1^a
    \end{aligned}
\end{align*}
and
\begin{align} \label{restrictionLeftOrdered}
    \alpha &= \frac{u_1^c}{v_1^c} = \dots = \frac{u_m^c}{v_m^c} \,, & \frac{1}{\beta} &=\frac{v_{n+1}^a}{w_{n+1}^a} = \dots = \frac{v_1^a}{w_1^a} \,.
\end{align}
Requiring that no obvious singularities arise,\footnote{As an example, $\mathbb{W}_{0,0,m,n}$ contains the subspace attained by setting $u_1^c=0$. We can solve \eqref{restrictionLeftOrdered} by $v_i^c=\frac{v_1^c}{u_1^c}u_i^c$, but this looks singular at $u_1^c = 0$, while $u_i^c=\frac{u_1^c}{v_1^c}v_i^c$ behaves nicely as $0 \leq u_1^c \leq v_1^c \leq 1$.} we choose
\begin{align} \label{identification1}
    \begin{aligned}
        u_i^c &= \alpha v_i^c \,,& w_i^a &= \beta v_i^a \,.
    \end{aligned}
\end{align}
Explicitly, we write
\begin{align*}
    \alpha =& \frac{u_1^c}{v_1^c} \,, & \beta =&  \frac{1 - \sum_{i=1}^{m}w^c_i}{1 - \sum_{i=1}^m v^c_i} \,.
\end{align*}
Finally, the potential we are interested in reads
\begin{align*}
\Omega_{0,0,m,n}^a(a,b,c,\gamma_1,\dots,\gamma_m,\delta_n,\dots,\delta_1) &= \Omega_{m+n}^a(a,b,c,\gamma_1,\dots,\gamma_m,\delta_n,\dots,\delta_1)\Big|_{\mathbb{W}_{0,0,m,n}} = \mu_{m,n}I_{m,n} \,,
\end{align*}
with
\begin{align*}
    \begin{aligned}
        \mu_{m,n} &= \mu\Big|_{\mathbb{W}_{0,0,m,n}}\,, & I_{m,n} &= I_{m+n}\Big|_{\mathbb{W}_{0,0,m,n}}\,.
    \end{aligned}
\end{align*}
Solving the closure constraint for $\vec{q}_{a,1}$, i.e.
\begin{align*}
    \vec{q}_{a,1}=1-\sum_{i=1}^m \vec{q}_{c,i} - \sum_{i=1}^n \vec{q}_{a,i} \,,
\end{align*}
yields the measure
\begin{align*}
    \begin{aligned}
        \mu_{m,n} &= (\alpha p_{a,c} + p_{b,c})^{m-1}(p_{a,b} + \beta p_{a,c})^n \bigg[ \sum_{j=2}^{n+1} \frac{1-v_{1}^a}{v_{1}^a v_1^c} v_j^a  \\
         &\times(p_{a,c} du_1^c \wedge dv_1^c \wedge dw_1^c \wedge \dots \wedge \widehat{dv_j^c} \wedge \dots dv_m^c \wedge dw_m^c \wedge du_{2}^a \wedge dv_{2}^a \wedge \dots \wedge du_{n+1}^a \wedge dv_{n+1}^a+\\
         &+ p_{b,c} du_1^c \wedge dv_1^c \wedge dw_1^c \wedge \dots \wedge \widehat{du_j^c} \wedge \dots dv_m^c \wedge dw_m^c \wedge du_{2}^a \wedge dv_{2}^a \wedge \dots \wedge du_{n+1}^a \wedge dv_{n+1}^a) + \\
         &+p_{b,c} dv_1^c \wedge dw_1^c \wedge \dots dv_m^c \wedge dw_m^c \wedge du_{2}^a \wedge dv_{2}^a \wedge \dots \wedge du_{n+1}^a \wedge dv_{n+1}^a \bigg] \,,
    \end{aligned}
\end{align*}
where the symbol $\,\,\widehat{.}\,\,$ denotes omission and

\begin{align*}
    I_{m,n} =& \exp[\text{Tr}[P Q_{m,n}^T] + \lambda(|Q_{m,n}^{12}|p_{a,b} + |Q_{m,n}^{13}|p_{a,c} + |Q_{m,n}^{23}|p_{b,c})] \,,
\end{align*}
for
\begin{align} \label{QLeftRestriction}
    Q_{m,n} =& \begin{pmatrix}
        -1 & 0 & 0 & \alpha v_1^c  & \dots & \alpha v_m^c& u_{n+1}^a & \dots & u_1^a \\
            0 & -1 & 0 & v_1^c  & \dots &v_m^c & v_{n+1}^a & \dots & v_1^a  \\
            0 & 0 & -1 & w_1^c  & \dots & w_m^c& \beta v_{n+1}^a & \dots & \beta v_1^a
    \end{pmatrix}
\end{align}
and $P$ as defined in \eqref{PQ}. From now on we will omit the arguments of $\Omega_{0,0,m,n}^a$, as they should be clear from the subscript.

The domain $\mathbb{W}_{0,0,m,n}$ \eqref{leftOrderedDomain} is vastly more complicated than the domains discussed in the lower order examples. In particular, the chains of (in)equalities introduce new types of boundaries. In the following, we will categorize the boundaries according to whether the differential form $\Omega_{0,0,m,n}^a$ evaluates to an $A_\infty$-term, gluing term, zero, or the boundary turns out to be a higher codimension boundary. Of course, the latter does not play a role in Stokes' theorem and therefore does not contribute to the proof. In principle, saturating any inequality in \eqref{leftOrderedDomain} leads to a boundary, but some inequalities might seem to be missing in this categorization. This is simply because they are already accounted for in some other boundary. For example, the $uw$-chain can almost entirely be derived from the $uv$- and $vw$-chain.

\paragraph{$A_\infty$-terms.} The same boundaries that were present in the left-ordered $N=1$ example yield $A_\infty$-terms, with the boundary $w_1^c=1$ taking a more general form, see boundary 5.
\begin{itemize}
    \item Boundary 1: At this boundary
    \begin{align*}
        u_i^\bullet=0 \,.
    \end{align*}
    \begin{figure}[h!]
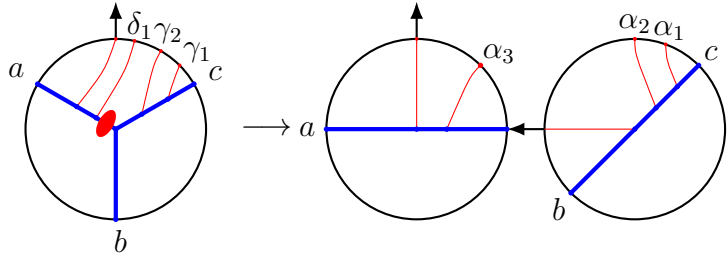

        \centering

        \caption{An example of boundary 1 contributing to $\mathcal{V}(a,\mathcal{V}(b,c,\alpha_1,\alpha_2),\alpha_3)$.}
        \label{B1}
    \end{figure}
    If $u_i^a=0$, the $uv$-chain becomes
    \begin{align*}
        0 \leq \frac{u_m^c}{v_m^c} \leq \dots \leq \frac{u_1^c}{v_1^c} \leq \frac{u_{n+1}^a}{v_{n+1}^a} \leq \dots \leq \frac{0}{v_{i}^a} \leq \dots \leq \frac{u^a_1}{v_1^a} \leq \infty \,,
    \end{align*}
    which forces $u_j^c=0$ for $j=1,\dots,m$ and $u_j^a=0$ for $j \geq i$. This leads to a higher codimension boundary. However, if $u_i^c=0$ for $i=1,\dots,m$, leading to $\alpha=0$, and the other $u$-variables nonzero, we find an $A_\infty$-term. After the change of coordinates
    \begin{align*}
    \begin{aligned}
        v_i^c &\rightarrow u^1_i\,, & w_i^c &\rightarrow v^1_{i} \,, & & \text{for } i=1,\dots,m\,,\\
        u_i^a &\rightarrow u^2_{n+2-i} \,, & v_{i}^a &\rightarrow u^1_{m+1} v^2_{n+2-i} \,, & w_i^a \rightarrow v_{m+1}^1 v_{n+2-i}^2\,, \, & \text{for } i=1,\dots,n+1\,,
    \end{aligned}
    \end{align*}
    this boundary is identified as $\mathbb{V}_{n} \times \mathbb{V}_{m}$ and 
\begin{align*}
        \boxed{\int_{\partial\mathbb{W}_{0,0,m,n}}\Omega_{0,0,m,n}^a|_{u_i^c=0} \sim \mathcal{V}(a,\mathcal{V}(b,c,\alpha_1,\dots,\alpha_m),\alpha_{m+1},\dots,\alpha_{m+n})}
\end{align*}
    on this boundary, with the exception of $\mathcal{V}(a,\mathcal{V}(b,c),\alpha_1,\dots,\alpha_{n})$, since the recipe required at least one element of $V^\ast$ to be attached to the $c$-leg. An example of a disk diagram at this boundary is shown in Fig. \ref{B1}.

    \item Boundary 2: At this boundary
    \begin{align*}
        u_1^c=v_1^c \,.
    \end{align*}
    The $uv$-chain of (in)equalities then becomes
    \begin{align*}
        1 \leq& \frac{u_{n+1}^a}{v_{n+1}^a} \leq \dots \leq \frac{u_1^a}{v_1^a} \leq \infty
    \end{align*}
    and the closure constraint requires $u_i^c=v_i^c$ for $i=1,\dots,m$ and $u_i^a=v_i^a$ for $i=1,\dots, n+1$. However, this describes a higher codimension boundary, except for $n=0$. In this case, we find, after the change of coordinates
    \begin{align*}
        \begin{aligned}
            v_i^c &\rightarrow u^2_i \,, & w_i^c &\rightarrow v^2_i \,, & & \text{for } i=1,\dots, m \,,\\
            u_1^a &\rightarrow u_{m+1}^2 \,, & v_1^a &\rightarrow u_{m+1}^2 \,, & w_1^a &\rightarrow v_{m+1}^2 \,,
        \end{aligned}
    \end{align*}
    that the boundary is identified as $\mathbb{V}_{m}\times\mathbb{V}_0$ and 
    \begin{align*}
        \boxed{\int_{\partial\mathbb{W}_{0,0,m,0}}\Omega_{0,0,m,0}^a|_{u_1^c=v_1^c} \sim \mathcal{V}(\mathcal{V}(a,b),c,\alpha_1,\dots,\alpha_{m})}
    \end{align*}
    on this boundary. An example of a disk diagram at this boundary is shown in Fig. \ref{B2}.
        \begin{figure}[h!]
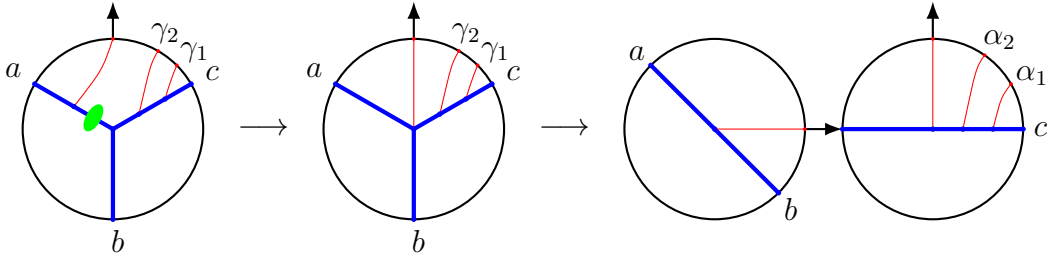

        \centering

        \caption{An example of boundary 2 contributing to $\mathcal{V}(\mathcal{V}(a,b),c,\alpha_1,\alpha_2)$.}
        \label{B2}
    \end{figure}

    \item Boundary 3: At this boundary
    \begin{align*}
        v_1^c&=w_1^c \,.
    \end{align*}
    The $vw$-chain of (in)equalities then becomes
    \begin{align*}
        1 &\leq \frac{v_2^c}{w_2^c} \leq \dots \leq \frac{v_m^c}{w_m^c} \leq \frac{v_{n+1}^a}{w_{n+1}^a} = \dots = \frac{v_1^a}{w_1^a} \leq \infty
    \end{align*}
    and the closure constraint requires $v_i^c=w_i^c$ for $i=1,\dots, m$ and $v_i^a=w_i^a$ for $i=1,\dots,n+1$. However, this describes a higher codimension boundary, except when $m=1$. In that case, we find after the change of coordinates
    \begin{align*}
        \begin{aligned}
            u_1^c &\rightarrow u_1^2 \,, & v_1^c &\rightarrow v_1^2 \,, & w_1^c &\rightarrow v_1^2 \,,\\
            u_i^a &\rightarrow u^2_{n+3-i} \,, & v_i^a &\rightarrow v^2_{n+3-i} \,, & & &\text{for } i=1,\dots,n+1 \,,\\
        \end{aligned}
    \end{align*}
    that the boundary is identified as $\mathbb{V}_{n+1} \times \mathbb{V}_{0}$ and  
    \begin{align*}
        \boxed{\int_{\partial\mathbb{W}_{0,0,1,n}}\Omega_{0,0,1,n}^a|_{v_1^c=w_1^c} \sim \mathcal{V}(a,\mathcal{V}(b,c),\alpha_1,\dots,\alpha_{n+1})}
    \end{align*}
    on this boundary. An example of a disk diagram at this boundary is shown in Fig. \ref{B3}.
\begin{figure}[h!]
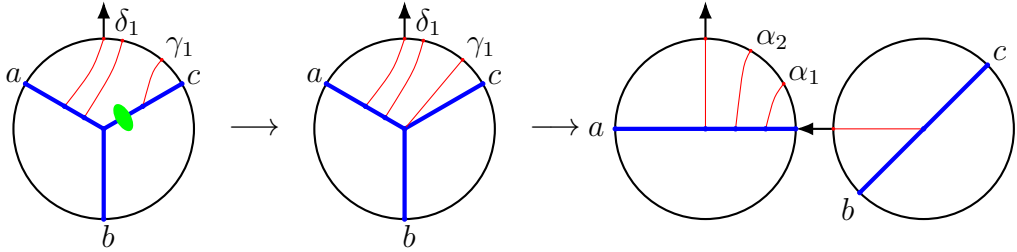

    \centering

    \caption{An example of boundary 3 contributing to $\mathcal{V}(a,\mathcal{V}(b,c),\alpha_1,\alpha_2)$.}
    \label{B3}
\end{figure}

    \item Boundary 4: At this boundary
    \begin{align*}
        w_i^\bullet=1 \,.
    \end{align*}
    If $w_i^a=1$, this requires both $\beta=1$ and $v_i^a=1$ and yields a higher codimension boundary. Next, we consider $w_i^c=1$. Then, the closure condition requires all other $ w$ variables to be zero, which leads to $\beta=0$. This yields a higher codimension boundary, except for $m=1$. After the change of coordinates
    \begin{align*}
        \begin{aligned}
            u_1^c &\rightarrow u^2_{1} \,, & v_1^c &\rightarrow v^2_{1} \,,\\
            u_i^a &\rightarrow u^2_{n+3-i} \,, & v_i^a &\rightarrow v^2_{n+3-i} \,, & \text{for } i=1,\dots,n+1 \,,            
        \end{aligned}
    \end{align*}
    the boundary is identified as $\mathbb{V}_{n+1}\times\mathbb{V}_{0}$ and 
    \begin{align*}
        \boxed{\int_{\partial\mathbb{W}_{0,0,1,n}}\Omega_{0,0,1,n}^a|_{w_1^c=1} \sim \mathcal{V}(a,b,\mathcal{U}(c,\alpha_1),\alpha_2,\dots,\alpha_{n+1})}
    \end{align*}
    on this boundary. An example of a disk diagram at this boundary is shown in Fig. \ref{B4}.
    \begin{figure}[h!]
        \centering

        \caption{An example of boundary 4 contributing to $\mathcal{V}(a,b,\mathcal{U}(c,\alpha_1),\alpha_2)$.}
        \label{B4}
    \end{figure}

    \item Boundary 5: At this boundary
    \begin{align*}
        w_i^\bullet &= 0\,.
    \end{align*}
        \begin{figure}[h!]
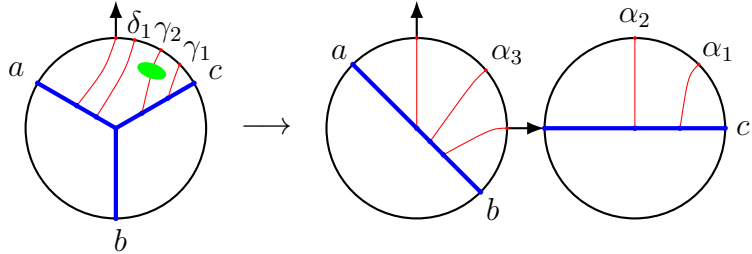

        \centering
        %
        \caption{An example of boundary 5 contributing to $\mathcal{V}(a,b,\mathcal{U}(c,\alpha_1,\alpha_2),\alpha_3)$.}
        \label{B5}
    \end{figure}
    If $w_i^c=0$, the $uw$-chain is equivalent to
    \begin{align*}
        \infty \geq \frac{w_1^c}{u_1^c} \geq \dots \geq \frac{0^{+}}{u_i^c} \geq \dots \geq \frac{w_m^c}{u_m^c} \geq \frac{w_{n+1}^a}{u_{n+1}^a} \geq \dots \geq \frac{w_1^a}{u_1^a} \geq 0\,,
    \end{align*}
    which forces $w_j^c=0$ for $j \leq i$ and $w_j^a=0$ for $j=1,\dots,n+1$. This leads to a higher codimension boundary. Next we consider $w_i^a=0$, which leads to $\beta=0$. For $m=1$ this is equivalent to boundary 4. After the change of coordinates
    \begin{align*}
    \begin{aligned}
        u_1^c &\rightarrow u^2_{1} u^1_1 \,,\\
        v_i^c &\rightarrow v^2_{1} u^1_i\,, & w_i^c &\rightarrow v^1_i \,, & \text{for } i=1,\dots,m \,,\\
        u_{i}^a &\rightarrow u^2_{n+3-i} \,, & v_{i}^a &\rightarrow v^2_{n+3-i} \,, & \text{for } i=1,\dots,n+1
    \end{aligned}
    \end{align*}
    the boundary is identified as $\mathbb{V}_{n+1}\times\mathbb{V}_{m-1}$ and 
    \begin{align*}
        \boxed{\int_{\partial\mathbb{W}_{0,0,m,n}}\Omega_{0,0,m,n}^a|_{w_{i}^a=0} \sim \mathcal{V}(a,b,\mathcal{U}(c,\alpha_,\dots,\alpha_{m}),\alpha_{m+1}\dots,\alpha_{m+n})}
    \end{align*}
    on this boundary, with the exception of $\mathcal{V}(a,b,\mathcal{U}(c,\alpha_1),\alpha_{2},\dots,\alpha_{n+1})$. An example of a disk diagram at this boundary is shown in Fig. \ref{B5}.
\end{itemize}

\paragraph{Gluing terms.}
\begin{itemize}
\item Boundary 6: At this boundary
\begin{align*}
    \frac{v_m^c}{w_m^c} &= \frac{v_{n+1}^a}{w_{n+1}^a} \,.
\end{align*}
For $m=1$ the $vw$-chain becomes
\begin{align*}
    0 \leq \frac{v_1^c}{w_1^c} = \frac{v_{n+1}^a}{w_{n+1}^a} = \dots = \frac{v_1^a}{w_1^a}  \leq \infty \,.
\end{align*}
\begin{figure}[h!]
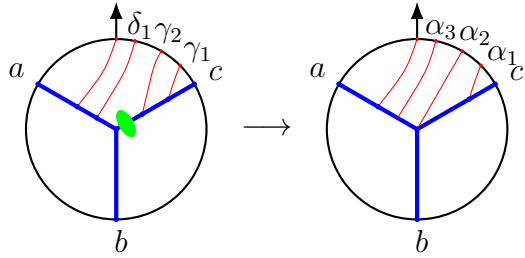

    \centering

    \caption{An example of boundary 6 contributing to a gluing term.}
    \label{B6}
\end{figure}
The closure constraint then requires $v_i^\bullet=w_i^\bullet$, so this boundary is equivalent to boundary 3 when $m=1$. For $m \neq 1$, $\int_{\partial\mathbb{W}_{0,0,m,n}}\Omega_{0,0,m,n}^a$ does not yield a familiar $A_\infty$-term and it does not vanish either. An example of a
disk diagram at this boundary is shown in Fig. \ref{B6}.

\item Boundary 7: At this boundary
\begin{align*}
    \frac{u_1^c}{v_1^c} &= \frac{u_{n+1}^a}{v_{n+1}^a}\,.
\end{align*}
For $n=0$ the $uv$-chain becomes
\begin{align*}
    0 \leq \frac{u_1^c}{v_1^c} = \dots = \frac{u_{m}^c}{v_{m}^c} = \frac{u_1^a}{v_1^a} \leq \infty \,.
\end{align*}
The closure constraint then requires $u_i^\bullet=v_i^\bullet$, so this boundary is equivalent to boundary 2 when $n=0$. For $n \neq 0$, $\int_{\partial\mathbb{W}_{0,0,m,n}}\Omega_{0,0,m,n}^a$ does not yield a familiar $A_\infty$-term and it does not vanish either. An example of a
disk diagram at this boundary is shown in Fig. \ref{B7}.
\begin{figure}[h!]
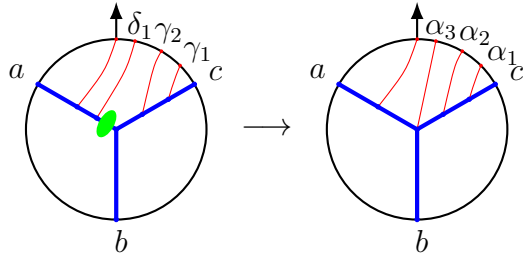

    \centering

    \caption{An example of boundary 7 contributing to a gluing term.}
    \label{B7}
\end{figure}
\end{itemize}

Fortunately, it turns out not to be necessary to explicitly evaluate $\Omega_{0,0,m,n}^a$ on boundary 6 and 7, as the contributions can be seen to cancel each other by construction. $\Omega_{0,0,m,n}^a$ on boundary 6 is found by restricting $\Omega_{m+n}^a$ to a submanifold by requiring
\begin{align*}
    \frac{u_1^c}{v_1^c}=&\dots=\frac{u_{m}^c}{v_m^c} \,, & \frac{v_m^c}{w_m^c}=\frac{v_{n+1}^a}{w_{n+1}^a}=\dots=\frac{v_1^a}{w_1^a} \,,
\end{align*}
while on boundary 7 one invokes
\begin{align*}
    \frac{u_1^c}{v_1^c}=&\dots=\frac{u_m^c}{v_m^c}=\frac{u_{n+1}^a}{v_{n+1}^a} \,, & \frac{v_{n+1}^a}{w_{n+1}^a}=\dots=\frac{v_1^a}{w_1^a} \,.
\end{align*}
Clearly, boundary 6 is equivalent to boundary 7 after the shift $m \rightarrow m+1$ and $n \rightarrow n-1$ and relabeling of the variables. Since \eqref{StokesToA_infty} sums over all $m,n$, such that $m+n=N$, all gluing terms cancel pairwise.

\paragraph{Zero measure terms.} Remember the definition $\vec{q}_{\bullet,i}=(u_i^\bullet,v_i^\bullet,w_i^\bullet)$ for the vectors that fill up the matrix $Q_D$. From \eqref{measure} it is clear that if two $q$-vectors are colinear , e.g. $\vec{q}_{c,i} = \xi \vec{q}_{c,j}$, with $\xi\in\mathbb{R}$, then $\mu_i^c \wedge \mu_j^c = 0$.
\begin{itemize}
\item Boundary 8: At this boundary
\begin{align*}
    \frac{v_i^c}{w_i^c} &= \frac{v_{i+1}^c}{w_{i+1}^c} \,, & \text{for } i=1,\dots ,m-1 \,,
\end{align*}
which makes the vectors $\vec{q}_{c,i}$ and $\vec{q}_{c,i+1}$ colinear and the measure in $\Omega_{0,0,m,n}^a$ vanishes on this boundary.

\item Boundary 9: At this boundary
\begin{align*}
    \frac{u_{i}^a}{v_{i}^a} &= \frac{u_{i+1}^a}{v_{i+1}^a} \quad \text{for } i=1,\dots, n \,,
\end{align*}
which makes the vectors $\vec{q}_{a,i}$ and $\vec{q}_{a,i+1}$ colinear. As a result the measure in $\Omega_{0,0,m,n}^a$ vanishes on this boundary.
\end{itemize}

\paragraph{Higher codimension boundaries.} Some types of boundaries are necessarily higher codimension boundaries: saturating one inequality leads to saturation of more inequalities and hence the resulting submanifold is parameterized by less than $2(m+n)$ independent coordinates. We have seen examples of this for the boundaries that also produce $A_\infty$-terms for specific values of $m$ and $n$. Here we present the higher codimension boundaries that are not discussed yet.

\begin{itemize}

\item Boundary 10: At this boundary
\begin{align*}
    v_i^\bullet=0 \,.
\end{align*}
The $vw$-chain becomes
\begin{footnotesize}
\begin{align*}
    \begin{aligned}
        0 \leq& \frac{v_1^c}{w_1^c} \leq \dots \leq \frac{0}{w_i^c} \leq \dots \leq \frac{v_m^c}{w_m^c} \leq \frac{v_{n+1}^a}{w_{n+1}^a} = \dots = \frac{v_1^a}{w_1^a} \leq \infty \,, \quad \text{or} \\
        0 \leq& \frac{v_1^c}{w_1^c} \leq \dots \leq \frac{v_m^c}{w_m^c} \leq \frac{v_{n+1}^a}{w_{n+1}^a} = \dots = \frac{0}{w_{i}^a} = \dots = \frac{v_1^a}{w_1^a}  \leq \infty \,, 
    \end{aligned}
\end{align*}
\end{footnotesize}\noindent
which implies $v_j^c=0$ for $j < i$ or all $v$-variables zero, respectively. In both cases one finds a higher codimension boundary. Only for $v_1^c=0$, the $vw$-chain is not responsible for a higher codimension boundary, but the condition
\begin{align*}
    0 \leq u_1^c \leq v_1^c \leq w_1^c \leq 1
\end{align*}
is, as it imposes $u_1^c=v_1^c=0$.

\item Boundary 11: At this boundary
\begin{align*}
    u_i^\bullet=1 \,.
\end{align*}
The closure constraint implies that all other $u$-variables vanish. If $m=1$ and $n=0$ and the boundary is given by $u_1^c=1$, the condition
\begin{align*}
    0 \leq u_1^c \leq v_1^c \leq w_1^c \leq 1
\end{align*}
 implies $u_1^c=v_1^c=w_1^c=1$.

\item Boundary 12: At this boundary
\begin{align*}
    v_i^\bullet=1 \,.
\end{align*}
The closure constraint implies that all other $v$-variables vanish. If $m=1$ and $n=0$ and the boundary is given by $v_1^c=1$, the condition
\begin{align*}
    0 \leq u_1^c \leq v_1^c \leq w_1^c \leq 1
\end{align*}
implies $v_1^c=w_1^c=1$.

\item Boundary 13: At this boundary
\begin{align*}
    \frac{u_m^c}{w_m^c}=\frac{u_{n+1}^a}{w_{n+1}^a}\,.
\end{align*}
Since
\begin{align*}
    \frac{u_m^c}{v_m^c}\frac{v_m^c}{w_m^c} =& \frac{u_m^c}{w_m^c} \leq \frac{u_{n+1}^a}{w_{n+1}^a} =\frac{u_{n+1}^a}{v_{n+1}^a}\frac{v_{n+1}^a}{w_{n+1}^a}
\end{align*}
and
\begin{align*}
    \frac{u_m^c}{v_m^c} \leq& \frac{u_{n+1}^a}{v_{n+1}^a} \,, \quad \frac{v_m^c}{w_m^c} \leq \frac{v_{n+1}^a}{w_{n+1}^a} \,,
\end{align*}
this boundary implies $\frac{u_m^c}{v_m^c} = \frac{u_{n+1}^a}{v_{n+1}^a}$ and $\frac{v_m^c}{w_m^c} = \frac{v_{n+1}^a}{w_{n+1}^a}$ and thus yields a higher codimension boundary.
\end{itemize}

With the above categorization of boundaries, we have established the equivalence of Stokes' theorem for $\Omega_{0,0,m,n}^a$ on $\mathbb{W}_{0,0,m,n}$ and the left-ordered $A_\infty$-relation, \eqref{StokesToA_infty},
that is,
\begin{align} \label{refinedStokes}
    0=\sum_{m+n=N}\int_{\mathbb{W}_{0,0,m,n}} d\Omega_{0,0,m,n}^a = \sum_{m+n=N}\int_{\partial\mathbb{W}_{0,0,m,n}} \Omega_{0,0,m,n}^a  \quad \Longleftrightarrow \quad A_\infty\text{-relations} \,. 
\end{align}
The same proof for the right-ordered $A_\infty$-relations can easily be inferred from \eqref{flip}, which relates potentials of the type $\Omega_{k,l,m,n}^a$ with $\Omega_{k,l,m,n}^c$. The disk diagram of the potential $\Omega_{m,0,0,n}^c$ is just the mirror image of the disk diagram for $\Omega_{0,0,m,n}$. It is easy to see from \eqref{generalA_infty} that the right-ordered $A_\infty$-terms can be obtained from the left-ordered ones by mirror symmetry as well.
\subsection{All order generalization: arbitrary ordering} \label{sec:allOrderings}
In the previous section, we considered the $A_\infty$-relations with a specific ordering. Now we turn to the $A_\infty$-relations with arbitrary ordering. These $A_\infty$-relations are given in \eqref{A} and the $A_\infty$-terms are presented in \eqref{generalA_infty}. The corresponding $Q$ matrices are given in \eqref{Qs}. As we explained in Sec. \ref{sec:recipe}, there are two types of diagrams and potentials: the $a$- and $c$- diagrams that correspond to potentials $\Omega_{k,l,m,n}^a$ and $\Omega_{k,l,m,n}^c$, respectively. Two examples of these types of diagrams are shown in Fig. \ref{SD6}. We also illustrated how the potentials related to the latter type of diagram can be obtained from the former in \eqref{flip}. Hence, we will thoroughly discuss the $a$-diagrams, after which we only state the results for the $c$-diagrams that contribute to the proof.

\begin{figure}
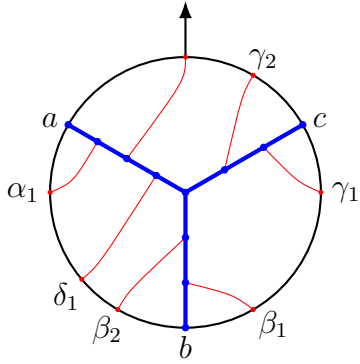

    \centering

    \caption{A disk diagram that corresponds to $\Omega_{1,2,2,1}^a(a,\alpha_1,\delta_1,\beta_2,b,\beta_1,\gamma_1,c,\gamma_2)$.}
    \label{generic}
\end{figure}

We consider the potentials $\Omega_{k,l,m,n}^a$ that we constructed in Sec. \ref{sec:recipe}. We note that the restriction \eqref{R} to $\mathbb{W}_{k,l,m,n} \subset \mathbb{U}_{N}$ can be solved in a multitude of ways. It will be a matter of convenience. In the left-ordered case, we required the coordinates to be nonsingular on all of $\mathbb{W}_{0,0,m,n}$, but it turns out that there is not a single chart that covers all of $\mathbb{W}_{k,l,m,n}$. This is easy to see from the equalities in the $uw$-chain that allows one to solve, for instance, $u_{l}^b=\frac{\alpha}{\beta}w_{l}^b$ or $w_{l}^b=\frac{\beta}{\alpha}u_{l}^b$, which both can become singular. Therefore, we will have to use various charts to describe the boundaries. The charts are constructed by choosing
\begin{align*}
    \begin{aligned}
        u_i^c &= \alpha v_i^c \quad\text{or}\quad v_i^c = \frac{u_i^c}{\alpha} \,, & \text{for } i=1,\dots,m \,,\\
        u_{i}^b &= \frac{\alpha}{\beta} w_{i}^b \quad\text{or}\quad w_{i}^b = \frac{\beta}{\alpha} u_{i}^b \,, & \text{for } i=1,\dots l \,,\\
        w_{i}^a &= \beta v_{i}^a \quad\text{or}\quad v_{i}^a = \frac{w_{i}^a}{\beta} \,, & \text{for } i=1,\dots,k+n+1 \,.
    \end{aligned}
\end{align*}
It will be clear from the context which chart was chosen. 

Like in the left-ordered case, here follows a categorization of the potentials evaluated on all boundaries of $\mathbb{W}_{k,l,m,n}$. Remember that for the left-ordered case $k=l=0$. These cases will be included in the following discussion, but not only as the left-ordered case: potentials with no lines attached to the $a$- and $b$-leg can still contribute to different orderings. One difference with the left-ordered case is that on the boundaries where we find $A_\infty$-terms, we find gluing terms as well, depending on the orientation of the red lines in the diagrams. Moreover, in the following categorization of $a$-diagrams we only consider potentials $\Omega_{k,l,m,n}^a$ and it is therefore not sufficient to prove \eqref{StokesToA_infty}. This categorization, however, is followed by a recipe for extracting the same information for the potentials $\Omega_{k,l,m,n}^c$ and a brief categorization of the $c$-diagrams that contribute to Stokes' theorem. This completes the proof.

\paragraph{$a$-diagrams.} The boundaries that yield $A_\infty$-terms are a bit more subtle than before. Namely, some boundaries that produce $A_\infty$-terms equally produce gluing terms, depending on the orientation of a particular line attached to one of the legs. This happens for boundaries where the green region is drawn on either one of the legs near the junction and for those diagrams we will show both options.
\begin{itemize}
    \item Boundary 1: At this boundary
    \begin{align*}
        u_{i}^\bullet=0 \,.
    \end{align*}
    If $u_i^a=0$, the $uv$-chain becomes
    \begin{align*}
        0 \leq& \frac{u_1^b}{v_1^b} \leq \dots \leq \frac{u_{l}^b}{v_{l}^b} \leq \frac{u_m^c}{v_m^c} = \dots = \frac{u_1^c}{v_1^c} \leq \frac{u_{k+n+1}^a}{v_{k+n+1}^a} \leq \dots \leq \frac{0}{v_i^a} \leq \dots \leq \frac{u_1^a}{v_1^a}\leq \infty\,,
    \end{align*}
    which leads to $u_j^c=0$ for $j=1,\dots,m$, $u_j^b=0$ for $j=1,\dots,l$ and $u_j^a=0$ for $j\geq i$. This yields a higher codimension boundary. Next we consider $u_i^c=0$ for $i=1,\dots,m$ and $u_i^b=0$ for $i=1,\dots,l$, which leads to $\alpha=0$. After the change of coordinates
    \begin{align*}
        \begin{aligned}
            v_i^c &\rightarrow u^1_i \,, & w_i^c &\rightarrow v^1_i \,, & & \text{for } i=1,\dots,m \,,\\
            v_i^b &\rightarrow u^1_{m+l+2-i} \,, & w_i^b &\rightarrow v^1_{m+l+2-i} \,, & & \text{for } i=1,\dots, l \,,\\
            u_{i}^a &\rightarrow u^2_{n+k+2-i} \,, & v_{i}^a &\rightarrow u^1_{m+1}v^2_{n+k+2-i} \,, & w_i^a &\rightarrow v_{m+1}^1v_{n+k+2-i}^2\,, & \text{for } i=1,\dots,k+n+1 \,,
        \end{aligned}
    \end{align*}
    the boundary is identified as $\mathbb{V}_{k+n}\times\mathbb{V}_{m+l}$ and
    \begin{align*}
        \boxed{
        \begin{aligned}
        &\int_{\partial\mathbb{W}_{k,l,m,n}}\Omega_{k,l,m,n}^{a}|_{u_i^c=u_i^b=0} \sim\\
        &\sim\mathcal{V}(\bullet,\dots,\bullet,a,\bullet,\dots,\bullet,\mathcal{V}(\bullet,\dots,\bullet,b,\bullet,\dots,\bullet,c,\bullet,\dots,\bullet),\bullet,\dots,\bullet)
        \end{aligned}}
    \end{align*}
    on this boundary, with the exception of $\mathcal{V}(\bullet,\dots,\bullet,a,\bullet,\dots,\bullet,\mathcal{V}(b,c),\bullet,\dots,\bullet)$. An example of a disk diagram at this boundary is shown in Fig. \ref{BB1}.
    \begin{figure}[h!]
        \centering

        \caption{An example of boundary 1 contributing to $\mathcal{V}(a,\alpha_1,\alpha_2,\mathcal{V}(\alpha_3,b,\alpha_4,\alpha_5,c,\alpha_6))$.}
        \label{BB1}
    \end{figure}

    \item Boundary 2: At this boundary
    \begin{align*}
        u_1^c=v_1^c \,.
    \end{align*}
    \begin{figure}[h!]
        \centering
        %
        \caption{An example of boundary 2 contributing to $\mathcal{V}(\mathcal{V}(a,b),\alpha_1,c,\alpha_2)$.}
        \label{BB2}
    \end{figure}
    This is only a boundary if $l=0$, in which case it is a higher codimension boundary unless $k=n=0$, like in the left-ordered case. The $uv$-chain then reads
    \begin{align*}
        1= \frac{u_m^c}{v_m^c} \leq \dots \leq \frac{u_1^c}{v_1^c} \leq \frac{u_1^a}{v_1^a}\,.
    \end{align*}
    The closure constraint now forces $u_i^\bullet=v_i^\bullet$. Then, after a change of coordinates
    \begin{align*}
        \begin{aligned}
            v_i^c &\rightarrow u^2_i \,, & w_i^c &\rightarrow v^2_i \,, & &\text{for } i=1,\dots,m \,,\\
            u_1^a &\rightarrow u^2_{m+1} \,, & w_1^a &\rightarrow v_{m+1}^2 \,,
        \end{aligned}
    \end{align*}
    the boundary is identified as $\mathbb{V}_m\times\mathbb{V}_0$ and
    \begin{align*}
        \boxed{\int_{\partial\mathbb{W}_{0,0,m,0}}\Omega_{0,0,m,0}^{a}|_{u_1^c=v_1^c} \sim \mathcal{V}(\mathcal{V}(a,b),\bullet,\dots,\bullet,c,\bullet,\dots,\bullet)}
    \end{align*}
    on this boundary. Since the line connected to the output arrow can only have one orientation, there is only one disk diagram at this boundary shown in Fig. \ref{BB2}.

    \item Boundary 3: At this boundary
    \begin{align*}
        v_1^c=w_1^c\,.
    \end{align*}
    \begin{figure}[h!]
    \centering

    \caption{An example of boundary 3 contributing to $\mathcal{V}(a,\alpha_1,\alpha_2,\mathcal{V}(b,c),\alpha_3)$.}
    \label{BB3}
\end{figure}
    The $vw$-chain becomes
    \begin{align*}
        1 =& \frac{v_1^c}{w_1^c} \leq \dots \leq \frac{v_m^c}{w_m^c} \leq \frac{v_{k+n+1}^a}{w_{k+n+1}^a} = \dots = \frac{v_1^a}{w_1^a}\leq \frac{v_{l}^b}{w_{l}^b} \leq \dots \leq \frac{v_1^b}{w_1^b} \leq \infty \,.
    \end{align*}
    and implies $v_i^\bullet \geq w_i^\bullet$. The closure constraint then leads to $v_i^\bullet=w_i^\bullet$, which gives a higher codimension boundary, except for $l=0, m=1$. Then after the change of coordinates
    \begin{align*}
        \begin{aligned}
            u_1^c &\rightarrow u^2_1 \,, & v_1^c &\rightarrow v^2_1 \,,\\
            u_i^a &\rightarrow u^2_{k+n+3-i} \,, & v_i^a &\rightarrow v^2_{k+n+3-i} \,, & \text{for } i=1,\dots,k+n+1\,,\\
        \end{aligned}
    \end{align*}
    the boundary is identified as $\mathbb{V}_{k+n+1}\times\mathbb{V}_0$ and  
    \begin{align*}
        \boxed{\int_{\partial\mathbb{W}_{k,0,1,n}}\Omega_{k,0,1,n}^{a}|_{v_1^c=w_1^c} \sim \mathcal{V}(\bullet,\dots,\bullet,a,\bullet,\dots,\bullet,\mathcal{V}(b,c),\bullet,\dots,\bullet)}
    \end{align*}    
    on this boundary. An example of a disk diagram at this boundary contributing to the $A_\infty$-term is shown in Fig. \ref{BB3}, while Fig. \ref{BB3G} shows a disk diagram contributing to a gluing term.

\begin{figure}[h!]
    \centering

    \caption{An example of boundary 3 contributing to a gluing term.}
    \label{BB3G}
\end{figure}

    \item Boundary 4: At this boundary
    \begin{align*}
        v_1^b &= w_1^b \,.
    \end{align*}
    \begin{figure}[h!]
        \centering
        %
        \caption{An example of boundary 4 contributing to $\mathcal{V}(a,\alpha_1,\alpha_2,\alpha_3,\mathcal{V}(b,c))$.}
        \label{BB4}
    \end{figure}
    The $vw$-chain becomes
    \begin{align*}
         0 \leq& \frac{v_1^c}{w_1^c} \leq \dots \leq \frac{v_m^c}{w_m^c} \leq \frac{v_{k+n+1}^a}{w_{k+n+1}^a} =\dots= \frac{v_1^a}{w_1^a}\leq \frac{v_{l}^b}{w_{l}^b} \leq \dots \leq \frac{v_1^b}{w_1^b} =1
    \end{align*}
    and implies $v_i^\bullet \leq w_i^\bullet$ for $i=1,\dots,N+1$. The closure constraint then leads to $v_i^\bullet=w_i^\bullet$, which leads to a higher codimension boundary, except for $l=1$, $m=0$. Then, after the change of coordinates
    \begin{align*}
        \begin{aligned}
            u_1^b &\rightarrow u^2_1 \,, & v_1^b &\rightarrow v^2_1 \,,\\
            u_i^a &\rightarrow u^2_{k+n+3-i} \,, & v_i^a &\rightarrow v^2_{k+n+3-i} \,, & \text{for } i=1,\dots,k+n+1 \,,\\
        \end{aligned}
    \end{align*}
    the boundary is identified as $\mathbb{V}_{k+n+1}\times\mathbb{V}_0$ and  
    \begin{align*}
        \boxed{\int_{\partial\mathbb{W}_{k,1,0,n}}\Omega_{k,1,0,n}^{a}|_{v_1^b=w_1^b} \sim \mathcal{V}(\bullet,\dots,\bullet,a,\bullet,\dots,\bullet,\mathcal{V}(b,c),\bullet,\dots,\bullet)}
    \end{align*}    
    on this boundary. An example of a disk diagram at this boundary contributing to the $A_\infty$-term is shown in Fig. \ref{BB4}, while Fig. \ref{BB4G} shows a disk diagram contributing to a gluing term.
    \begin{figure}[h!]
        \centering

        \caption{An example of boundary 4 contributing to a gluing term.}
        \label{BB4G}
    \end{figure}

    \item Boundary 5: At this boundary
    \begin{align*}
        w_i^\bullet=1\,.
    \end{align*}
    The closure constraint forces all other $w$-variables to be zero. This yields a higher codimenion boundary, except for $m=1$ and $w_1^c=1$, which leads to $\beta=0$. After the change of coordinates
    \begin{align*}
        \begin{aligned}
            u_1^c &\rightarrow u^2_{l+1} \,, & v_1^c &\rightarrow v^2_{l+1}\,,\\
            u_i^b & \rightarrow u^2_{i} \,, & v_i^b &\rightarrow v^2_{i} \,, & \text{for } i=1,\dots,l\,,\\
            u_{i}^a &\rightarrow u^2_{k+n+l+3-i} \,, & v_i^a &\rightarrow v^2_{k+n+l+3-i} \,, & \text{for } i=1,\dots,k+n+1 \,.
        \end{aligned}
    \end{align*}
    the boundary is identified as $\mathbb{V}_{k+l+n+1}\times\mathbb{V}_0$. The canonical ordering of the $q$-vectors in $Q$ in the resulting nested disk diagram is different than the canonical ordering of the corresponding $A_\infty$-term. We therefore rewrite the matrix $Q$ as
    \begin{align*}
        Q &= (
            \vec{q}_a , \vec{q}_b , \vec{q}_c , \vec{q}_{c,1} \vec{q}_{b,_1} , \dots , \vec{q}_{b,l} , \vec{q}_{a,k+n+1}, \dots , \vec{q}_{a,1}
        ) \,.
    \end{align*}
    It can now be seen that 
    \begin{align*}
        \boxed{\int_{\partial\mathbb{W}_{k,l,1,n}}\Omega_{k,l,1,n}^{a}|_{w_1^c=1} \sim \mathcal{V}(\bullet,\dots,\bullet,a,\bullet,\dots,\bullet,b,\bullet,\dots,\bullet,\mathcal{U}(c,\bullet),\bullet,\dots,\bullet)}
    \end{align*}
    and
    \begin{align*}
        \boxed{\int_{\partial\mathbb{W}_{k,l,1,n}}\Omega_{k,l,1,n}^{a}|_{w_1^c=1} \sim \mathcal{V}(\bullet,\dots,\bullet,a,\bullet,\dots,\bullet,b,\bullet,\dots,\bullet,\mathcal{U}(\bullet,c),\bullet,\dots,\bullet)}
    \end{align*}
    on this boundary. An example of a disk diagram at this boundary is shown in Fig. \ref{BB5}.

    \begin{figure}[h!]
        \centering

        \caption{An example of boundary 5 contributing to $\mathcal{V}(a,\alpha_1,\alpha_2,\alpha_3,b,\alpha_4,\mathcal{U}(\alpha_5,c))$.}
        \label{BB5}
    \end{figure}
    
    \item Boundary 6: At this boundary
    \begin{align*}
        w_i^\bullet=0 \,.
    \end{align*}
    For $m=1$ and $w_1^c=1$, this is the same as boundary 5. Otherwise, if $w_i^c=0$, the $uw$-chain is equivalent to
    \begin{align*}
        0 \geq& \frac{w_1^a}{u_1^a} \geq \dots \geq \frac{w_{k+n+1}^a}{u_{k+n+1}^a} \geq \frac{w_{1}^b}{u_{1}^b} = \dots = \frac{w_l^b}{u_l^b} \geq \frac{w_{m}^a}{u_{m}^c} \geq \dots \geq \frac{0^+}{u_i^c} \geq \dots \geq \frac{w_1^c}{u_1^c}\geq \infty \,,
    \end{align*}
    which forces $w_j^c=0$ for $j>i$ and $w_k^a=w_k^b=0$ for any $k$. This leads to a higher codimension boundary. Only when we consider $w_i^a=0$ for $i=1,\dots,k+n+1$ and $w_i^b=0$ for $i=1,\dots,l$, we find an $A_\infty$-term. Then, after the change of coordinates
    \begin{align*}
        \begin{aligned}
            u_1^c &\rightarrow u^2_{l+1} u^1_1 \,,\\
            v_i^c &\rightarrow v^2_{l+1} u^1_i \,, & w_i^c & \rightarrow v^1_i \,, & \text{for } i=1,\dots,m\\
            u_i^b &\rightarrow u^2_{i} \,, & v_i &\rightarrow v^2_{i} \,, & \text{for } i=1,\dots, l \,,\\
            u_{i}^a & \rightarrow u^2_{k+l+n+3-i} \,, & v_{i}^a & \rightarrow v^2_{k+l+n+3-i} \,, & \text{for } i=1,\dots,k+n+1
        \end{aligned}
    \end{align*}
    the boundary is identified as $\mathbb{V}_{k+l+n+1}\times\mathbb{V}_{m-1}$. We also change the matrix $Q$, such that it corresponds to the canonical ordering for nested vertices. It then reads
    \begin{align*}
        Q &= (
            \vec{q}_a , \vec{q}_b , \vec{q}_c , \vec{q}_{c,1} , \dots , \vec{q}_{c,m} , \vec{q}_{b,1} , \dots , \vec{q}_{b,l} , \vec{q}_{a,k+n+1} , \dots , \vec{q}_{a,1}
        ) \,.
    \end{align*}
    It can now be seen that 
    \begin{align*}
        \boxed{\int_{\partial\mathbb{W}_{k,l,m,n}}\Omega_{k,l,m,n}^{a}|_{w_{i}^a=w_i^b=0} \sim \mathcal{V}(\bullet,\dots,\bullet,a,\bullet,\dots,\bullet,b,\bullet,\dots,\bullet,\mathcal{U}(\bullet,\dots,\bullet,c,\bullet,\dots,\bullet),\bullet,\dots,\bullet)}
    \end{align*}
    on this boundary, with the exception of $\mathcal{V}(\bullet,\dots,\bullet,a,\bullet,\dots,\bullet,b,\bullet,\dots,\bullet,\mathcal{U}(c,\bullet),\bullet,\dots,\bullet)$ and $\mathcal{V}(\bullet,\dots,\bullet,a,\bullet,\dots,\bullet,b,\bullet,\dots,\bullet,\mathcal{U}(\bullet,c),\bullet,\dots,\bullet)$. An example of a disk diagram at this boundary is shown in Fig. \ref{BB6}.

    \begin{figure}[h!]
        \centering

        \caption{An example of boundary 6 contributing to $\mathcal{V}(a,\alpha_1,\alpha_2,\alpha_3,b,\alpha_4,\mathcal{U}(\alpha_5,c,\alpha_6))$.}
        \label{BB6}
    \end{figure}

    \item Boundary 7: At this boundary
    \begin{align*}
        v_i^\bullet=1\,.
    \end{align*}
    \begin{figure}[h!]
        \centering
        %
        \caption{An example of boundary 7 contributing to $\mathcal{V}(a,\alpha_1,\alpha_2,\mathcal{U}(b,\alpha_3),\alpha_4,c,\alpha_5)$.}
        \label{BB7}
    \end{figure}

    The closure constraint forces all other $v$-variables to be zero. This yields a higher codimension boundary, except for $l=1$ and $v_1^b=1$. Then, after the change of variables
    \begin{align*}
        \begin{aligned}
            u_i^c &\rightarrow u^2_i \,, & w_i^c &\rightarrow v^2_i \,, & \text{for } i=1,\dots,m \,,\\
            u_1^b &\rightarrow u^2_{m+1} \,, & w_1^b &\rightarrow v^2_{m+1} \,,\\
            u_i^a &\rightarrow u^2_{k+m+n+3-i} \,, & w_i^a &\rightarrow v^2_{k+m+n+3-i} \,, & \text{for } i=1,\dots,k+n+1
        \end{aligned}
    \end{align*}
    the boundary is identified as $\mathbb{V}_{k+m+n+1}\times\mathbb{V}_0$. We also change the matrix $Q$, such that it corresponds to the canonical ordering for nested vertices. It then reads
    \begin{align*}
        Q &= (
            \vec{q}_a , \vec{q}_b , \vec{q}_c , \vec{q}_{b,1}, \vec{q}_{c,1} , \dots , \vec{q}_{c,m} , \vec{q}_{a,k+n+1} , \dots , \vec{q}_{a,1}
        ) \,.
    \end{align*}
    It can now be seen that 
    \begin{align*}
        \boxed{\int_{\partial\mathbb{W}_{k,1,m,n}}\Omega_{k,1,m,n}^{a}|_{v_{1}^b=1} \sim \mathcal{V}(\bullet,\dots,\bullet,a,\bullet,\dots,\bullet,\mathcal{U}(b,\bullet),\bullet,\dots,\bullet,c,\bullet,\dots,\bullet)}
    \end{align*}
    and
    \begin{align*}
        \boxed{\int_{\partial\mathbb{W}_{k,1,m,n}}\Omega_{k,1,m,n}^{a}|_{v_{1}^b=1} \sim \mathcal{V}(\bullet,\dots,\bullet,a,\bullet,\dots,\bullet,\mathcal{U}(\bullet,b),\bullet,\dots,\bullet,c,\bullet,\dots,\bullet)}
    \end{align*}
    on this boundary. An example of a disk diagram at this boundary is shown in Fig. \ref{BB7}.

    \item Boundary 8: At this boundary 
    \begin{align*}
        v_i^\bullet=0 \,.
    \end{align*}
    \begin{figure}[h!]
        \centering

        \caption{An example of boundary 8 contributing to $\mathcal{V}(a,\alpha_1,\alpha_2,\mathcal{U}(\alpha_3,b,\alpha_4),\alpha_5,c,\alpha_6)$.}
        \label{BB8}
    \end{figure}
    The $vw$-chain becomes
    \begin{align*}
        \begin{aligned}
            0 &\leq \frac{v_1^c}{w_1^c} \leq \dots \leq \frac{v_m^c}{w_m^c} \leq \frac{v_{k+n+1}^a}{w_{k+n+1}^a} = \dots =\frac{0^+}{w_i^a} = \dots = \frac{v_1^a}{w_1^a}\leq \frac{v_{l}^b}{w_{l}^b} \leq \dots \leq \frac{v_1^b}{w_1^b} \leq \infty \,,\\
            0 &\leq \frac{v_1^c}{w_1^c} \leq \dots \leq \frac{v_m^c}{w_m^c} \leq \frac{v_{k+n+1}^a}{w_{k+n+1}^a} = \dots = \frac{v_1^a}{w_1^a}\leq \frac{v_{l}^b}{w_{l}^b} \leq \dots \leq \frac{0^+}{w_i^b} \dots \leq \frac{v_1^b}{w_1^b} \leq \infty
        \end{aligned}
    \end{align*}
    if $v_i^a=0$ and $v_i^b=0$, respectively. This forces all $v_i^a=0$ and $v_i^c=0$ in both cases, while in the latter we also have $v_j^b=0$ for $j>i$.  This leads to a higher codimension boundary. It is only when $v_i^c=0$ for $i=1,\dots,m$ and $v_i^a=0$ for $i=1,\dots,k+n+1$, that one finds an $A_\infty$-term. After the change of variables
     \begin{align*}
        \begin{aligned}
            u_{i}^c &\rightarrow u^2_{i} \,, & w_{i}^c &\rightarrow v^2_{i}\,, &&& \text{for } i=1,\dots,m \,,\\
            u_i^b &\rightarrow u^2_{m+1}u^1_i \,,&
            v_{i}^b &\rightarrow v^1_i \,, & w_{i}^b &\rightarrow v^2_{m+1}u^1_1\,, & \text{for } i=1,\dots,l \,,\\
            u_{i}^a &\rightarrow u^2_{k+m+n+3-i} \,, & w_{i}^a &\rightarrow v^2_{k+m+n+3-i}\,, &&& \text{for } i=1,\dots,k+n+1 \,,\\
        \end{aligned}
    \end{align*}
    the boundary is identified as $\mathbb{V}_{k+m+n+1}\times\mathbb{V}_{l-1}$. We also change the matrix $Q$, such that it corresponds to the canonical ordering for nested vertices. It then reads
    \begin{align*}
        Q &= (
            \vec{q}_a , \vec{q}_b , \vec{q}_c , \vec{q}_{b,1} , \dots , \vec{q}_{b,l} , \vec{q}_{c,1} , \dots , \vec{q}_{c,m} , \vec{q}_{a,k+n+1} , \dots , \vec{q}_{a,1}
        ) \,.
    \end{align*}
    It can now be seen that 
    \begin{align*}
        \boxed{\int_{\partial\mathbb{W}_{k,l,m,n}}\Omega_{k,l,m,n}^{a} \sim \mathcal{V}(\bullet,\dots,\bullet,a,\bullet,\dots,\bullet,\mathcal{U}(\bullet,\dots,\bullet,b,\bullet,\dots,\bullet),\bullet,\dots,\bullet,c,\bullet,\dots,\bullet)}
    \end{align*}
    on this boundary, with the exception of $\mathcal{V}(\bullet,\dots,\bullet,a,\bullet,\dots,\bullet,\mathcal{U}(b,\bullet),\bullet,\dots,\bullet,c,\bullet,\dots,\bullet)$ and $\mathcal{V}(\bullet,\dots,\bullet,a,\bullet,\dots,\bullet,\mathcal{U}(\bullet,b),\bullet,\dots,\bullet,c,\bullet,\dots,\bullet)$. An example of a disk diagram at this boundary is shown in Fig. \ref{BB8}.

\end{itemize}

\paragraph{Gluing terms.}
\begin{itemize}

    \item Boundary 9: 
    
    At this boundary
    \begin{align*}
        \frac{u_m^c}{v_m^c}=\frac{u_{l}^b}{v_{l}^b}\,.
    \end{align*}
    \begin{figure}[h!]
       \centering

       \caption{Two examples of boundary 9 contributing to a gluing term, with both orientations of $\beta_2$.}
       \label{BB9}
   \end{figure}
   $\int_{\partial\mathbb{W}_{k,l,m,n}}\Omega_{k,l,m,n}^{a}$ does not yield a familiar $A_\infty$-term on this boundary and it does not vanish either. An example of a disk diagram at this boundary is shown in Fig. \ref{BB9}.

   \item Boundary 10: At this boundary
    \begin{align*}
        \frac{u_1^c}{v_1^c}=\frac{u_{k+n+1}^a}{v_{k+n+1}^a} \,.
    \end{align*}
    For $k=l=n=0$ the $uv$-chain becomes
    \begin{align*}
        0 \leq \frac{u_m^c}{v_m^c} = \dots = \frac{u_1^c}{v_1^c} = \frac{u_{1}^a}{v_{1}^a} \leq \infty\,.
    \end{align*}
    The closure constraint then gives $u_i^\bullet=v_i^\bullet$, so this boundary is equivalent to boundary 2 when $k=l=n=0$ and does not yield  a gluing term, but an $A_\infty$-term instead. Otherwise, $\int_{\partial\mathbb{W}_{k,l,m,n}}\Omega_{k,l,m,n}^{a}$ does not yield a familiar $A_\infty$-term on this boundary and it does not vanish either. An example of a disk diagram at this boundary is shown in Fig. \ref{BB10}. Gluing terms corresponding to the same type as the left diagram cancel with gluing terms belonging to the type of diagrams on the left of Fig. \ref{BB9}. A special type of gluing term with no elements between the junction and the output arrow, i.e. $n=0$, is shown in Fig. \ref{BB10S}. This is the only type of gluing term that does not cancel with any other gluing term from the potential $\Omega_{k,l,m,n}^a$, but rather from the type $\Omega_{k,l,m,n}^c$, `gluing' them together.

       \begin{figure}[h!]
       \centering

       \caption{Two examples of boundary 10 contributing to a gluing term, with both orientations of $\delta_1$.}
       \label{BB10}
   \end{figure}

   \begin{figure}[h!]
       \centering
       %
       \caption{Special case of a gluing term coming from boundary 10. This gluing term is cancelled by a gluing term coming from a potential of the type $\Omega_{k,l,m,n}^c$.}
       \label{BB10S}
   \end{figure}

    \item Boundary 11: At this boundary
    \begin{align*}
        \frac{v_m^c}{w_m^c} =& \frac{v_{k+n+1}^a}{w_{k+n+1}^a}  \,.
    \end{align*}
    For $l=0$, $m=1$ the $vw$-chain reads
    \begin{align*}
         0 \leq& \frac{v_1^c}{w_1^c} = \frac{v_{k+n+1}^a}{w_{k+n+1}^a} = \dots = \frac{v_{1}^a}{w_{1}^a} \leq \infty\,.
    \end{align*}
    The closure constraint then gives $v_i^\bullet=w_i^\bullet$, so this boundary is equivalent to boundary 3 when $l=0$ and $m=1$, producing either a gluing term or an $A_\infty$-term. Otherwise, $\int_{\partial\mathbb{W}_{k,l,m,n}}\Omega_{k,l,m,n}^{a}$ does not yield a familiar $A_\infty$-term on this boundary and it does not vanish either. An example of a disk diagram at this boundary is shown in Fig. \ref{BB11}. Gluing terms corresponding to the type of diagrams on the left cancel with gluing terms belonging to the type of diagrams on the right of Fig. \ref{BB10}, while the type of diagrams on the right cancel with the diagrams of the type on the right of Fig. \ref{BB9}.

    \begin{figure}[h!]
       \centering

       \caption{Two examples of boundary 11 contributing to a gluing term, with both orientations of $\gamma_2$.}
       \label{BB11}
   \end{figure}

    \item Boundary 12: At this boundary
    \begin{align*}
        \frac{v_1^a}{w_1^a} &= \frac{v_{l}^b}{w_{l}^b} \,.
    \end{align*}
    This implies $w_{l}^b=\beta v_{l}^b$ and from 
    \begin{align*}
        \frac{u_{l}^b}{w_{l}^b} =& \frac{1}{\beta}\frac{u_{l}^b}{v_{l}^b} = \frac{\alpha}{\beta}
    \end{align*}
    we get $\frac{u_{l}^b}{v_{l}^b}=\alpha=\frac{u_m^c}{v_m^c}$, so we find that this boundary is equivalent to boundary 9. It is only when $m=0$ that boundary 9 does not exist and we have to consider this one. $\int_{\partial\mathbb{W}_{k,l,m,n}}\Omega_{k,l,m,n}^{a}$ does not yield a familiar $A_\infty$-term on this boundary and it does not vanish either. The disk diagrams at this boundary resemble the ones in Fig. \ref{BB9} when there are no lines attached to the $c$-leg. An example of a disk diagram at this
    boundary is shown in Fig. \ref{BB12}. Gluing terms corresponding to the type of diagrams on the right cancel with gluing terms belonging to the type of diagrams on the right of Fig. \ref{BB11}.

    \begin{figure}[h!]
       \centering

       \caption{Two examples of boundary 12 contributing to a gluing term, with both orientations of $\beta_2$.}
       \label{BB12}
   \end{figure}

    \item Boundary 13: At this boundary
    \begin{align*}
        \frac{u_1^b}{w_1^b} =& \frac{u_{k+n+1}^a}{w_{k+n+1}^a}\,.
    \end{align*}
    This implies $w_{k+n+1}^a=\frac{\beta}{\alpha}u_{k+n+1}^a$ and from
    \begin{align*}
        \frac{v_{k+n+1}^a}{w_{k+n+1}^a}=\frac{\alpha}{\beta}\frac{v_{k+n+1}^a}{u_{k+n+1}^a}=\frac{1}{\beta}
    \end{align*}
    we get $\frac{u_{k+n+1}^a}{v_{k+n+1}^a}=\alpha=\frac{u_1^c}{v_1^c}$, so we find that this boundary is equivalent to boundary 10. It is only when $m=0$ that boundary 10 does not exist and we have to consider this one. $\int_{\partial\mathbb{W}_{k,l,m,n}}\Omega_{k,l,m,n}^{a}$ does not yield a familiar $A_\infty$-term on this boundary and it does not vanish either. The disk diagrams at this boundary resemble the ones in Fig. \ref{BB10} when there are no lines attached to the $c$-leg. An example of a disk diagram at this
    boundary is shown in Fig. \ref{BB13}. Gluing terms corresponding to the type of diagrams on the left cancel with gluing terms belonging to the type of diagrams on the left of Fig. \ref{BB12}, while the type of diagrams on the right cancel with gluing terms corresponding to the left of Fig. \ref{BB11}.
       \begin{figure}[h!]
       \centering

       \caption{Two examples of boundary 13 contributing to a gluing term, with both orientations of $\delta_1$.}
       \label{BB13}
   \end{figure}

    \item Boundary 14: At this boundary
    \begin{align*}
        \frac{u_{m}^c}{w_{m}^c} =& \frac{u_{l}^b}{w_{l}^b}\,.
    \end{align*}
    This implies $u_{m}^c=\frac{\alpha}{\beta}w_{m}^c$ and from
    \begin{align*}
        \frac{u_m^c}{v_m^c}=\frac{\alpha}{\beta}\frac{w_m^c}{v_m^c}=\alpha
    \end{align*}
    we get $\frac{v_m^c}{w_m^c}=\frac{1}{\beta}=\frac{v_{k+n+1}^a}{w_{k+n+1}^a}$, so we find that this boundary is equivalent to boundary 11. Boundary 11 does not exist when $m=0$, but neither does this one. This means that this boundary is always equivalent to boundary 11.

\end{itemize}

\paragraph{Zero measure terms.}

\begin{itemize}
    \item Boundary 15: At this boundary
    \begin{align*}
        \frac{v_{i}^c}{w_{i}^c}=\frac{v_{i+1}^c}{w_{i+1}^c}\,, \quad \text{for } i=1,\dots,m-1 \,,
    \end{align*}
    which makes the vectors $\vec{q}_{c,i}$ and $\vec{q}_{c,i+1}$ colinear and the measure in $\Omega_{k,l,m,n}^{a}$ vanishes on this boundary.
    
    \item Boundary 16: At this boundary
    \begin{align*}
        \frac{v_{i}^b}{w_i^b}=\frac{v_{i+1}^b}{w_{i+1}^b}\,, \quad \text{for } i=1,\dots,l-1 \,,
    \end{align*}
    which makes the vectors $\vec{q}_{b,i}$ and $\vec{q}_{b,i+1}$ colinear and the measure in $\Omega_{k,l,m,n}^{a}$ vanishes on this boundary.
    
    \item Boundary 17: At this boundary
    \begin{align*}
        \frac{v_i^a}{w_i^a}=\frac{v_{i+1}^a}{w_{i+1}^a}\,, \quad \text{for } i=1,\dots,k+n\,,
    \end{align*}
    which makes the vectors $\vec{q}_{a,i}$ and $\vec{q}_{a,i+1}$ colinear and the measure in $\Omega_{k,l,m,n}^{a}$ vanishes on this boundary.
\end{itemize}

\paragraph{Higher codimension boundaries.}

\begin{itemize}
\item Boundary 18: At this boundary
\begin{align*}
    u_i^\bullet=1 \,.
\end{align*}
The closure constraint implies that all other $u$-variables are zero. If $m=1$ and $k=l=n=0$, the boundary is given by $u_1^c=1$, which implies $v_1^c=w_1^c=1$ through
\begin{align*}
    0 \leq u_1^c \leq v_1^c \leq w_1^c \leq 1\,.
\end{align*}

\item Boundary 19: At this boundary
\begin{align*}
    u_1^c =& w_1^c \,.
\end{align*}
The $uw$-chain becomes
\begin{align*}
    1=\frac{u_1^c}{w_1^c} \leq \dots \leq \frac{u_m^c}{w_m^c} \leq \frac{u_{l}^b}{w_{l}^b} = \dots = \frac{u_1^b}{w_1^b} \leq \frac{u_{k+n+1}^a}{w_{k+n+1}^a} \leq \dots \leq \frac{u_1^a}{w_1^a}  \,.
\end{align*}
The closure constraint implies that this is a higher codimension boundary, except when $m=1$ and $k+n=0$, in which case $u_i^\bullet=w_i^\bullet$. The $uv$-chain and $vw$-chain then read
\begin{align*}
    0 \leq \frac{u_1^b}{v_1^b} \leq \dots \leq \frac{u_l^b}{v_l^b} \leq \frac{u_1^c}{v_1^c} \leq \frac{u^a_{1}}{v_{1}^a} \leq \infty \,,\\
    0 \leq \frac{v_1^c}{w_1^c} \leq \frac{v_{1}^a}{w_{1}^a} \leq \frac{v_l^b}{w_l^b} \leq \dots \leq \frac{v_{1}^b}{w_{1}^b} \leq \infty
\end{align*}
and contradict each other when setting $u_i^\bullet=w_i^\bullet$, thus we find a higher codimension boundary.

\item Boundary 20: At this boundary
\begin{align*}
    u_{l}^b =& w_{l}^b \,.
\end{align*}
This is only a boundary when $m=0$. The $uw$-chain becomes
\begin{align*}
    1=& \frac{u_{l}^b}{w_{l}^b} = \dots = \frac{u_1^b}{w_1^b} \leq \frac{u_{k+n+1}^a}{w_{k+n+1}^a} \leq  \dots \leq \frac{u_1^a}{w_1^a}\leq \infty\,,
\end{align*}
This yields a higher codimension boundary, unless $k=n=0$, in which case this boundary is equivalent to boundary boundary 13.

\item Boundary 21: At this boundary
    \begin{align*}
        u_1^b &= v_1^b \,.
    \end{align*}
    The $uv$-chain becomes
    \begin{align*}
        1 = \frac{u_1^b}{v_1^b} = \dots = \frac{u_l^b}{v_l^b} \leq \frac{u_m^c}{v_m^c} = \dots = \frac{u_1^c}{v_1^c} \leq \frac{u_{k+n+1}^a}{v_{k+n+1}^a} \leq \dots \leq \frac{u_1^a}{v_1^a}\leq \infty\,.
    \end{align*}
    which implies $u_i^\bullet \geq v_i^\bullet$. The closure constraint leads to a higher codimension boundary, except when $l=1$, $k=n=0$. However, in this case the $uw$-chain and $vw$-chain become
    \begin{align*}
        0 &\leq \frac{u_1^c}{w_1^c} \leq \dots \leq \frac{u_{m}^c}{w_{m}^c} \leq \frac{u_1^b}{w_1^b} \leq \frac{u_{1}^a}{w_1^a} \leq \infty \,,\\
        0 &\leq \frac{v_1^c}{w_1^c} \leq \dots \leq \frac{v_m^c}{w_m^c} \leq \frac{v_{1}^a}{w_{1}^a} \leq \frac{v_{1}^b}{w_{1}^b} \leq \infty
    \end{align*}
   and contradict each other when setting $u_i^\bullet=v_i^\bullet$. Thus, we find a higher codimension boundary.

\item Boundary 22: At this boundary
\begin{align*}
    \frac{u_m^c}{w_m^c}=\frac{u_{k+n+1}^a}{w_{k+n+1}^a}\,.
\end{align*}
This is a boundary only if $l=0$. Since
\begin{align*}
    \frac{u_m^c}{v_m^c}\frac{v_m^c}{w_m^c} =& \frac{u_{m}^c}{w_{m}^c} \leq \frac{u_{k+n+1}^a}{w_{k+n+1}^a} =\frac{u_{k+n+1}^a}{v_{k+n+1}^a}\frac{v_{k+n+1}^a}{w_{k+n+1}^a}
\end{align*}
and
\begin{align*}
    \frac{u_m^c}{v_m^c} \leq& \frac{u_{k+n+1}^a}{v_{k+n+1}^a} \,, \quad \frac{v_m^c}{w_m^c} \leq \frac{v_{k+n+1}^a}{w_{k+n+1}^a} \,,
\end{align*}
this implies $\frac{u_m^c}{v_m^c} = \frac{u_{k+n+1}^a}{v_{k+n+1}^a}$ and $\frac{v_m^c}{w_m^c} = \frac{v_{k+n+1}^a}{w_{k+n+1}^a}$ and thus yields a higher codimension boundary.

\item Boundary 23: At this boundary
\begin{align*}
    \frac{u_{l}^b}{v_{l}^b}=\frac{u_{k+n+1}^a}{v_{k+n+1}^a}\,.
\end{align*}
This is only a boundary for $m=0$. Since
\begin{align*}
    \frac{u_l^b}{w_l^b}/\frac{v_l^b}{w_l^b}=\frac{u_l^b}{v_l^b}\leq \frac{u_{k+n+1}^a}{v_{k+n+1}^a}=\frac{u_{k+n+1}^a}{w_{k+n+1}^a}/\frac{v_{k+n+1}^a}{w_{k+n+1}^a}
\end{align*}
and
\begin{align*}
    \begin{aligned}
        \frac{u_l^b}{w_l^b} &\leq \frac{u_{k+n+1}^a}{w_{k+n+1}^a} \,, & \frac{v_{k+n+1}^a}{w_{k+n+1}^a} &\leq\frac{v_l^b}{w_l^b} \,,
    \end{aligned}
\end{align*}
this implies $\frac{u_l^b}{w_l^b} = \frac{u_{k+n+1}^a}{w_{k+n+1}^a} $ and  $\frac{v_{k+n+1}^a}{w_{k+n+1}^a} = \frac{v_l^b}{w_l^b}$ and thus yields a higher codimension boundary.
\end{itemize}

As can be seen from the disk diagrams for $A_\infty$-relations in Fig. \ref{A_inftyDiagrams} or in the corresponding expression \eqref{generalA_infty}, \eqref{Qs}, 
\begin{align*}
    \mathcal{V}(\bullet,\dots,\bullet,\mathcal{V}(\bullet,\dots,\bullet,a,\bullet,\dots,\bullet,b,\bullet,\dots,\bullet),\bullet,\dots,\bullet,c,\bullet,\dots,\bullet)
\end{align*}
and
\begin{align*}
    \mathcal{V}(\bullet,\dots,\bullet,a,\bullet,\dots,\bullet,\mathcal{V}(\bullet,\dots,\bullet,b,\bullet,\dots,\bullet,c,\bullet,\dots,\bullet),\bullet,\dots,\bullet)
\end{align*}
are related to each other by reversing the nested boundary ordering and swapping $a \leftrightarrow c$ and so are 
\begin{align*}
    \mathcal{V}(\bullet,\dots,\bullet,a,\bullet,\dots,\bullet,b,\bullet,\dots,\bullet,\mathcal{U}(\bullet,\dots,\bullet,c,\bullet,\dots,\bullet),\bullet,\dots,\bullet)
\end{align*}
and
\begin{align*}
    \mathcal{V}(\bullet,\dots,\bullet,\mathcal{U}(\bullet,\dots,\bullet,a,\bullet,\dots,\bullet),\bullet,\dots,\bullet,b,\bullet,\dots,\bullet,c,\bullet,\dots,\bullet) \,.
\end{align*}
One can take the vectors $\vec{q}^{\,\,1}_i=(q^1_{u,i},q^1_{v,i},q^1_{w,i})$ and $\vec{q}^{\,\,2}_i=(q^2_{u,i},q^2_{v,i},q^2_{w,i})$ from the first $A_\infty$-term of both of the pairs mentioned and replace them by $\vec{q}^{\,\,'1}_i=(q^1_{w,1},q^1_{v,i},q^1_{u,i})$ and $\vec{q}^{\,\,'2}=(q^2_{w,i},q^2_{v,i},q^2_{u,i})$. The expressions for the entries of these vectors can be found in \eqref{Qs}. Since, after reversing the nested boundary ordering, the labeling of the vectors $\vec{q}^{\,\,'1}$ and $\vec{q}^{\,\,'2}$ does not match the ordering of the corresponding $A_\infty$-terms in the matrix $Q$, the ordering needs to be adjusted. The matrix becomes
\begin{align*}
    Q &= (
         \vec{q}_a , \vec{q}_b , \vec{q}_c , \vec{q}^{\,\,'1}_{r} , \dots , \vec{q}^{\,\,'1}_1 , \vec{q}^{\,\,'2}_{s} , \dots , \vec{q}^{\,\,'2}_1
    )\,.
\end{align*}
Lastly, one applies the $\mathbb{Z}_2$-transformation on both $u^1$ and $v^1$ and $u^2$ and $v^2$ variables in the first pair, while in the second pair one only applies the transformation on the $u^2$ and $v^2$ variables. One could say that the $A_\infty$-terms in these pairs are mirror images of each other. Similarly,
\begin{align*}
    \mathcal{V}(\bullet,\dots,\bullet,a,\bullet,\dots,\bullet,\mathcal{U}(\bullet,\dots,\bullet,b,\bullet,\dots,\bullet),\bullet,\dots,\bullet,c,\bullet,\dots,\bullet)
\end{align*}
simply returns to an $A_\infty$-term of the same type after applying these operations, so these $A_\infty$-terms are mirror images of themselves.

The potentials $\Omega_{k,l,m,n}^a$ and $\Omega_{m,l,k,n}^c$ are related through almost the same operations: one reverses the boundary ordering and swaps $a \leftrightarrow c$ and $\vec{q}_{a,i} \leftrightarrow \vec{q}_{c,i}$. This is not yet the same as for the $A_\infty$-terms, but after evaluating the potentials on the boundary, one can apply a $\mathbb{Z}_2$-transformation on the variables $u^1,v^1$ and/or $u^2,v^2$. It is now easy to see that the boundaries in the categorization above that gave $A_\infty$-terms, will give the $A_\infty$-terms related by the above relation on the same boundaries, but for the potential $\Omega_{m,l,k,n}^c$. Moreover, the gluing terms coming from $\Omega_{m,l,k,n}^c$ can be related in a similar way and can be easily seen to vanish among themselves or with the gluing terms from $\Omega_{k,l,m,n}^a$. The boundaries that yielded zero measure terms and higher codimension boundary terms will do so again and they will therefore not contribute. We will now list the boundaries that produce $A_\infty$-terms and gluing terms for $\Omega_{m,l,k,n}^c$. For this, remember that the matrix $Q$ is given by \eqref{Q'} for these potentials.

\paragraph{$c$-diagrams.}
\begin{itemize}
    \item Boundary 1: At this boundary
    \begin{align*}
        u_{i}^\bullet=0 \,.
    \end{align*}
    This leads to a higher codimension boundary, except when $u_i^c=0$ for $i=1,\dots,m$ and $u_i^b=0$ for $i=1,\dots,l$, in which case it leads to $\alpha=0$. After the change of coordinates
    \begin{align*}
        \begin{aligned}
            v_i^c &\rightarrow u^1_i \,, & w_i^c &\rightarrow v^1_i \,, & & \text{for } i=1,\dots,m \,,\\
            v_i^b &\rightarrow u^1_{m+l+2-i} \,, & w_i^b &\rightarrow v^1_{m+l+2-i} \,, & & \text{for } i=1,\dots, l \,,\\
            u_{i}^a &\rightarrow  u^2_{n+k+2-i} \,, & v_{i}^a &\rightarrow u^1_{m+1}v^2_{n+k+2-i} \,, & w_i^a &\rightarrow v_{m+1}^1v_{n+k+2-i}^2\,, \, & \text{for } i=1,\dots,k+n+1 \,,
        \end{aligned}
    \end{align*}
    and a $\mathbb{Z}_2$-transformation on both integration domains, the boundary is identified as $\mathbb{V}_{k+n}\times\mathbb{V}_{m+l}$. We also change the matrix $Q$,such that it corresponds to the canonical ordering for nested vertices. It then reads
    \begin{align*}
        Q =& (\vec{q}_a,\vec{q}_b,\vec{q}_c,\vec{q}_{b,1}^{\,\,'},\dots,\vec{q}_{b,l}^{\,\,'},\vec{q}_{c,m}^{\,\,'},\dots,\vec{q}_{c,1}^{\,\,'},\vec{q}_{a,1}^{\,\,'},\dots,\vec{q}_{a,k+n+1}^{\,\,'}) \,.
    \end{align*}
    It can now be seen that
    \begin{align*}
        \boxed{
        \begin{aligned}
        &\int_{\partial\mathbb{W}_{m,l,k,n}}\Omega_{m,l,k,n}^c|_{u_i^c=u_i^b=0} \sim\\
        &\sim\mathcal{V}(\bullet,\dots,\bullet,\mathcal{V}(\bullet,\dots,\bullet,a,\bullet,\dots,\bullet,b,\bullet,\dots,\bullet),\bullet,\dots,\bullet,c,\bullet,\dots,\bullet)
        \end{aligned}}
    \end{align*}
    on this boundary, with the exception of $\mathcal{V}(\bullet,\dots,\bullet,\mathcal{V}(a,b),\bullet,\dots,\bullet,c,\bullet,\dots,\bullet)$. An example of a disk diagram at this boundary is shown in Fig. \ref{BB1C}.
    \begin{figure}[h!]
        \centering

        \caption{An example of boundary 1 contributing to $\mathcal{V}(\mathcal{V}(\alpha_1,a,\alpha_2,\alpha_3,\alpha_4,b,\alpha_5),\alpha_6,c)$.}
        \label{BB1C}
    \end{figure}

    \item Boundary 2: At this boundary
    \begin{align*}
        u_1^c=v_1^c \,.
    \end{align*}
    \begin{figure}[h!]
        \centering
        %
        \caption{An example of boundary 2 contributing to $\mathcal{V}(\alpha_1,a,\alpha_2,\mathcal{V}(b,c))$.}
        \label{BB2C}
    \end{figure}
    This is only a boundary if $l=0$, in which case it is a higher codimension boundary unless $k=n=0$, like in the left-ordered case. The $uv$-chain then reads
    \begin{align*}
        1= \frac{u_m^c}{v_m^c} \leq \dots \leq \frac{u_1^c}{v_1^c} \leq \frac{u_1^a}{v_1^a}\,.
    \end{align*}
    The closure constraint now forces $u_i^\bullet=v_i^\bullet$. Then, after a change of coordinates
    \begin{align*}
        \begin{aligned}
            v_i^c &\rightarrow u^2_i \,, & w_i^c &\rightarrow v^2_i \,, & \text{for } i=1,\dots,m \,,\\
            u_1^a &\rightarrow u^2_{m+1} \,, & w_1^a &\rightarrow v^2_{m+1} \,,
        \end{aligned}
    \end{align*}
    and a $\mathbb{Z}_2$-transformation on the remaining variables, the boundary is identified as $\mathbb{V}_m\times\mathbb{V}_0$ . We also change the matrix $Q$, such that it corresponds to the canonical ordering for nested vertices. It then reads
    \begin{align*}
        Q =& (\vec{q}_a,\vec{q}_b,\vec{q}_c,\vec{q}_{a,1}^{\,\,'},\vec{q}_{c,m}^{\,\,'},\dots,\vec{q}_{c,1}^{\,\,'}) \,.
    \end{align*}
    It can now be seen that
    \begin{align*}
        \boxed{\int_{\partial\mathbb{W}_{0,0,m,0}}\Omega_{m,0,0,0}^c|_{u_1^c=v_1^c} \sim \mathcal{V}(a,\bullet,\dots,\bullet,\mathcal{V}(b,c),\bullet,\dots,\bullet)}
    \end{align*}
    on this boundary. Since the line connected to the output arrow can only have one orientation, there is only one disk diagram at this boundary shown in Fig. \ref{BB2C}.

    \item Boundary 3: At this boundary
    \begin{align*}
        v_1^c=w_1^c\,.
    \end{align*}
    \begin{figure}[h!]
    \centering

    \caption{An example of boundary 3 contributing to $\mathcal{V}(\alpha_1,\mathcal{V}(a,b),\alpha_2,c)$.}
    \label{BB3C}
\end{figure}
    The $vw$-chain becomes
    \begin{align*}
        1 =& \frac{v_1^c}{w_1^c} \leq \dots \leq \frac{v_m^c}{w_m^c} \leq \frac{v_{k+n+1}^a}{w_{k+n+1}^a} = \dots = \frac{v_1^a}{w_1^a}\leq \frac{v_{l}^b}{w_{l}^b} \leq \dots \leq \frac{v_1^b}{w_1^b} \leq \infty \,.
    \end{align*}
    and implies $v_i^\bullet \geq w_i^\bullet$. The closure constraint then leads to $v_i^\bullet=w_i^\bullet$, which gives a higher codimension boundary, except for $l=0, m=1$. Then after the change of coordinates
    \begin{align*}
        \begin{aligned}
            u_1^c &\rightarrow u^2_1 \,, & v_1^c &\rightarrow v^2_1 \,,\\
            u_i^a &\rightarrow u^2_{k+n+3-i} \,, & v_i^a &\rightarrow v^2_{k+n+3-i} \,, & \text{for } i=1,\dots,k+n+1\,,\\
        \end{aligned}
    \end{align*}
    and a $\mathbb{Z}_2$-transformation on the remaining variables, the boundary is identified as $\mathbb{V}_{k+n+1}\times\mathbb{V}_0$. We also change the matrix $Q$, such that it corresponds to the canonical ordering for nested vertices. It then reads
    \begin{align*}
        Q =& (\vec{q}_a,\vec{q}_b,\vec{q}_c,\vec{q}_{a,1}^{\,\,'},\dots,\vec{q}_{a,k+n+1}^{\,\,'},\vec{q}_{c,1}^{\,\,'}) \,.
    \end{align*}
    It can now be seen that
    \begin{align*}
        \boxed{\int_{\partial\mathbb{W}_{1,0,k,n}}\Omega_{1,0,k,n}^c|_{v_1^c=w_1^c} \sim \mathcal{V}(\bullet,\dots,\bullet,\mathcal{V}(a,b),\bullet,\dots,\bullet,c,\bullet,\dots,\bullet)}
    \end{align*}    
    on this boundary. An example of a disk diagram at this boundary contributing to the $A_\infty$-term is shown in Fig. \ref{BB3C}, while Fig. \ref{BB3GC} shows a disk diagram contributing to a gluing term.

\begin{figure}[h!]
    \centering

    \caption{An example of boundary 3 contributing to a gluing term.}
    \label{BB3GC}
\end{figure}

    \item Boundary 4: At this boundary
    \begin{align*}
        v_1^b &= w_1^b \,.
    \end{align*}
    \begin{figure}[h!]
        \centering
        %
        \caption{An example of boundary 4 contributing to $\mathcal{V}(\mathcal{V}(a,b),\alpha_1,\alpha_2,\alpha_3)$.}
        \label{BB4C}
    \end{figure}
    The $vw$-chain becomes
    \begin{align*}
         0 \leq& \frac{v_1^c}{w_1^c} \leq \dots \leq \frac{v_m^c}{w_m^c} \leq \frac{v_{k+n+1}^a}{w_{k+n+1}^a} =\dots= \frac{v_1^a}{w_1^a}\leq \frac{v_{l}^b}{w_{l}^b} \leq \dots \leq \frac{v_1^b}{w_1^b} =1
    \end{align*}
    and implies $v_i^\bullet \leq w_i^\bullet$. The closure constraint then leads to $v_i^\bullet=w_i^\bullet$, which leads to a higher codimension boundary, except for $l=1$, $m=0$. Then, after the change of coordinates
    \begin{align*}
        \begin{aligned}
            u_1^b &\rightarrow u^2_1 \,, & v_1^b &\rightarrow v^2_1 \,,\\
            u_i^a &\rightarrow u^2_{k+n+3-i} \,, & v_i^a &\rightarrow v^2_{k+n+3-i} \,, & \text{for } i=1,\dots,k+n+1 \,,\\
        \end{aligned}
    \end{align*}
    and a $\mathbb{Z}_2$-transformation on the remainging coordinates, the boundary is identified as $\mathbb{V}_{k+n+1}\times\mathbb{V}_0$. We also change the matrix $Q$, such that it corresponds to the canonical ordering for nested vertices. It then reads
    \begin{align*}
        Q =& (\vec{q}_a,\vec{q}_b,\vec{q}_c,\vec{q}_{a,1}^{\,\,'},\dots,\vec{q}_{a,k+n+1}^{\,\,'},\vec{q}_{b,1}^{\,\,'}) \,.
    \end{align*}
    It can now be seen that
    \begin{align*}
        \boxed{\int_{\partial\mathbb{W}_{0,1,k,n}}\Omega_{0,1,k,n}^c|_{v_1^b=w_1^b} \sim \mathcal{V}(\bullet,\dots,\bullet,\mathcal{V}(a,b),\bullet,\dots,\bullet,c,\bullet,\dots,\bullet)}
    \end{align*}    
    on this boundary. An example of a disk diagram at this boundary contributing to the $A_\infty$-term is shown in Fig. \ref{BB4C}, while Fig. \ref{BB4GC} shows a disk diagram contributing to a gluing term.
    \begin{figure}[h!]
        \centering

        \caption{An example of boundary 4 contributing to a gluing term.}
        \label{BB4GC}
    \end{figure}

    \item Boundary 5: At this boundary
    \begin{align*}
        w_i^\bullet=1\,.
    \end{align*}
    The closure constraint forces all other $w$-variables to be zero. This yields a higher codimenion boundary, except for $m=1$ and $w_1^c=1$, which leads to $\beta=0$. After the change of coordinates
    \begin{align*}
        \begin{aligned}
            u_1^c &\rightarrow u^2_{l+1} \,, & v_1^c &\rightarrow v^2_{l+1}\,,\\
            u_i^b & \rightarrow u^2_{i} \,, & v_i^b &\rightarrow v^2_{i} \,, & \text{for } i=1,\dots,l\,,\\
            u_{i}^a &\rightarrow u^2_{k+n+l+3-i} \,, & v_i^a &\rightarrow v^2_{k+n+l+3-i} \,, & \text{for } i=1,\dots,k+n+1 \,.
        \end{aligned}
    \end{align*}
    and a $\mathbb{Z}_2$-transformation on both integration domains, the boundary is identified as $\mathbb{V}_{k+l+n+1}\times\mathbb{V}_0$. We also change the matrix $Q$, such that it corresponds to the canonical ordering for nested vertices. It then reads
    \begin{align*}
        Q =& (\vec{q}_a,\vec{q}_b,\vec{q}_c,\vec{q}_{c,1}^{\,\,'},\vec{q}_{a,1}^{\,\,'},\dots,\vec{q}_{a,k+n+1}^{\,\,'},\vec{q}_{b,l},\dots,\vec{q}_{b,1}^{\,\,'}) \,.
    \end{align*}
    It can now be seen that
    \begin{align*}
        \boxed{\int_{\partial\mathbb{W}_{1,l,k,n}}\Omega_{1,l,k,n}^c|_{w_1^c=1} \sim \mathcal{V}(\bullet,\dots,\bullet,\mathcal{U}(a,\bullet),\bullet,\dots,\bullet,b,\bullet,\dots,\bullet,c,\bullet,\dots,\bullet)}
    \end{align*}
    and
    \begin{align*}
        \boxed{\int_{\partial\mathbb{W}_{1,l,k,n}}\Omega_{1,l,k,n}^c|_{w_1^c=1} \sim \mathcal{V}(\bullet,\dots,\bullet,\mathcal{U}(\bullet,a),\bullet,\dots,\bullet,b,\bullet,\dots,\bullet,c,\bullet,\dots,\bullet)}
    \end{align*}
    on this boundary. An example of a disk diagram at this boundary is shown in Fig. \ref{BB5C}.

    \begin{figure}[h!]
        \centering

        \caption{An example of boundary 5 contributing to $\mathcal{V}(\mathcal{U}(a,\alpha_1),\alpha_2,b,\alpha_3,\alpha_4,c)$.}
        \label{BB5C}
    \end{figure}
    
    \item Boundary 6: At this boundary
    \begin{align*}
        w_i^\bullet=0 \,.
    \end{align*}
    For $m=1$ and $w_1^c=1$, this is the same as boundary 5. Otherwise, if $w_i^c=0$, the $uw$-chain is equivalent to
    \begin{align*}
        0 \geq& \frac{w_1^a}{u_1^a} \geq \dots \geq \frac{w_{k+n+1}^a}{u_{k+n+1}^a} \geq \frac{w_{1}^b}{u_{1}^b} = \dots = \frac{w_l^b}{u_l^b} \geq \frac{w_{m}^a}{u_{m}^c} \geq \dots \geq \frac{0^+}{u_i^c} \geq \dots \geq \frac{w_1^c}{u_1^c}\geq \infty \,,
    \end{align*}
    which forces $w_j^c=0$ for $j>i$ and $w_k^a=w_k^b=0$ for any $k$. This leads to a higher codimension boundary. Only when we consider $w_i^a=0$ for $i=1,\dots,k+n+1$ and $w_i^b=0$ for $i=1,\dots,l$, we find an $A_\infty$-term. Then, after the change of coordinates
    \begin{align*}
        \begin{aligned}
            u_1^c &\rightarrow u^2_{l+1} u^1_1 \,,\\
            v_i^c &\rightarrow v^2_{l+1} u^1_i \,, & w_i^c & \rightarrow v^1_i \,, & \text{for } i=1,\dots,m\\
            u_i^b &\rightarrow u^2_{i} \,, & v_i &\rightarrow v^2_{i} \,, & \text{for } i=1,\dots, l \,,\\
            u_{i}^a & \rightarrow u^2_{k+l+n+3-i} \,, & v_{i}^a & \rightarrow v^2_{k+l+n+3-i} \,, & \text{for } i=1,\dots,k+n+1
        \end{aligned}
    \end{align*}
    and a $\mathbb{Z}_2$-transformation on the remaining variables, the boundary is identified as $\mathbb{V}_{k+l+n+1}\times\mathbb{V}_{m-1}$. We also change the matrix $Q$, such that it corresponds to the canonical ordering for nested vertices. It then reads
    \begin{align*}
        Q =& (\vec{q}_a,\vec{q}_b,\vec{q}_c,\vec{q}_{c,m}^{\,\,'},\dots,\vec{q}_{c,1}^{\,\,'},\vec{q}_{a,1}^{\,\,'},\dots,\vec{q}_{a,k+n+1}^{\,\,'},\vec{q}_{b,l}^{\,\,'},\dots,\vec{q}_{b,1}^{\,\,'}) \,.
    \end{align*}
    It can now be seen that
    \begin{align*}
        \boxed{\int_{\partial\mathbb{W}_{m,l,k,n}}\Omega_{m,l,k,n}^c|_{w_{i}^a=w_i^b=0} \sim \mathcal{V}(\bullet,\dots,\bullet,\mathcal{U}(\bullet,\dots,\bullet,a,\bullet,\dots,\bullet),\bullet,\dots,\bullet,b,\bullet,\dots,\bullet,c,\bullet,\dots,\bullet)}
    \end{align*}
    on this boundary, with the exception of $\mathcal{V}(\bullet,\dots,\bullet,\mathcal{U}(a,\bullet),\bullet,\dots,\bullet,b,\bullet,\dots,\bullet,c,\bullet,\dots,\bullet)$ and $\mathcal{V}(\bullet,\dots,\bullet,\mathcal{U}(\bullet,a),\bullet,\dots,\bullet,b,\bullet,\dots,\bullet,c,\bullet,\dots,\bullet)$. An example of a disk diagram at this boundary is shown in Fig. \ref{BB6C}.

    \begin{figure}[h!]
        \centering

        \caption{An example of boundary 6 contributing to $\mathcal{V}(\mathcal{V}(a,\alpha_1,\alpha_2),\alpha_3,b,\alpha_4,\alpha_5,c)$.}
        \label{BB6C}
    \end{figure}

    \item Boundary 7: At this boundary
    \begin{align*}
        v_i^\bullet=1\,.
    \end{align*}
    \begin{figure}[h!]
        \centering
        %
        \caption{An example of boundary 7 contributing to $\mathcal{V}(a,\alpha_1,\alpha_2,\mathcal{U}(b,\alpha_3),\alpha_4,c)$.}
        \label{BB7C}
    \end{figure}

    The closure constraint forces all other $v$-variables to be zero. This yields a higher codimension boundary, except when $l=1$ and $v_1^b=1$. Then, after the change of variables
    \begin{align*}
        \begin{aligned}
            u_i^c &\rightarrow u^2_i \,, & w_i^c &\rightarrow v^2_i \,, & \text{for } i=1,\dots,m \,,\\
            u_1^b &\rightarrow u^2_{m+1} \,, & w_1^b &\rightarrow v^2_{m+1} \,,\\
            u_i^a &\rightarrow u^2_{k+m+n+3-i} \,, & w_i^a &\rightarrow v^2_{k+m+n+3-i} \,, & \text{for } i=1,\dots,k+n+1
        \end{aligned}
    \end{align*}
    and after a $\mathbb{Z}_2$-transformation on the remaining variables, the boundary is identified as $\mathbb{V}_{k+m+n+1}\times\mathbb{V}_0$. We also change the matrix $Q$, such that it corresponds to the canonical ordering for nested vertices. It then reads
    \begin{align*}
        Q =& (\vec{q}_a,\vec{q}_b,\vec{q}_c,\vec{q}_{b,1}^{\,\,'},\vec{q}_{a,1}^{\,\,'},\dots,\vec{q}_{a,k+n+1}^{\,\,'},\vec{q}_{c,m}^{\,\,'},\dots,\vec{q}_{c,1}^{\,\,'}) \,.
    \end{align*}
    It can now be seen that
    \begin{align*}
        \boxed{\int_{\partial\mathbb{W}_{m,1,k,n}}\Omega_{m,1,k,n}^c|_{v_{1}^b=1} \sim \mathcal{V}(\bullet,\dots,\bullet,a,\bullet,\dots,\bullet,\mathcal{U}(b,\bullet),\bullet,\dots,\bullet,c,\bullet,\dots,\bullet)}
    \end{align*}
    and
    \begin{align*}
        \boxed{\int_{\partial\mathbb{W}_{m,1,k,n}}\Omega_{m,1,k,n}^c|_{v_{1}^b=1} \sim \mathcal{V}(\bullet,\dots,\bullet,a,\bullet,\dots,\bullet,\mathcal{U}(\bullet,b),\bullet,\dots,\bullet,c,\bullet,\dots,\bullet)}
    \end{align*}
    on this boundary. An example of a disk diagram at this boundary is shown in Fig. \ref{BB7C}.

    \item Boundary 8: At this boundary 
    \begin{align*}
        v_i^\bullet=0 \,.
    \end{align*}
    \begin{figure}[h!]
        \centering

        \caption{An example of boundary 8 contributing to $\mathcal{V}(a,\alpha_1,\alpha_2,\mathcal{V}(\alpha_3,b,\alpha_4),\alpha_5,c)$.}
        \label{BB8C}
    \end{figure}
    The $vw$-chain becomes
    \begin{align*}
        \begin{aligned}
            0 &\leq \frac{v_1^c}{w_1^c} \leq \dots \leq \frac{v_m^c}{w_m^c} \leq \frac{v_{k+n+1}^a}{w_{k+n+1}^a} = \dots =\frac{0^+}{w_i^a} = \dots = \frac{v_1^a}{w_1^a}\leq \frac{v_{l}^b}{w_{l}^b} \leq \dots \leq \frac{v_1^b}{w_1^b} \leq \infty \,,\\
            0 &\leq \frac{v_1^c}{w_1^c} \leq \dots \leq \frac{v_m^c}{w_m^c} \leq \frac{v_{k+n+1}^a}{w_{k+n+1}^a} = \dots = \frac{v_1^a}{w_1^a}\leq \frac{v_{l}^b}{w_{l}^b} \leq \dots \leq \frac{0^+}{w_i^b} \dots \leq \frac{v_1^b}{w_1^b} \leq \infty
        \end{aligned}
    \end{align*}
    if $v_i^a=0$ and $v_i^b=0$, respectively. This forces all $v_i^a=0$ and $v_i^c=0$ in both cases, while in the latter we also have $v_j^b=0$ for $j>i$.  This leads to a higher codimension boundary. It is only when $v_i^c=0$ for $i=1,\dots,m$ and $v_i^a=0$ for $i=1,\dots,k+n+1$, that one finds an $A_\infty$-term. After the change of variables
     \begin{align*}
        \begin{aligned}
            u_{i}^c &\rightarrow u^2_{i} \,, & w_{i}^c &\rightarrow v^2_{i}\,, &&& \text{for } i=1,\dots,m \,,\\
            u_i^b &\rightarrow u^2_{m+1}u^1_i \,,&
            v_{i}^b &\rightarrow v^1_i \,, & w_{i}^b &\rightarrow v^2_{m+1}u^1_1\,, & \text{for } i=1,\dots,l \,,\\
            u_{i}^a &\rightarrow u^2_{k+m+n+3-i} \,, & w_{i}^a &\rightarrow v^2_{k+m+n+3-i}\,, &&& \text{for } i=1,\dots,k+n+1 \,,\\
        \end{aligned}
    \end{align*}
    and a $\mathbb{Z}_2$-transformation on the $u^2$- and $v^2$-variables, the boundary is identified as $\mathbb{V}_{k+m+n+1}\times\mathbb{V}_{l-1}$. We also change the matrix $Q$, such that it corresponds to the canonical ordering for nested vertices. It then reads
    \begin{align*}
        Q =& (\vec{q}_a,\vec{q}_b,\vec{q}_c,\vec{q}_{b,1}^{\,\,'},\dots,\vec{q}_{b,l}^{\,\,'},\vec{q}_{a,1}^{\,\,'},\dots,\vec{q}_{a,k+n+1}^{\,\,'},\vec{q}_{c,m}^{\,\,'},\dots,\vec{q}_{c,1}^{\,\,'}) \,.
    \end{align*}
    It can now be seen that
    \begin{align*}
        \boxed{\int_{\partial\mathbb{W}_{m,l,k,n}}\Omega_{m,l,k,n}^{c} \sim \mathcal{V}(\bullet,\dots,\bullet,a,\bullet,\dots,\bullet,\mathcal{U}(\bullet,\dots,\bullet,b,\bullet,\dots,\bullet),\bullet,\dots,\bullet,c,\bullet,\dots,\bullet)}
    \end{align*}
    on this boundary, with the exception of $\mathcal{V}(\bullet,\dots,\bullet,a,\bullet,\dots,\bullet,\mathcal{U}(b,\bullet),\bullet,\dots,\bullet,c,\bullet,\dots,\bullet)$ and $\mathcal{V}(\bullet,\dots,\bullet,a,\bullet,\dots,\bullet,\mathcal{U}(\bullet,b),\bullet,\dots,\bullet,c,\bullet,\dots,\bullet)$. An example of a disk diagram at this boundary is shown in Fig. \ref{BB8C}.

\end{itemize}

\paragraph{Gluing terms.}
\begin{itemize}

    \item Boundary 9: 
    
    At this boundary
    \begin{align*}
        \frac{u_m^c}{v_m^c}=\frac{u_{l}^b}{v_{l}^b}\,.
    \end{align*}
    \begin{figure}[h!]
       \centering

       \caption{Two examples of boundary 9 contributing to a gluing term, with both orientations of $\beta_2$.}
       \label{BB9C}
   \end{figure}
   $\int_{\partial\mathbb{W}_{k,l,m,n}}\Omega_{k,l,m,n}^c$ does not yield a familiar $A_\infty$-term on this boundary and it does not vanish either. An example of a disk diagram at this boundary is shown in Fig. \ref{BB9C}.

   \item Boundary 10: At this boundary
    \begin{align*}
        \frac{u_1^c}{v_1^c}=\frac{u_{k+n+1}^a}{v_{k+n+1}^a} \,.
    \end{align*}
    For $k=l=n=0$ the $uv$-chain becomes
    \begin{align*}
        0 \leq \frac{u_m^c}{v_m^c} = \dots = \frac{u_1^c}{v_1^c} = \frac{u_{1}^a}{v_{1}^a} \leq \infty\,.
    \end{align*}
    The closure constraint then gives $u_1^\bullet=v_1^\bullet$, so this boundary is equivalent to boundary 2 when $k=l=n=0$ and does not yield  a gluing term, but an $A_\infty$-term instead. Otherwise, $\int_{\partial\mathbb{W}_{m,l,k,n}}\Omega_{m,l,k,n}^c$ does not yield a familiar $A_\infty$-term on this boundary and it does not vanish either. An example of a disk diagram at this boundary is shown in Fig. \ref{BB10C}. Gluing terms corresponding to the type of diagrams on the right cancel with gluing terms
    belonging to the type of diagrams on the right of Fig. \ref{BB9C}. A special type of gluing term with no elements between the junction and the output arrow, i.e. $n=0$, is shown in Fig. \ref{BB10SC}. This is the only type of gluing term that does not cancel with any other gluing term from the potential $\Omega_{m,l,k,n}^c$, but rather from the type $\Omega_{k,l,m,n}^a$, `gluing' them together. In particular, it cancels with the type of diagrams depicted in Fig. \ref{BB10S}.

       \begin{figure}[h!]
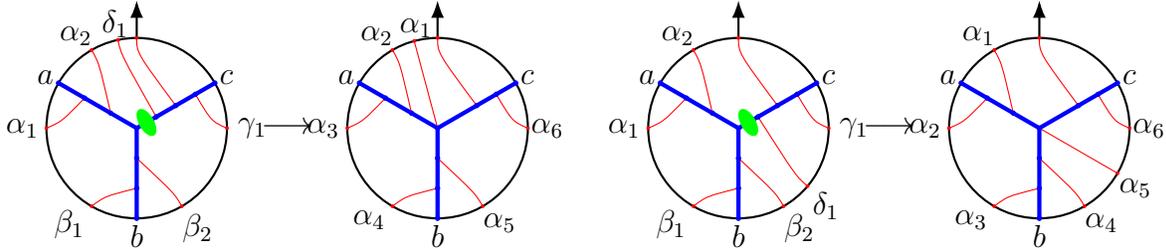

       \centering

       \caption{Two examples of boundary 10 contributing to a gluing term, with both orientations of $\delta_1$.}
       \label{BB10C}
   \end{figure}

   \begin{figure}[h!]
       \centering
       %
       \caption{Special case of a gluing term coming from boundary 10. This gluing term is cancelled by a gluing term coming from a potential of the type $\Omega_{k,l,m,n}^a$.}
       \label{BB10SC}
   \end{figure}

    \item Boundary 11: At this boundary
    \begin{align*}
        \frac{v_m^c}{w_m^c} =& \frac{v_{k+n+1}^a}{w_{k+n+1}^a}  \,.
    \end{align*}
    For $l=0$, $m=1$ the $vw$-chain reads
    \begin{align*}
         0 \leq& \frac{v_1^c}{w_1^c} = \frac{v_{k+n+1}^a}{w_{k+n+1}^a} = \dots = \frac{v_{1}^a}{w_{1}^a} \leq \infty\,.
    \end{align*}
    The closure constraint then gives $v_i^\bullet=w_i^\bullet$, so this boundary is equivalent to boundary 3 when $l=0$ and $m=1$, producing either a gluing term or an $A_\infty$-term. Otherwise, $\int_{\partial\mathbb{W}_{m,l,k,n}}\Omega_{m,l,k,n}^c$ does not yield a familiar $A_\infty$-term on this boundary and it does not vanish either. An example of a disk diagram at this boundary is shown in Fig. \ref{BB9C}. Gluing terms corresponding to the type of diagrams on the left cancel with gluing terms belonging to the type of diagrams on the left of Fig. \ref{BB11C}, while gluing terms corresponding to the type of diagrams on the right cancel with gluing terms belonging to the type of diagrams on the left of Fig. \ref{BB10C}.

    \begin{figure}[h!]
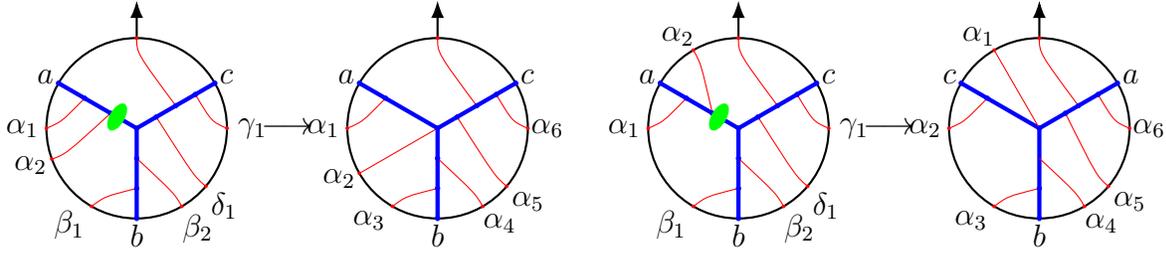

       \centering

       \caption{Two examples of boundary 11 contributing to a gluing term, with both orientations of $\gamma_2$.}
       \label{BB11C}
   \end{figure}

    \item Boundary 12: At this boundary
    \begin{align*}
        \frac{v_1^a}{w_1^a} &= \frac{v_{l}^b}{w_{l}^b} \,.
    \end{align*}
    This implies $w_{l}^b=\beta v_{l}^b$ and from 
    \begin{align*}
        \frac{u_{l}^b}{w_{l}^b} =& \frac{1}{\beta}\frac{u_{l}^b}{v_{l}^b} = \frac{\alpha}{\beta}
    \end{align*}
    we get $\frac{u_{l}^b}{v_{l}^b}=\alpha=\frac{u_m^c}{v_m^c}$, so we find that this boundary is equivalent to boundary 9. It is only when $m=0$ that boundary 9 does not exist and we have to consider this one. $\int_{\partial\mathbb{W}_{m,l,k,n}}\Omega_{m,l,k,n}^c$ does not yield a familiar $A_\infty$-term on this boundary and it does not vanish either. The disk diagrams at this boundary resemble the ones in Fig. \ref{BB9C} when there are no lines attached to the $c$-leg. An example of a disk diagram at this
    boundary is shown in Fig. \ref{BB12C}. Gluing terms corresponding to the type of diagrams on the left cancel with gluing terms belonging to the type of diagrams on the left of Fig. \ref{BB11C}.

    \begin{figure}[h!]
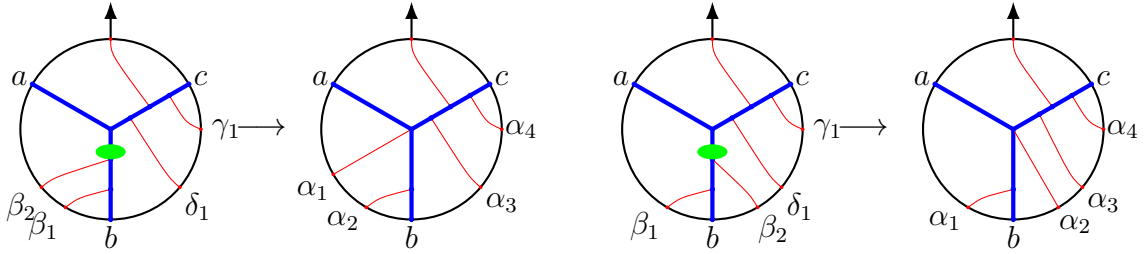

       \centering

       \caption{Two examples of boundary 12 contributing to a gluing term, with both orientations of $\beta_2$.}
       \label{BB12C}
   \end{figure}

    \item Boundary 13: At this boundary
    \begin{align*}
        \frac{u_1^b}{w_1^b} =& \frac{u_{k+n+1}^a}{w_{k+n+1}^a}\,.
    \end{align*}
    This implies $w_{k+n+1}^a=\frac{\beta}{\alpha}u_{k+n+1}^a$ and from
    \begin{align*}
        \frac{v_{k+n+1}^a}{w_{k+n+1}^a}=\frac{\alpha}{\beta}\frac{v_{k+n+1}^a}{u_{k+n+1}^a}=\frac{1}{\beta}
    \end{align*}
    we get $\frac{u_{k+n+1}^a}{v_{k+n+1}^a}=\alpha=\frac{u_1^c}{v_1^c}$, so we find that this boundary is equivalent to boundary 10. It is only when $m=0$ that boundary 10 does not exist and we have to consider this one. $\int_{\partial\mathbb{W}_{m,l,k,n}}\Omega_{m,l,k,n}^c$ does not yield a familiar $A_\infty$-term on this boundary and it does not vanish either. The disk diagrams at this boundary resemble the ones in Fig. \ref{BB10C} when there are no lines attached to the $c$-leg. An example of a disk diagram at this
    boundary is shown in Fig. \ref{BB13C}. Gluing terms corresponding to the type of diagrams on the left cancel with gluing terms belonging to the type of diagrams on the right of Fig. \ref{BB11C}, while gluing terms corresponding to the type of diagrams on the right cancel with gluing terms belonging to the type of diagrams on the right of Fig. \ref{BB12C}.
       \begin{figure}[h!]
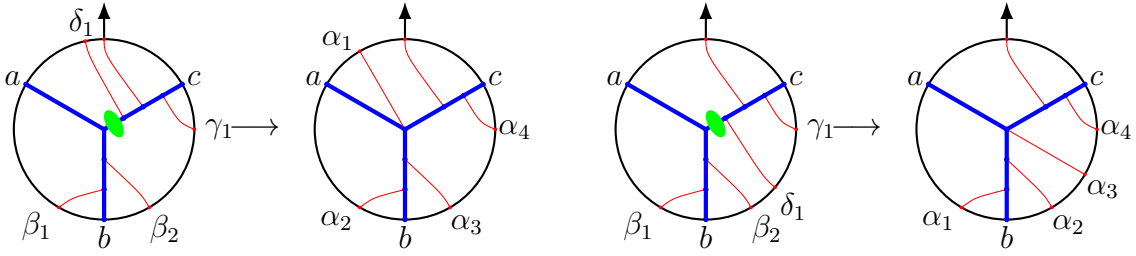

       \centering

       \caption{Two examples of boundary 13 contributing to a gluing term, with both orientations of $\delta_1$.}
       \label{BB13C}
   \end{figure}

    \item Boundary 14: At this boundary
    \begin{align*}
        \frac{u_{m}^c}{w_{m}^c} =& \frac{u_{l}^b}{w_{l}^b}\,.
    \end{align*}
    This implies $u_{m}^c=\frac{\alpha}{\beta}w_{m}^c$ and from
    \begin{align*}
        \frac{u_m^c}{v_m^c}=\frac{\alpha}{\beta}\frac{w_m^c}{v_m^c}=\alpha
    \end{align*}
    we get $\frac{v_m^c}{w_m^c}=\frac{1}{\beta}=\frac{v_{k+n+1}^a}{w_{k+n+1}^a}$, so we find that this boundary is equivalent to boundary 11. Boundary 11 does not exist when $m=0$, but neither does this one. This means that this boundary is always equivalent to boundary 11.

\end{itemize}

\section{Conclusions and discussion}
\label{sec:conclusions}
It has already been understood that there are tight links between higher spin gravities and deformation quantization. The first link is obvious: higher spin algebras are associative algebras resulting from deformation quantization of coadjoint orbits that correspond to irreducible representations of the space-time symmetry algebras (often-times, $so(d,2)$). Therefore, the product is given by the Kontsevich formula (in fact, Fedosov's construction suffices since the coadjoint orbits are symplectic manifolds). 

The second link to deformation quantization is more subtle. Any higher spin algebra determines the free equations of motion. The interactions are due to a certain Hochschild cohomology group being nontrivial and the next one, which contains obstructions, being trivial. In the simplest case of the Weyl algebra $A_1$, it is the group $HH^2(A_1, A^*_1)$ that leads to Chiral Theory. The group $HH^2(A_1, A^*_1)$ is one dimensional and the cocycle can be obtained from Shoikhet--Tsygan--Kontsevich's formality \cite{Kontsevich:1997vb, Tsygan,Shoikhet:2000gw}. However, a nontrivial cocycle is not yet a vertex, it only justifies its existence. There does not seem any simple way to generate any vertices directly from the formality.\footnote{The cubic vertex can formally be written in a factorized form with the help of the cocycle \cite{Sharapov:2019vyd}, but this form is forbidden by additional physical assumptions, e.g. the existence of the smooth flat limit. } 

Another link to formality is the very form of the vertices: they are represented by graphs similar to Kontsevich's ones with certain weights. Since the Poisson structure $\epsilon^{AB}$ is constant for our case, there are no genuine bulk vertices and all the graphs have legs on the boundary. These graphs can be re-summed to give the final result presented in Sec. \ref{sec:CD}, see also \cite{Sharapov:2022awp,Sharapov:2022wpz,Sharapov:2022nps}. Lastly, in this paper, we managed to prove the $A_\infty$-relations via the Stokes theorem, which is a method typical for formality theorems thanks to Kontsevich. 

The arguments here and above suggest that there is a bigger picture where (Shoikhet--Tsygan--)Kontsevich formality occupies the first two floors. While this structure is yet to be found, a more specific problem is to construct new theories (or recast the old ones, e.g. conformal higher spin gravity) along the lines of this paper, i.e., to find appropriate configuration spaces for vertices and $A_\infty$-relations.

The observation \cite{Sharapov:2022wpz,Sharapov:2022nps} that Chiral Theory is essentially a Poisson sigma-model\footnote{That the $A_\infty$-algebra of Chiral theory is a pre-Calabi-Yau one implies that its symmetrization (essentially, by inserting the fields $\omega$ and $C$ into the structure maps) is a usual commutative Poisson structure (note that pre-Calabi-Yau structure is a noncommutative analog of the Poisson structure, see e.g. \cite{IYUDU202163, kontsevich2021pre}). Therefore, the equations of motion have automatically the form of those of a Poisson sigma-model, i.e. $d\omega_k=\tfrac12\partial_k\pi^{ij}(C)\,\omega_i\,\omega_j$, $dC^i=\pi^{ij}(C)\,\omega_j$, where we introduced notation $\omega_k$ and $C^i$ for the fields, which also stresses that they live in the spaces dual to each other. } -- determined by a (noncommutative) Poisson structure -- may also lead to new insight into the problem of higher spin theories. It is plausible that all of them are Poisson sigma-models too, at least at the formal level. In this regard let us note that Poisson sigma-models of a different kind have already appeared in the higher spin literature, see e.g. \cite{Engquist:2005yt,Arias:2015wha,Arias:2016agc}.

Thus, the main conclusions of the paper are: (i) there has to exist a formality that extends (Shoikhet--Tsygan--)Kontsevich formality; (ii) Chiral Theory's vertices are its elementary consequences; (iii) there should exist a two-dimensional topological model that explains all of the above at the physics' level of rigour, similar to how the Poisson sigma-model is related to the Kontsevich formality theorem \cite{Cattaneo:1999fm}. It would be interesting to give these observations more solid support in the future.

\section*{Acknowledgments}
\label{sec:Aknowledgements}
We would like to thank the anonymous Referee for making many valuable suggestions to improve the manuscript. 
The work of E. S. and R. van D. was partially supported by the European Research Council (ERC) under the European Union’s Horizon 2020 research and innovation programme (grant agreement No 101002551). A. Sh. gratefully acknowledges the financial support from the São Paulo Research Foundation (FAPESP), grant 2022/13596-8, and  the Foundation for the Advancement of Theoretical Physics and Mathematics ``BASIS''.

\appendix

\footnotesize
\providecommand{\href}[2]{#2}\begingroup\raggedright\endgroup

\end{document}